\documentclass[prb, reprint, 9pt, superscriptaddress,notitlepage, nofootinbib, longbibliography,
floatfix]{revtex4-2}
\usepackage{natbib}
\usepackage{ulem}
\usepackage{braket}
\usepackage{float}
\usepackage{enumerate}
\usepackage[colorlinks]{hyperref}
\usepackage{xcolor}
\usepackage{graphicx}% Include figure files
\usepackage{dcolumn}% Align table columns on decimal point
\usepackage{bm}% bold math
\usepackage{amsmath}
\usepackage{amssymb}
\usepackage{lipsum}
\usepackage{mathtools}
\usepackage{comment}
\usepackage{subcaption}

\renewcommand{\raggedright}{\leftskip=0pt \rightskip=0pt plus 0cm}
\usepackage{tikz}
\newcommand{\beq}{\begin{equation}}
\newcommand{\eeq}{\end{equation}}
\newcommand{\beqn}{\begin{eqnarray}}
\newcommand{\eeqn}{\end{eqnarray}}

\definecolor{LinkColor}{rgb}{0,0,1}
\hypersetup{
	colorlinks=true,
	citecolor=LinkColor,
	linkcolor=LinkColor,
	urlcolor=LinkColor
}

\begin{document}
\newcommand{\jianhao}[1]{ { \color{violet} \small (\textsf{JHZ}) \textsf{\textsl{#1}} }}
\newcommand{\yy}[1]{ { \color{red} \small (\textsf{YY}) \textsf{\textsl{#1}} }}
\newcommand{\shijun}[1]{ { \color{olive} \small (\textsf{SS}) \textsf{\textsl{#1}} }}

\newcommand{\peter}[1]{{\color{blue} peter:~#1}}

\title{Spacetime duality between sequential and measurement-feedback circuits}

\author{Tsung-Cheng Lu}
\email{tclu@umd.edu}
\affiliation{Joint Center for Quantum Information and Computer Science,
University of Maryland, College Park, Maryland 20742, USA}

\author{Sarang Gopalakrishnan}
\email{sgopalakrishnan@princeton.edu}
\affiliation{Department of Electrical and Computer Engineering,
Princeton University, Princeton, NJ 08544, USA}

\author{Yizhi You}
\email{y.you@northeastern.edu}
\affiliation{Department of Physics, Northeastern University, Boston, MA, 02115, USA}

\date{\today}
\begin{abstract}

Two prevalent approaches for preparing long-range entangled quantum states are (i) linear-depth sequential unitary (SU) circuits, which apply local unitary gates sequentially, and (ii) constant-depth measurement-feedback (MF) circuits, which employ mid-circuit measurements and conditional feedback based on measurement outcomes. Here, we establish that a broad class of SU and MF circuits are dual to each other under a spacetime rotation. We investigate this spacetime duality in the preparation of various long-range entangled states, including GHZ states, topologically ordered states, and fractal symmetry-breaking states. As an illustration, applying a spacetime rotation to a linear-depth SU circuit that implements a non-invertible Kramers-Wannier duality—originally used to prepare a 1D GHZ state—yields a constant-depth MF circuit that implements a $\mathbb{Z}_2$ symmetry gauging map, which equivalently prepares the GHZ state. Leveraging this duality, we further propose experimental protocols that require only a constant number of qubits to measure unconventional properties of 1D many-body states. These include (i) measurement of disorder operators, which diagnose the absence of spontaneous symmetry breaking, and (ii) postselection-free detection of measurement-induced long-range order, which emerges in certain symmetry-protected topological phases. We also show that measurement-induced long-range order provides a lower bound for strange correlators, which may be of independent interest.

\end{abstract}

\maketitle

	{
		\hypersetup{linkcolor=blue}
		\tableofcontents
	}

\section{Motivation}

Long-range entangled quantum states are central to many-body physics. They provide valuable resources for fault-tolerant quantum information processing, as well as underlie exotic phenomena such as fractionalization and quantum criticality. Preparing such states on a quantum computer, starting from simple unentangled states, typically requires deep circuits. This requirement follows from the Lieb-Robinson bound, which forbids the emergence of long-range order or entanglement under finite-depth local unitary evolution~\cite{lieb_robinson_1972,Hastings_topological_bound_2006}. Notably, such limitation can be circumvented with measurement, a fundamentally non-unitary operation. Inspired by the frameworks of measurement-based quantum computation~\cite{Briegel_01,Raussendorf_05} and quantum error correction~\cite{shor_1995,steane_1996,gottesman1997stabilizer,Kitaev:1997wr,Dennis_2002}, recent works have shown that a broad class of long-range entangled states can be prepared from the short-range entangled ones in constant depth with \textit{measurement-feedback} (MF) circuits \cite{piroli2021quantum,tantivasadakarn2021long,bravyi2022adaptive,lu2022measurement,verresen2021efficiently,tantivasadakarn2023hierarchy,nat_shortest_nonabelian_2023,iqbal2023creation,foss2023experimental,iqbal2023topological,satzinger2021realizing,tantivasadakarn2023hierarchy,Yabo_2023_measurement,lu2023mixed,buhrman2024state,Piroli_approximate_2024,stephen2024preparing,sahay2024finite,smith2024constant,zhang2024characterizing,sahay2025classifying}. In MF circuits, a finite-depth local unitary circuit is applied, followed by measurements on a subset of qubits and conditional feedback unitary gates based on the outcomes. Crucially, generating long-range entanglement in constant depth - an apparent violation of locality - is allowed because the feedback implicitly involves long-range classical communication. As such, MF circuits provide a time-efficient state-preparation protocol at the expense of increasing spatial overhead as some fraction of the physical qubits are measured out and do not participate in the final long-range entangled state.

 An alternative paradigm, which is space-efficient but involves temporal overhead, is the \textit{sequential unitary} (SU) circuits \cite{PhysRevLett.95.110503, PhysRevA.75.032311,PhysRevA.103.042613,pichler2017universal, PhysRevResearch.3.033002, PRXQuantum.4.030334, gopalakrishnan2023push, chen2024sequential,chen2024sequential2,seiberg2024majorana,tantivasadakarn2024string,vanhove2024duality,chen2024quantum,hu2025preparing}. SU circuits apply local unitary transformations to patches, strips, or other sub-regions of a system in a sequential manner. They preserve the entanglement area law because each qubit is only acted upon by a finite number of gates, yet the linear depth of the circuit is sufficient to reorganize long-range correlations and entanglement patterns. Therefore, SU circuits can create nontrivial many-body states such as symmetry-breaking GHZ states or topologically ordered states, generate symmetry defects, and implement non-invertible symmetry transformations \cite{chen2024quantum,chen2024sequential2,chen2024sequential,vanhove2024duality}. In addition, SU circuits provide a framework of holographic simulation, in which states on $O(N)$ qubits can be simulated while only having access to $\ll N$ physical qubits \cite{PhysRevA.103.042613,PhysRevResearch.3.033002,PRXQuantum.4.030334}.

\begin{figure*}[t]
		\centering
		\begin{subfigure}{\textwidth}
		\includegraphics[width=0.88\textwidth]{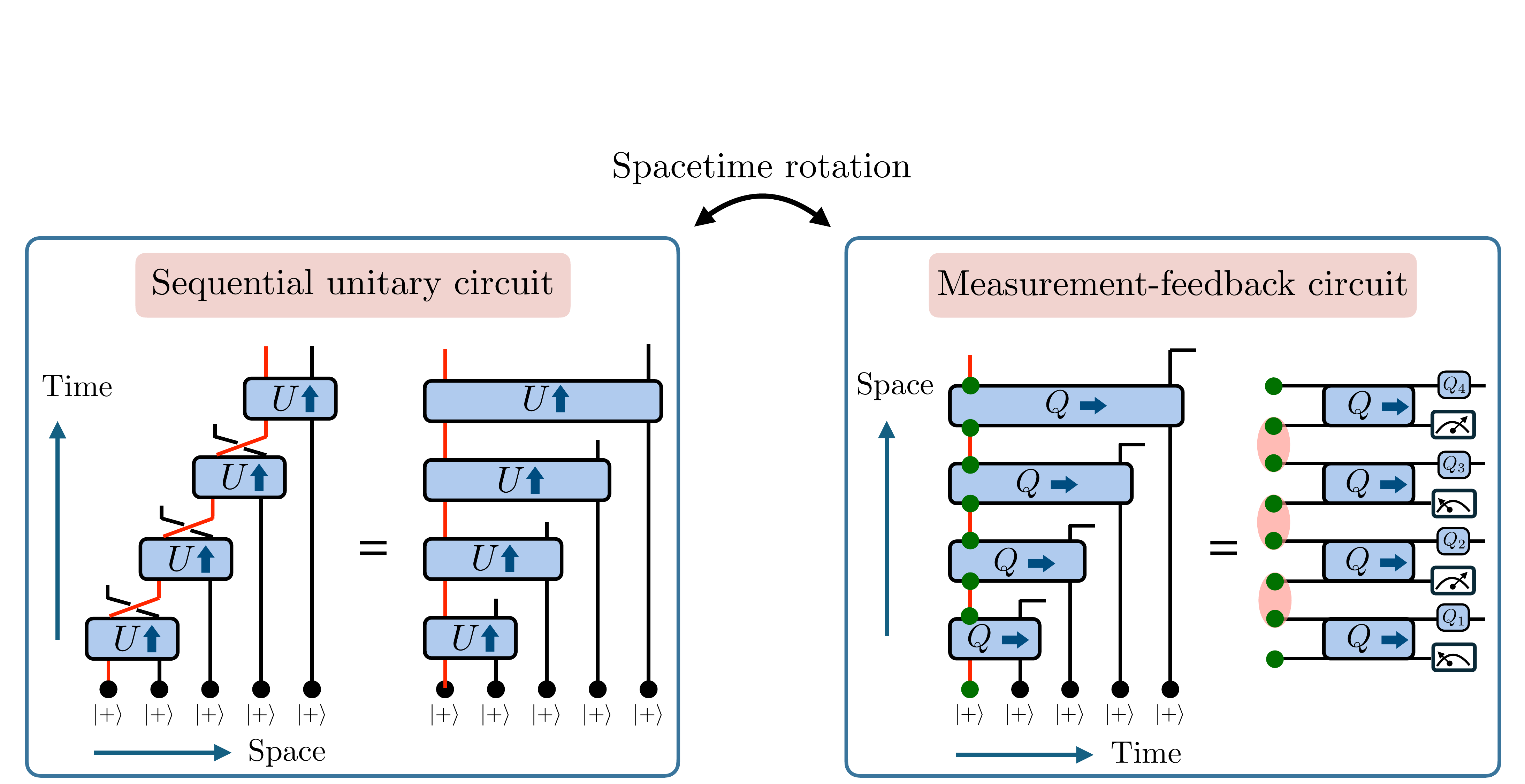}
	\end{subfigure}\caption{Left: with the addition of SWAP gates, a standard sequential circuit composed of geometrically local unitary gates is equivalent to a circuit where a single qubit (whose worldline is marked in red) sequentially interacts with the remaining qubits. Right: the spacetime-rotated view of the same circuit, where the initial state (marked by the green dots) is evolved with a depth-1 circuit composed of $Q$ gates, followed by projecting some of the qubits to $\ket{+}$. When a `measurement-feedback’ symmetry exists, the projection can be deterministically achieved via measurement and feedback conditioned on the measurement outcomes.}
		\label{fig:main}
	\end{figure*}

Although MF circuits and SU circuits are superficially distinct protocols, we show that many of them are in fact related by a simple spacetime rotation \cite{Akila_duality_2016,bertini2018exact,bertini2019entanglement,bertini2019exact,piroli2020exact,klobas2021entanglement,Chalker_2021_duality,garratt2021local,garratt2021manybody,Khemani_2021_duality,Lu_2021_duality,Khemani_2022_duality,Kos2023circuitsofspacetime,Abanin_2023_spacetime,bruno_full_counting_2023,Ho_2025,bertini2025_dual_unitary,stephen2024universal,zhang2024holographic} (Fig.\ref{fig:main}). We illustrate this correspondence with several examples of non-trivial many-body states, including the GHZ symmetry-breaking state, fractal symmetry-breaking state, and toric-code topologically ordered state. Notably, non-invertible symmetry transformations \cite{seiberg2024majorana,shao2023s,KW_Sahand_2024}  (such as the Kramers–Wannier duality and its generalization) in an SU circuit \cite{chen2024sequential2,vanhove2024duality} corresponds to symmetry-gauging operations in the spacetime-dual MF circuit \cite{tantivasadakarn2021long}, thereby unifying two seemingly distinct concepts in many-body physics. In addition to these conceptual contributions, we utilize the spacetime duality to develop efficient sequential protocols for measuring nonlocal observables, such as disorder operators and measurement-induced entanglement, that would otherwise be challenging to measure in a scalable way.

The rest of this work is organized as follows. In Sec.~\ref{setup} we introduce the spacetime duality between SU circuits and MF circuits, with the 1d GHZ state being a primary example. In Sec.~\ref{higherD} we generalize the spacetime duality to higher spatial dimensions. In Sec.~\ref{applications} we discuss protocols for measuring certain unconventional physical properties of quantum states by leveraging spacetime duality. Finally in Sec.~\ref{discuss} we conclude with a discussion of potential extensions of our results.

\section{Spacetime duality}\label{setup}
This section is organized as follows. First, we outline the circuit-level mapping from sequential unitary circuits to shallow non-unitary circuits with post-selected measurements under a spacetime rotation. Second, we illustrate this mapping with a one-dimensional example: the preparation of the GHZ state. In the context of this example, we show that the Kramers-Wannier duality transformation in sequential unitary circuits, under a spacetime rotation, can be mapped onto the symmetry gauging in post-selected circuits. We further show that the gauge structure of the resulting post-selected circuit enables one to bypass postselection by a feedback conditioned on the measurement outcomes. Finally, we apply the duality in reverse, connecting measurement-feedback circuits to their spacetime-dual sequential circuits, which are unitary under the conditions that we identify.

\subsection{From sequential unitary circuits to PEPS}\label{sec:1d}
We first observe that any sequential circuit can be simply rearranged into a shallow circuit involving \emph{post-selected} measurements on a subset of qubits. This rearrangement is depicted in Fig.~\ref{fig:main}. The sequential circuit is first rearranged into one in which some subsystem (e.g. the leftmost qubit of the figure) interacts repeatedly with an ``environment'': at each time step, a new environment qubit, initialized in some reference state $\ket{+}$, is brought in, interacts with the subsystem, and does not participate in the subsequent dynamics. When the subsystem is post-selected to a particular outcome, the state of these environment qubits can be exactly understood as a matrix product state (MPS), in which the subsystem represents the bond space and the environment qubits represent the ``physical'' space.

Now we interchange the roles of space and time. The leftmost, ``bond-space'' qubit in the sequential circuit maps to a collection of symmetric Bell pairs in the initial state of the rotated circuit. These Bell pairs are then evolved with a depth-1 circuit composed of local (generally non-unitary) $Q$ gates, followed by post-selecting the qubits at every other site to $\ket{+}$. In fact, this is exactly the construction of a Projected Entangled Pair State (PEPS) \cite{peps_review_2021}: each unit cell contains two qubits, which form Bell pairs with neighboring cells, and the local Q gate, together with the subsequent projection, serves as a linear map that compresses the two-qubit Hilbert space to an effective single-qubit Hilbert space within each unit cell.

While it is well established that (i) a sequential circuit builds an MPS and (ii) an MPS can be viewed as a spatial slice of a PEPS, the construction in Fig.~\ref{fig:main} makes the underlying spacetime rotation explicit, clarifying the discussions that follow.

\subsection{Sequential circuits for 1d GHZ state}\label{sec:GHZ}

We begin by providing an overview of the sequential circuit that prepares the 1d GHZ state exhibiting $\mathbb{Z}_2$ symmetry breaking. For concreteness, let's consider the transverse-field Ising model in 1d: $-\sum_i Z_i Z_{i+1} - g \sum_i X_i$ with $g\geq 0$. It exhibits a symmetric phase ($g>1$) and a symmetry-breaking phase ($g<1$) where the $\mathbb{Z}_2$ symmetry ($\prod_i X_i$) is spontaneously broken. The representative fixed-point states for the symmetric phase and the symmetry-breaking phase are $\ket{++...+}$ and the GHZ state $|00..0\rangle + |11...1 \rangle$, respectively. While these two states cannot be connected by finite-depth local unitary circuits, they can be connected by a linear-depth sequential circuit $\mathcal{U}$ \cite{Hsieh_sequential_2019,chen2024sequential}: 

\begin{equation}\label{eq:1dGHZ_before_swap}
\begin{split}
\mathcal{U}  =   \prod_{i=1}^{N-1} &u_{i,i+1} =u_{N-1,N}  \dots   u_{1,2},\\ 
\text{with } & u_{i,i+1} =   e^{-i\frac{\pi}{4} X_{i+1}} e^{-i\frac{\pi}{4} Z_i Z_{i+1}} .
\end{split}
\end{equation}  
To understand the action of the gate $u_{i,i+1}$, we may express it as

\begin{equation}
\begin{split}
 \ket{0}  \bra{0}_{i}  e^{-i\frac{\pi}{4} X_{i+1}  } e^{-i\frac{\pi}{4} Z_{i+1} } + \ket{1}  \bra{1}_{i}  e^{-i\frac{\pi}{4} X_{i+1}  } e^{i\frac{\pi}{4} Z_{i+1} }
\end{split}
\end{equation}
Namely, $u_{i,i+1}$ can be regarded as a controlled gate to align the two qubits in the computational basis: if the $i$-th qubit is in $\ket{0}$, one applies $e^{-i\frac{\pi}{4} X_{i+1}  } e^{-i\frac{\pi}{4} Z_{i+1}  }$, which can map $\ket{+}$ to $\frac{1-i}{\sqrt{2}}\ket{0}$ for the $(i+1$)-th qubit. On the other hand, if the $i$-th qubit is in $\ket{1}$, one applies $e^{-i\frac{\pi}{4} X_{i+1}  } e^{i\frac{\pi}{4} Z_{i+1}  }$, which can map $\ket{+}$ to $\frac{1-i}{\sqrt{2}}\ket{1}$ for the $(i+1$)-th qubit. As such, the sequential applications of $u$ gate aligns all the qubits in the computational basis, resulting in the symmetry-breaking GHZ state.

Before implementing the spacetime rotation, we would like to rearrange the circuit geometry such that a single qubit sequentially interacts with the rest of the qubits. This will have the advantage that the spacetime-rotated counterpart is a shallow circuit, as shown in Fig.\ref{fig:main}. To this end, we define $U_{i,i+1} =  \text{SWAP}_{i,i+1}u_{i,i+1}$ so that the sequential circuit reads $  \mathcal{U}=\prod_i (\text{SWAP}_{i,i+1} U_{i,i+1}$), i.e. the action of the $U$ gate is followed by a SWAP gate. Under the redefinition of the circuit via the sequential SWAP gates, the overall sequential circuit can alternatively be rearranged into the form

\begin{equation}\label{eq:1d_GHZ}
\includegraphics[width=5.8cm]{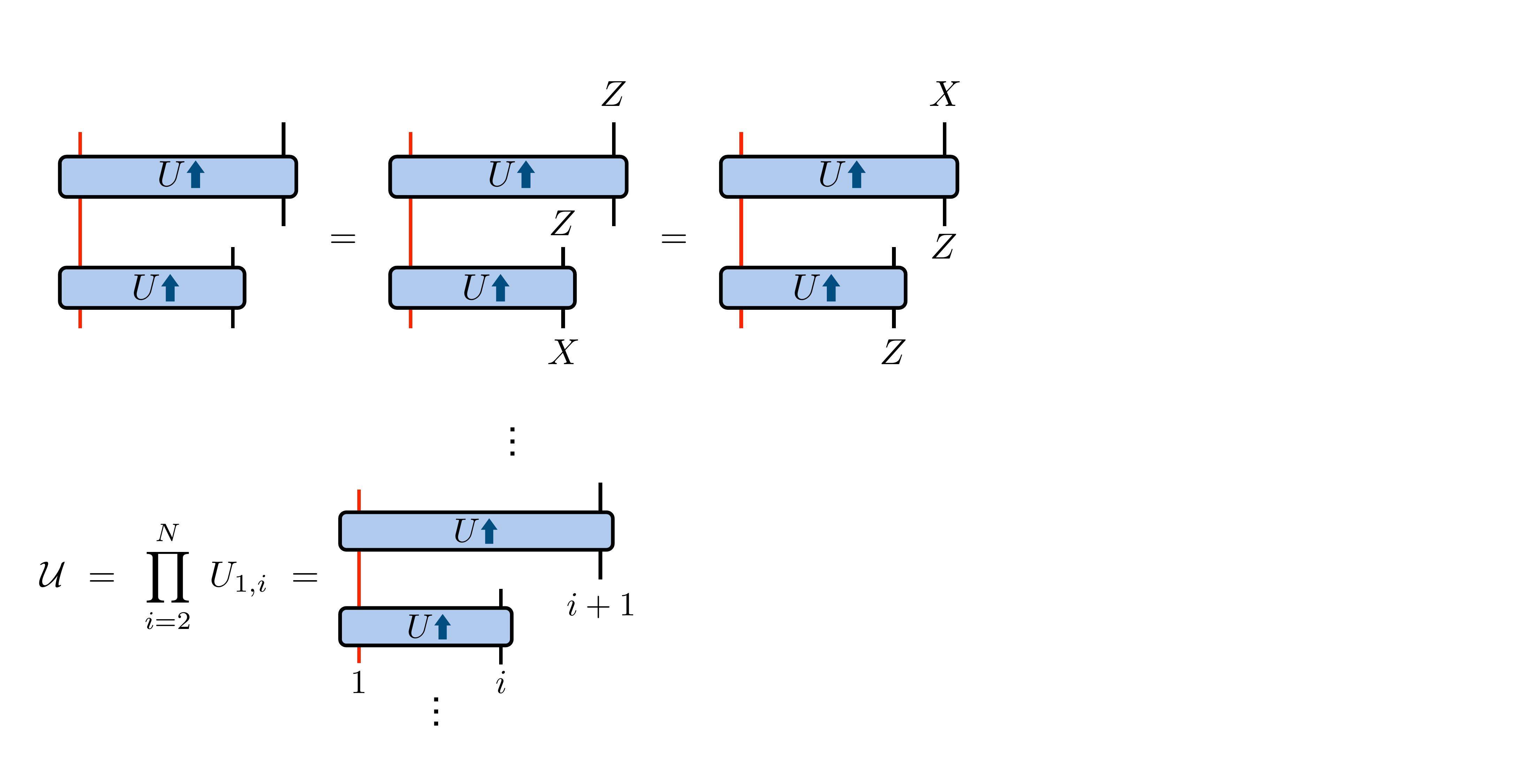}
\end{equation}
\begin{align}\label{unitary1}
\text{with   }  U_{1,i} = \text{SWAP}_{1,i}(e^{-i\frac{\pi}{4} X_{i}} e^{-i\frac{\pi}{4} Z_1 Z_{i}}).
\end{align}
Namely, the first qubit sequentially interacts with the \(i\)-th qubit via \(U_{1,i}\), which can be seen by following the red worldline of the first qubit (see Fig. \ref{fig:main} left panel) \footnote{Technically, one should add a SWAP gate between the $(N-1)$-th qubit and the $N$-th qubit in the circuit in Eq.\ref{eq:1dGHZ_before_swap} so that it can be rearranged to the form in Eq.\ref{eq:1d_GHZ}. However, this consideration is irrelevant as this SWAP gate acts trivially on the GHZ state.}.

Finally, we remark that the sequential circuit in Eq.\ref{eq:1d_GHZ} implements the Kramers-Wannier (KW) duality transformation \cite{shao2023s,seiberg2024majorana,KW_Sahand_2024}, which accomplishes the following mapping of bulk operators under the conjugation of the circuit\footnote{To be precise, $\mathcal{U}X_i \mathcal{U}^{\dagger}  =  Z_iZ_{i+1}$ holds for $i=2, 3, ..., N$ with $N+1 \equiv 1$, and $\mathcal{U}Z_i Z_{i+1} \mathcal{U}^{\dagger} = X_{i+1}$ holds for $i=2,3,...,N-1$.}: 
\begin{equation}\label{eq:non-invert}
\begin{split}
&\mathcal{U}X_i \mathcal{U}^{\dagger}  =  Z_iZ_{i+1}, \\
   &  \mathcal{U}Z_i Z_{i+1} \mathcal{U}^{\dagger} = X_{i+1}
\end{split}
\end{equation}
Hence, it realizes a non-invertible symmetry transformation that can map between a $\mathbb{Z}_2$ symmetric phase and a $\mathbb{Z}_2$ symmetry-broken phase in the transverse-field Ising model (see Appendix.\ref{appendix:KW} for a brief review and clarification on this non-invertible symmetry). As we will see later, the operator-mapping rule under the KW transformation will have important implications under a spacetime rotation (e.g., see Eq.\ref{eq:ghz_KW_stabilizer} and the discussion thereafter).

\subsection{Spacetime rotation of sequential circuits: the dual-$Q$ circuits}\label{sec:1d_dual_Q}

\begin{comment}
    
    \begin{figure}
\includegraphics[width=0.48\textwidth]{dualnew.png}
    \caption{a) Dual Q-circuit: The process begins by preparing the input state as a collection of EPR pairs between even and odd sites. A \(Q\)-gate (the space-time dual of the sequential unitary) is then applied that entangles them. Here, we insert an additional SWAP gate that exchanges the even and odd sublattices, following our preferred notation.
    Finally, a projection is performed on all even sites (solid red dots) to the \(X = 1\) basis state, resulting in a GHZ state. b) The space time duality for the two-body gate, which consists of control gate combined with a SWAP gate.}
    \label{dualitynew}
\end{figure} 

\end{comment}

With the rearranged sequential circuit, we now discuss the emergence of a shallow unitary-projective circuit via a spacetime rotation. To begin with, we observe that the input qubits in the spacetime-rotated circuit (depicted by green dots in Fig.\ref{fig:main} right panel) are given by different time slices of the first qubit in the original sequential circuit. In particular, in the rotated circuit, the two qubits connected by a red bond form a Bell pair ($ \propto  \ket{00}+\ket{11}$) because in the sequential circuit, the first qubit evolves trivially over that time duration. As such, the initial state $\ket{\psi_0}$ consists of an extensive number of Bell pairs, along with two dangling qubits on the two ends: $\ket{\psi_0} =   \ket{+}_1 \otimes_{i=1}^{\frac{\tilde{N}}{2} -1 } \frac{1}{\sqrt{2}}(\ket{00}+ \ket{11})_{2i,2i+1}   \otimes  \ket{+}_{\tilde{N}}$ with $\tilde{N}=2(N-1)$ \footnote{Note that fixing the $\tilde{N}$-th qubit to be $\ket{+}$ corresponds to fixing the upper endpoint of the red worldline to be $\ket{+}$, in which case the output is the $(N-1)$-qubit GHZ state.}.

The initial state in the rotated circuit evolves under a layer of the two-body $Q$ gate, obtained by a spacetime rotation on the $U$ gate. Notably, the controlled-SWAP structure of the $U$ gate (Eq.\ref{unitary1}) implies the dual unitarity, so that the $Q$ gate remains unitary. (See Ref.\cite{bertini2025_dual_unitary} for review on dual unitary circuits and their applications.) This can be observed from the following figure, with the red arrows penetrating the four-leg tensor indicating the direction of time.

\begin{equation}
\includegraphics[width=7cm]{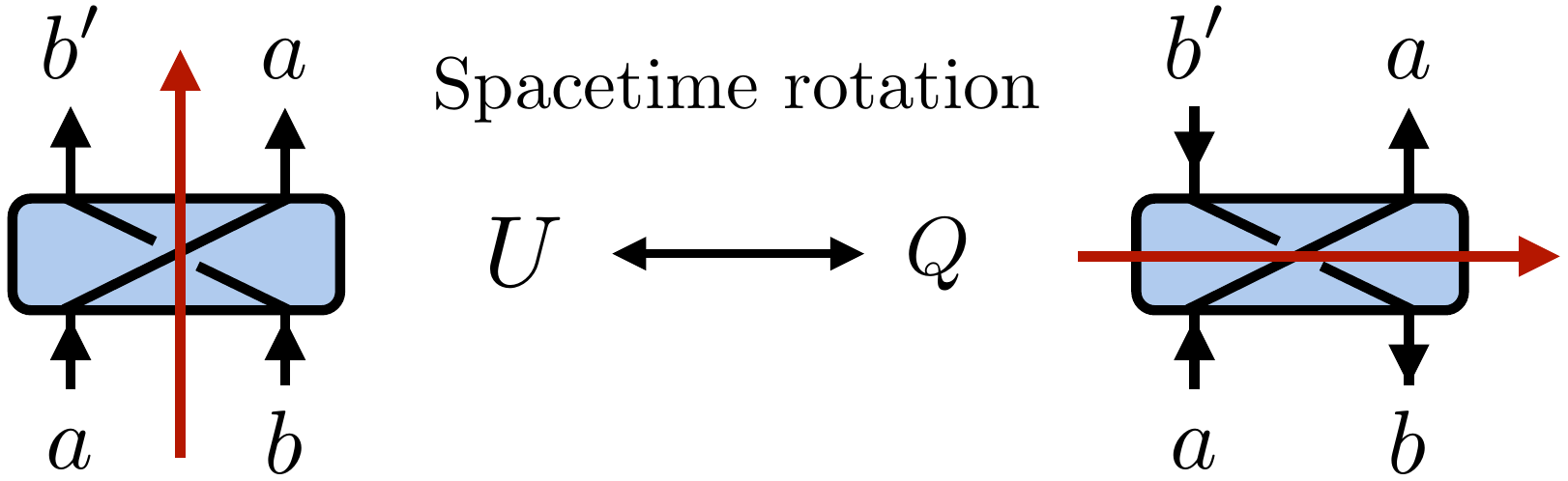}
\end{equation}
For the $U$ gate on the left, the worldline labeled by $a$ indicates the controlled qubit, and the other worldline labels the target qubit, which is acted by a single-qubit operation conditioned on the controlled qubit. Under a spacetime rotation, the direction of the worldline of the target qubit is reversed, which corresponds to taking a partial transpose (see Appendix.\ref{app:dualcircuit} for details). This allows us to deduce the explicit form of the unitary $Q$ gate: 
\begin{align}\label{eq:1d_q}
Q_{i,j} = \text{SWAP}_{i,j}(e^{- i\frac{\pi}{4} Z_i Z_{j}} e^{ -i\frac{\pi}{4} X_{j}}   ) 
\end{align}
which is essentially the same as the $U$ gate defined in Eq.\ref {unitary1}, up to a partial transpose on the target qubit (the qubit $j$ in $Q_{i,j}$ ).  

The extensive applications of these $Q$ gates give the \textit{dual-\( Q \) circuit}:
\begin{align} \label{eq:qgate}
\mathcal{Q} = \prod_i Q_{2i-1,2i} 
\end{align}
which outputs a short-range entangled state $\ket{\psi}  = \mathcal{Q}  \ket{\psi_0}$. Finally, all the qubits on odd sites are projected to $\ket{+}$ since they are fixed by the initial state in the sequential circuit (see Fig.\ref{fig:main} right panel). This generates the GHZ state on even sites, which precisely matches the GHZ state generated by the original sequential circuit.

As the mapping from the sequential circuit to its rotated counterpart is simply a rearrangement of a circuit diagram, it is not surprising that they both generate the same GHZ state. However, the spacetime rotation provides a fresh insight. As we will discuss in Sec.\ref{sec:KW}, $\ket{\psi}$ generated by the dual-$Q$ circuit is essentially a cluster-state SPT \cite{cluster_spt_2012} (up to a layer of SWAP gates between the even sites and odd sites), which is known to produce the GHZ-type long-range order upon measuring a subset of qubits \cite{Briegel_01,tantivasadakarn2021long,lu2022measurement}. Alternatively, the process in the rotated circuit can be understood as a $\mathbb{Z}_2$ symmetry gauging - the extensive applications of the $Q$ gates leads to a $\mathbb{Z}_2$ gauge theory, and a subsequent projection on the gauge fields freezes their dynamics, thereby producing a true long-range (GHZ-type) order among the matter fields. This shallow unitary-projective protocol is essentially the symmetry gauging described in Ref.\cite{tantivasadakarn2021long}. As such, the spacetime duality unifies two seemingly distinct concepts: the Kramers-Wannier (KW) duality via sequential unitary circuits \cite{seiberg2024majorana,KW_Sahand_2024,chen2024sequential} and symmetry gauging via shallow circuits \cite{tantivasadakarn2021long}. Finally, as will be discussed in Sec.\ref{sec:gauge_symmetry}, the projection in the rotated circuit can be accomplished by Pauli-X measurements, and due to the gauge symmetry, the unwanted measurement outcomes can be compensated by a subsequent feedback unitary. Therefore, the output GHZ state can be deterministically prepared without postselection.

Before moving on to the next section, we briefly remark on the dual unitarity of the $U$ gate. As we have discussed in this section, a unitary controlled gate followed by a SWAP gate is always mapped to a unitary under a spacetime rotation. This property is especially helpful for applying a spacetime rotation to the sequential circuits for preparing ground states of stabilizer codes. In these models, one sequentially applies certain controlled gates to generate the desired stabilizers. As such, the $Q$ gate obtained by spacetime rotating these controlled gates must be unitary as well. On the other hand, if one considers a more general sequential unitary circuit, we show that the corresponding $Q$ gate can be implemented by introducing ancilla Bell pairs followed by Bell projection. This is based on the insight that the $Q$ gate is related to the unitary $U$ gate by a partial transpose, and the partially-transposed gate can be physically implemented with ancilla Bell pairs and Bell projection (see Appendix~\ref{app:dualcircuit} for details).

\subsection{KW duality and symmetry gauging}\label{sec:KW}
Here we provide a detailed exposition of the symmetry gauging perspective in the rotated (shallow) circuit \cite{tantivasadakarn2021long}, and strengthen its connection to the non-invertible KW duality in the original sequential circuit. 

First, to manifest the symmetry gauging perspective in the rotated circuit, we apply the extensive SWAP gates between even and odd sites after the $Q$ gates, so that the output short-range entangled state reads 

\begin{equation}\label{eq:psi_prime_output}
 \ket{\psi'} =  \prod_i (\text{SWAP}_{2i-1,2i}  Q_{2i-1,2i}  )\ket{\psi_0}.
\end{equation}

\begin{figure}[t]
\includegraphics[width=0.3\textwidth]{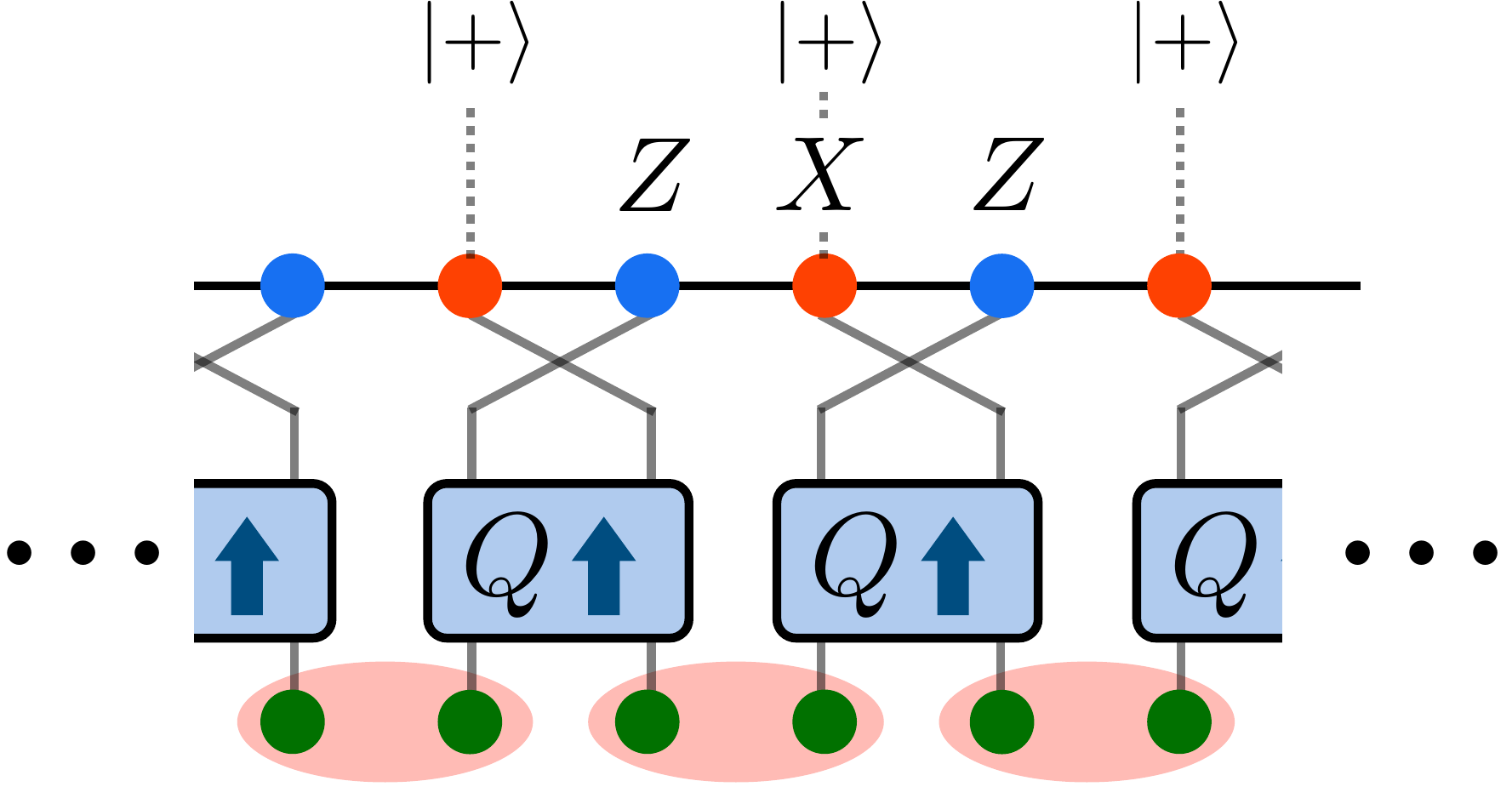}
    \caption{The output of the shallow dual-$Q$ unitary circuit with additional SWAP gates is a cluster state stabilized by $Z_{i-1} X_iZ_{i+1}$. A subsequent projection of the qubits at red sites to $\ket{+}$  generates the GHZ state on blue sites.} \label{eq:dual_Q}
\end{figure}

Correspondingly, the subsequent projection to $\ket{+}$ will be performed on even sites (the red qubits in Fig.\ref{eq:dual_Q}). As one can check, $ \ket{\psi'}$ is exactly a cluster state stabilized by the $Z_{i-1} X_iZ_{i+1}$ stabilizer in the bulk, which can be interpreted as a $\mathbb{Z}_2$ gauge theory. In this formulation, the \(\mathbb{Z}_2\) matter fields (blue qubits) reside on the odd sites while the gauge fields (red qubits) reside on the even sites. The condition \(Z_{2i} X_{2i+1} Z_{2i+2} = 1\)  acts as a \(\mathbb{Z}_2\) Gauss law, enforcing gauge invariance, and the constraint \(Z_{2i-1} X_{2i} Z_{2i+1} = 1\) ensures the minimal coupling between the matter fields and gauge fields. As such, the $Q$-gates acting on Bell pairs serve as a gauging operator that minimally couples the $Z_2$ matter fields to the $Z_2$ gauge fields. In particular, with the minimal coupling constraint \(Z_{2i-1} X_{2i} Z_{2i+1} = 1\) between the matter fields at site $2i-1$ and $2i+1$, projecting the gauge field in between onto $\ket{+}$ (so that $X_{2i}=1 $) implies $Z_{2i-1} Z_{2i+1} =1$ for all odd sites, thereby generating the long-range GHZ order among the matter fields. Alternatively, the cluster state may be regarded as a state in the Higgs phase of the $\mathbb{Z}_2$ gauge theory as described in Ref.~\cite{verresen2022higgs}, in which a projection on matter fields leads to a symmetry-breaking long-range order of the matter fields.

Having demonstrated the symmetry-gauging perspective in the rotated circuit, we now elucidate its intrinsic connection to the KW duality in the sequential circuit \cite{seiberg2024majorana,KW_Sahand_2024,chen2024sequential}. In particular, we show that the emergence of the cluster-state stabilizers in the rotated circuit follows directly from the operator mapping rules of the KW duality.

First, we recall the operator mapping under the KW transformation via the sequential circuit (Eq.\ref{eq:non-invert}): $X_i \to Z_{i}Z_{i+1}  $ and $Z_{i}Z_{i+1} \to X_{i+1}$. In particular, this implies $\mathcal{U}  =   Z_{i}Z_{i+1} \mathcal{U} X_i  =X_{i+1} \mathcal{U} Z_{i}Z_{i+1}$. This relation can be diagrammatically represented as 

\begin{equation}\label{eq:ghz_KW_stabilizer}
\includegraphics[width=7.6cm]{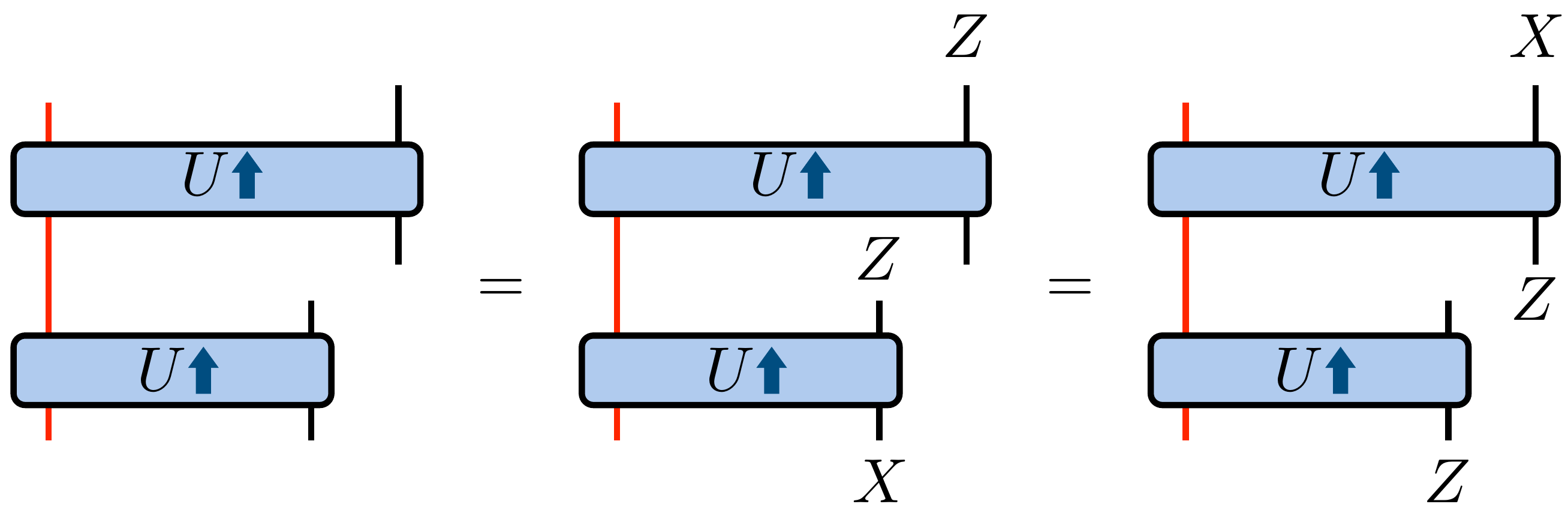}
\end{equation}

Under a spacetime rotation, the input/output qubits of the original sequential circuit transform into the output qubits on the odd/even sites of the dual $Q$-circuit. This immediately implies that the output state of the dual $Q$-circuit satisfies the following constraint \footnote{Alternatively, these two constraints can be derived by conjugating the Bell-pair constraints, i.e. $Z_{2i } Z_{2i+1} = 1 $ and $Y_{2i } Y_{2i+1} = 1 $, with the dual-$Q$ circuit in Eq.\ref{eq:qgate}.} :
\begin{align}\label{eq:cluster}
X_{2i-1}Z_{2i}Z_{2i+2} =1, ~Z_{2i-1} Z_{2i+1}X_{2i+2}=1. 
\end{align}
Under an exchange between the even and odd sites via $\prod_i \text{SWAP}_{2i-1,2i}$ (see Fig.\ref{eq:dual_Q}), these constraints become the $Z_{i-1} X_i Z_{i+1}=1$ constraint on three consecutive sites, thereby indicating the cluster-state order, or equivalently, the \(\mathbb{Z}_2\) gauge theory. This perspective therefore provides a very straightforward connection between the KW duality in the sequential circuit and the symmetry-gauging perspective under a spacetime rotation.

\subsection{Measurement feedback from gauge symmetry}\label{sec:gauge_symmetry}

As we have discussed above, spacetime-rotating a sequential circuit leads to a shallow unitary-projective circuit where a subset of qubits are projected to $\ket{+}$. While this can be implemented by Pauli-X measurement and postselection on the $+1$ measurement outcome, post-selection is not a scalable approach on quantum devices. Here we will discuss how to bypass the post-selection issue via a feedback operation conditioned on the measurement outcome, so that the rotated unitary-projective circuit can be transformed into a measurement-feedback (MF) circuit, in which the GHZ state can be deterministically prepared.  While this protocol itself is not a new result \cite{Briegel_01,tantivasadakarn2021long,lu2022measurement}, we show how the existence of such an MF protocol is enabled by the gauge symmetry of the output cluster state before projection (i.e. Eq.\ref{eq:psi_prime_output}), hence providing a new perspective.

We first recall that the rotated circuit generates a cluster state that can be interpreted as a gauged \( \mathbb{Z}_2 \) theory  (Eq.\ref{eq:psi_prime_output}), and the following condition can be regarded as a gauge symmetry
\begin{align}
  Z_{2i}X_{2i+1}Z_{2i+2} |\psi'\rangle=|\psi'\rangle 
\end{align}
where $X_{2i+1}$ represents the matter field and $Z_{2i}, Z_{2i+2}$ represent the gauge fields. Alternatively, the above equation can be recast to

\begin{align}
  Z_{2i}Z_{2i+2} |\psi'\rangle= X_{2i+1} | \psi'\rangle. 
\end{align}
This manifests the fact that the $ZZ$ action on the two adjacent gauge fields can be compensated for by a single $X$ action on the matter field in between. For instance, if the outcome of the Pauli-X measurement on the 2nd qubit and the 8th qubit is $\ket{-}$, which amounts to having a Pauli-Z defect on these two sites, one can then apply a string operator $X_3X_5X_7$ to clean out the $Z$ defects, thereby deterministically generating the GHZ state.

\subsection{From MF circuits back to SU circuits}\label{sec:reverse}
In the previous discussion, we have demonstrated that a sequential unitary (SU) circuit can be mapped to a shallow, non-unitary circuit involving postselection (or measurement-feedback). However, this mapping is not inherently reversible: while a unitary gate transforms into a non-unitary gate under spacetime duality, the inverse transformation—converting a non-unitary gate back into a unitary one—is not always guaranteed. This raises an intriguing question: can we establish a two-way duality that maps an MF circuit back to an SU circuit? Under what conditions does such a two-way duality hold?

In this section, we introduce a two-way duality between SU circuits and MF circuits. We draw inspiration from Ref.~\cite{lu2022measurement,stephen2024preparing,sahay2024finite,smith2024constant,zhang2024characterizing,sahay2025classifying}, which explores the many-body states that can be prepared with MF circuits based on the tensor-network formalism \cite{peps_review_2021}. Specifically, it was shown that the symmetry of local tensors, dubbed measurement-feedback (MF) symmetry, translates to the preparability of the tensor-network states. We will show that such a symmetry exactly guarantees the two-way duality between SU circuits and MF circuits. As an application, this insight illustrates the connection between the Kramer-Wannier duality implemented by an SU circuit and the MF symmetry in an MF circuit. Below, we will limit our discussion to 1d examples, but the two-way duality can be naturally extended to higher dimensions. This is illustrated in Appendix.\ref{appendix:reverse_toric}, focusing on the 2d toric-code topological order, and we expect it to be generalizable to any abelian quantum double models \cite{Kitaev:1997wr}.

To discuss the two-way duality, we first briefly summarize a key insight of Ref.\cite{zhang2024characterizing}, focusing on matrix product states (MPS), i.e. the 1d tensor networks. To begin, we first prepare many small clusters of three qubits along a 1d line in constant depth \footnote{This can be straightforwardly generalized to qudits and the clusters along the 1d line need not be identical.}:

\begin{equation}\label{eq:1d_tensor}
\includegraphics[width=5cm]{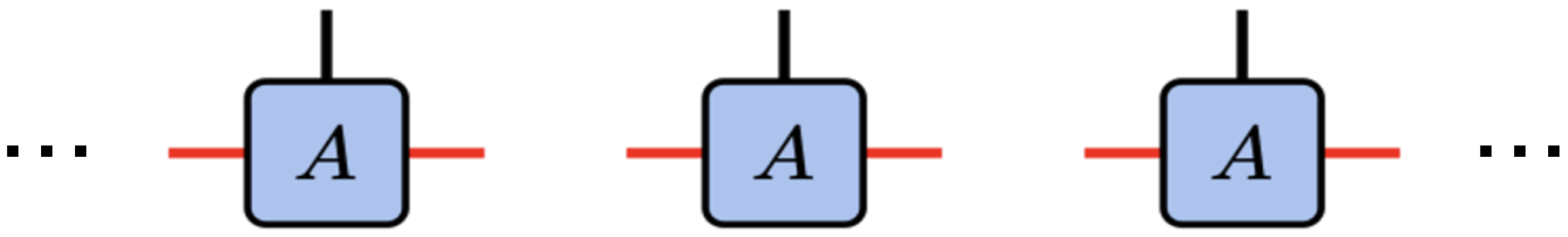}
\end{equation}

The wavefunction in the computational basis of each cluster is described by a tensor $A_{ij}^{\alpha}$, where $\alpha$ labels the basis of the top (physical) qubit, and $i, j,$ label the basis for the two (virtual) qubits on the two red legs. To glue together the decoupled clusters, we extensively perform the Bell-basis measurements on two neighboring virtual qubits belonging to two adjacent clusters: 

\begin{equation}\label{eq:1d_bell_fusion}
\includegraphics[width=5cm]{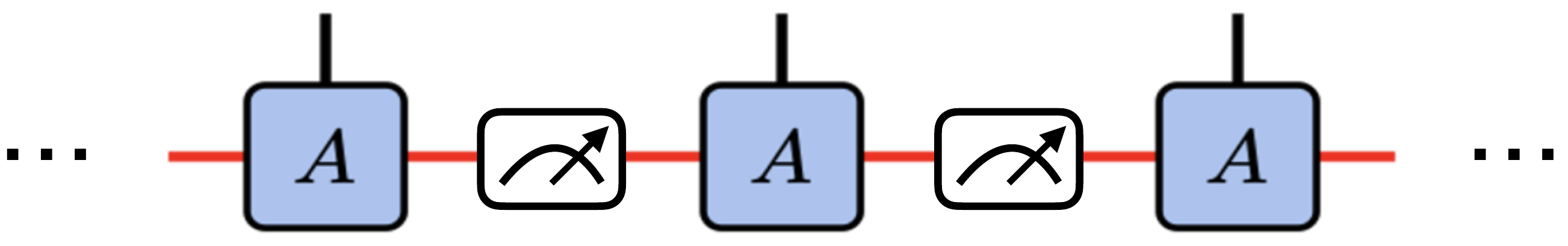}
\end{equation}

The Bell-basis measurement will project the two measured qubits into one of the four possible Bell pairs: $ X^a Z^b \frac{1}{\sqrt{2}} (\ket{00} + \ket{11})$ where $X, Z$ are the Pauli-X, Z operators for one of the qubits and $a, b \in \{ 0,1\}$. When the outcome is $(a,b)=(0,0)$ for all Bell-basis measurements, all the virtual qubits are contracted correctly, leading to the desired MPS that describes the physical qubits (denoted by the black legs). However, the unwanted outcomes, represented by single-qubit Pauli defects (i.e. $X, Y, Z$), will appear generically 

\begin{equation}
\includegraphics[width=5.5cm]{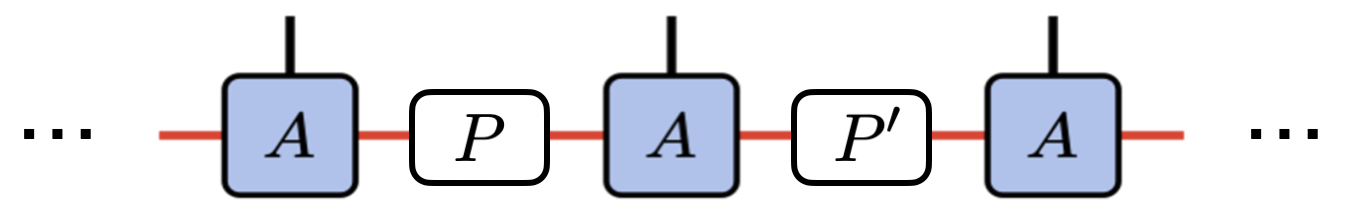}
\end{equation}
so the essential question is whether those defects can be removed by applying certain unitary feedback. The answer is affirmative if the local tensor satisfies the so-called measurement-feedback (MF) symmetry \cite{zhang2024characterizing,sahay2024finite,smith2024constant,stephen2024preparing}:
\begin{equation}\label{eq:MF}
\includegraphics[width=5cm]{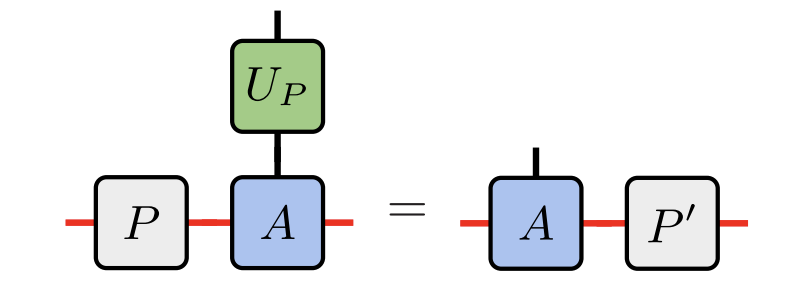}
\end{equation} 

This condition indicates that a Pauli defect on a given bond can be moved across the local tensor $A$, at the expense of adding single-site unitary operations on the physical legs. Therefore, by applying appropriate unitary feedback based on the measurement outcomes, all the defects in the bulk of the MPS can be annihilated, with potential defects existing on the boundary of the MPS, which can be annihilated with a unitary operation as well.

Ref.\cite{zhang2024characterizing} further shows that the MF symmetry implies a particular structure of a local tensor:  
\begin{equation}\label{eq:MF_symmetry}
\includegraphics[width=5cm]{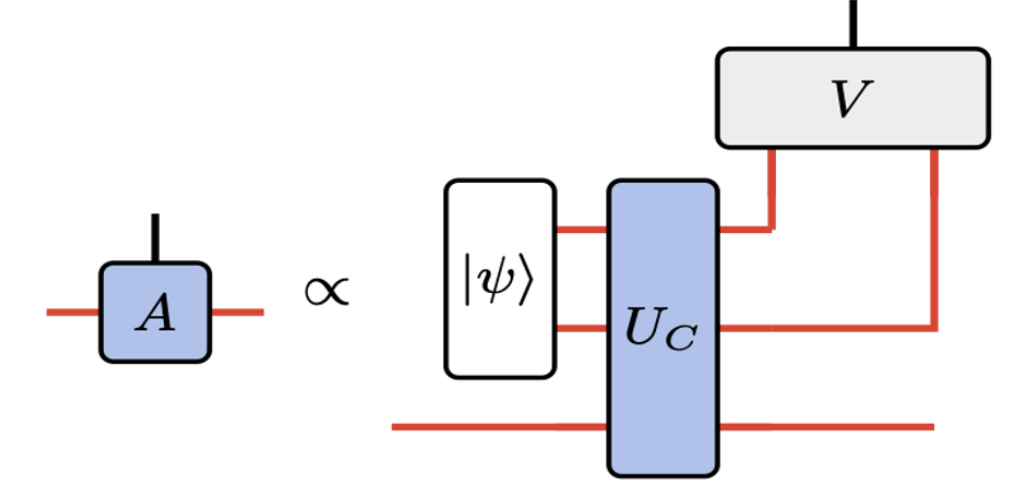}
\end{equation}
$\ket{\psi}$ is any initial state, $V$ is an isometry, and $U_c$ is a Clifford unitary, which maps Pauli operators to Pauli operators. For the case where there is no isometry $V$ and the dimensions of $\ket{\psi}$ and the physical (black) leg are identical, Eq.\ref{eq:MF_symmetry} implies that the corresponding local tensor can simply be understood as a Clifford gate with a certain input $\ket{\psi}$:

\begin{equation}
\includegraphics[width=4cm]{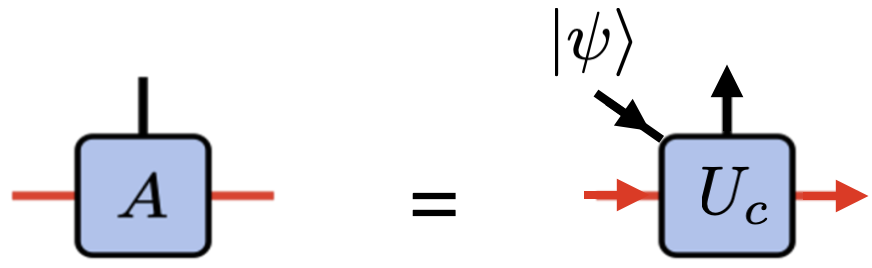}
\end{equation}
Notably, exchanging the roles of space and time allows the original MF circuit to be reinterpreted as a sequential unitary circuit:

\begin{equation}
\includegraphics[width=4cm]{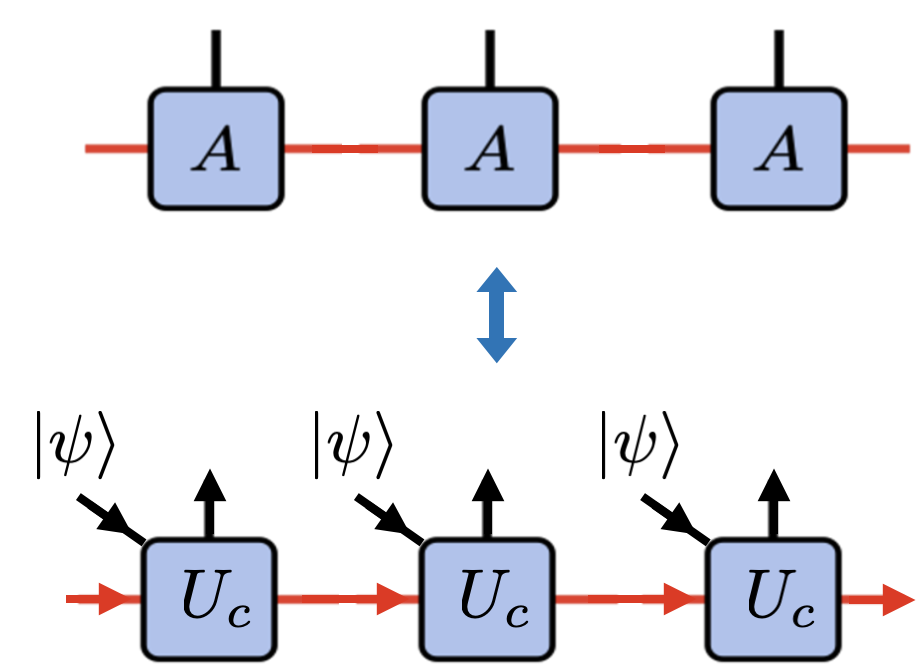}
\end{equation}
Namely, the constant-depth MF circuit that generates an MPS can be understood as a sequential unitary circuit where a single qubit sequentially interacts with qubits at different sites that are initialized at the state $\ket{\psi}$.

We now illustrate the formalism above with the example of the GHZ state. To begin with, we consider a three-qubit cluster $\sum_{i,j,\alpha} A^{\alpha}_{ij} \ket{ i,j,\alpha  }$, where $ \ket{ i,j,\alpha  }$ is the computational (Pauli-Z) basis, and $A^{\alpha}_{ij}$ is the MPS tensor defined as $A^{\alpha}_{ij}=  \delta_{i\alpha}\delta_{j\alpha}$. Namely, $A^{\alpha}_{ij}$ takes the value 1 when all three indices are the same, and takes the value $0$ otherwise. We can prepare multiple identical clusters in parallel along a 1d lattice (see Eq.\ref{eq:1d_tensor}). After the extensive Bell-basis measurement as in Eq.\ref{eq:1d_bell_fusion}, if all the measurement outcomes correspond to the Bell pair $\ket{00}+ \ket{11}$, the virtual legs can be contracted correctively, giving rise to the GHZ state \( |000 \dots\rangle + |111  \dots\rangle \) on the physical legs. Importantly, all possible measurement outcomes can be corrected via feedback due to the MF symmetry of the local tensor $A$: 

\begin{equation}\label{eq:1d_GHZ_MF_symmetry}
\includegraphics[width=7.4cm]{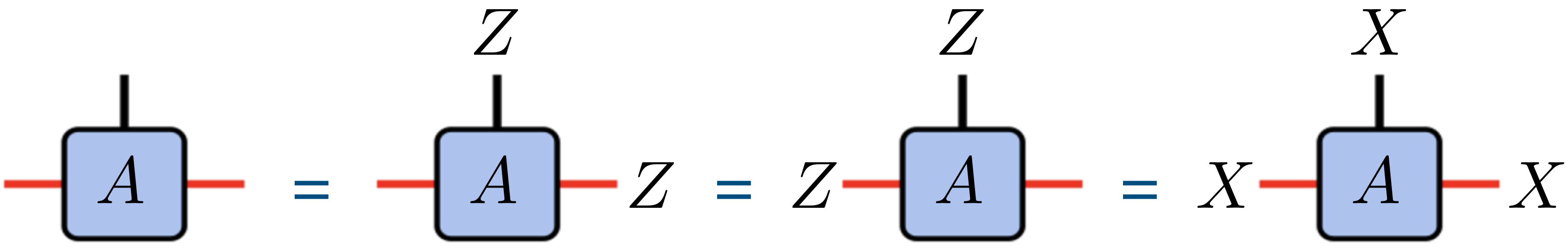}
\end{equation}

Notably, the MF symmetry above leads to the emergence of the Kramers-Wannier (KW) duality under a spacetime rotation. To see this, we first the interpreted the tensor $A_{ij}^{\alpha}$ as a two-qubit Clifford unitary (defined in Eq.\ref{unitary1}) with one of the input qubits fixed to be $\ket{+}$. The KW duality that maps an input Pauli-X to two Pauli-Zs in the output can then be understood through the process depicted in Eq.\ref{eq:1d_KW_ghz}, which is enabled by the MF symmetry involving Pauli-Z operators:  

\begin{equation}\label{eq:1d_KW_ghz}
\includegraphics[width=7.5cm]{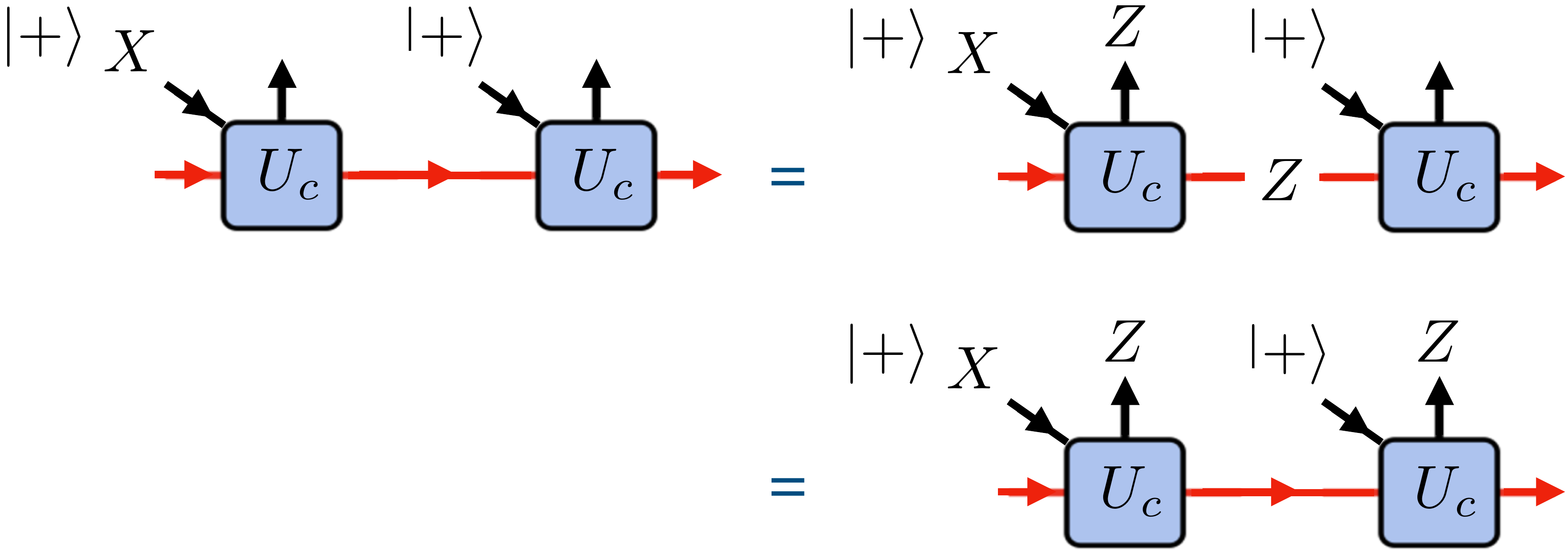}
\end{equation}
Here, the input Pauli-X is first mapped to the two Pauli-Zs (one on the output physical leg and one on the bond space of the MPS). The Pauli-Z on the bond space can be further conveyed to the physical leg at the next site.  

\begin{center}

\begin{table*}
\renewcommand\arraystretch{1.4}
\begin{tabular}{| l | l |}
\hline
\textbf{Sequential Unitary (SU) Circuit} & \textbf{Measurement-Feedback (MF) Circuit} \\ \hline
Input and output states & Odd and even sites of output qubits \\ \hline
Time evolution of the first qubit &  Bell pairs as input state  \\ \hline
Non-invertible duality transformation & Stabilizers in the output state $\psi_{\text{out}}$ before projective measurements \\ \hline
\end{tabular}
\captionsetup{justification=centering}
\caption{A Summary for Spacetime Duality between SU Circuits and MF Circuits}\label{table1}
\end{table*}
\end{center}

On the other hand, the last condition of Eq.\ref{eq:1d_GHZ_MF_symmetry}, which involves three Pauli-Xs, corresponds to the local $\mathbb{Z}_2$ symmetry conservation: 

\begin{equation}\label{symmetry}
\includegraphics[width=5cm]{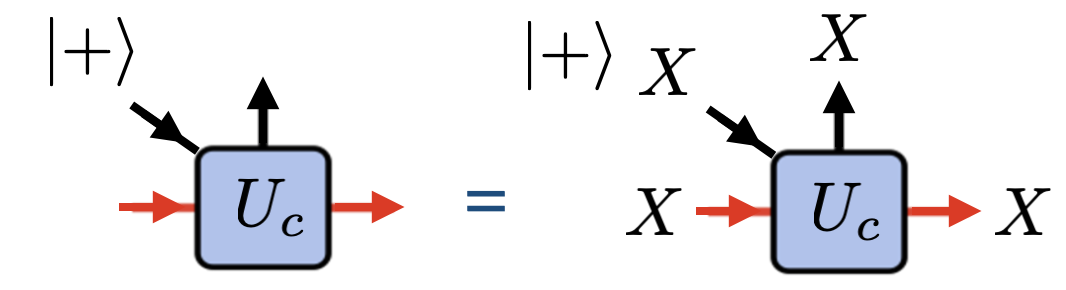}
\end{equation}
which is a manifestation that the local Clifford gate preserves the $\mathbb{Z}_2$ symmetry.

Finally, we emphasize that the MF symmetry can be satisfied even when the MPS represents a non-fixed-point or non-stabilizer state. A simple example is the deformed GHZ state $\ket{\psi(\beta)} \propto e^{\beta \sum_{i=2}^{N-1} X_i} \ket{\text{GHZ}_N}$ discussed in Ref.\cite{sahay2025classifying}, where $ \ket{\text{GHZ}_N} $ is defined as the $N$ qubit GHZ state. $\ket{\psi(\beta)}$ preserves the $\mathbb{Z}_2$ symmetry $\prod_i X_i$ and exhibits a non-decaying two-point correlation $\langle Z_i Z_{j \neq i} \rangle = \text{sech}(2\beta)^2$, indicating a GHZ-type long-range order for all finite $\beta$. As shown in Ref.\cite{sahay2025classifying}, $\ket{\psi(\beta)}$ can be prepared by an MF circuit for any $\beta$ due to the existence of the MF symmetry (Eq.\ref{eq:MF}) in its MPS representation. Such MF symmetry, with the structure theorem in Eq.\ref{eq:MF_symmetry}, then immediately implies the existence of a corresponding sequential unitary circuit under a spacetime rotation. Therefore, the MF-symmetry-based perspective developed in this section provides a useful framework for exploring spacetime duality in the preparation of general non-fixed-point states.

\section{Generalization of spacetime duality to higher dimensions}
\label{higherD}

In this section, we explore the duality between sequential unitary circuits and measurement-feedback circuits in higher dimensions. We will discuss the preparation of 2d GHZ state (Sec.\ref{sec:2dising}) and toric-code topological order (Sec.\ref{sec:2dtoric_main}), and some details regarding spacetime rotation and the explicit form of their associated rotated circuits are presented in Appendix.\ref{app:2dghz} and Appendix.\ref{app:2dtoric}, respectively. The discussion on fractal symmetry-breaking state and its related concepts, such as Kramers-Wannier duality and symmetry gauging with fractal symmetry, is presented in Appendix.\ref{sec:fractal}.

Below, we first summarize the essential steps of spacetime duality, built on our earlier discussion (see Table \ref{table1} for a succinct summary):

1. Begin with a sequential unitary circuit where the first unit cell sequentially entangles with the remaining unit cells. 

2. By applying a spacetime rotation to the sequential circuit, we obtain a dual Q-circuit. If the spacetime-rotated gates are unitary, the dual Q-circuit is then a finite-depth unitary circuit. If the gates are non-unitary, the dual Q-circuit can be viewed as a combination of finite-depth unitary operations and projections (see Appendix.\ref{app:dualcircuit}).

3. The dual-Q circuit acts on a doubled Hilbert space consisting of a layer of system qubits and a layer of ancilla qubits, where each system qubit and its corresponding ancilla qubit form a Bell pair, coming from the worldline of the first qubit (unit cell) at various time slices of the sequential circuit. Applying the dual Q-circuit produces an output state $\psi_{\text{out}}$, where the system and ancilla qubits now correspond to the output and input states, respectively, of the original sequential circuit.

4. Perform a projective measurement on the ancilla qubits to match the initial state of the sequential circuit. The post-measurement state then reproduces the wavefunction generated by the sequential circuit.

\subsection{2d GHZ state}\label{sec:2dising}
Here, we first discuss a sequential circuit for preparing the GHZ state in 2d, and then present the corresponding measurement-feedback circuit under a spacetime rotation.

\begin{figure}[h!]
\includegraphics[width=0.3\textwidth]{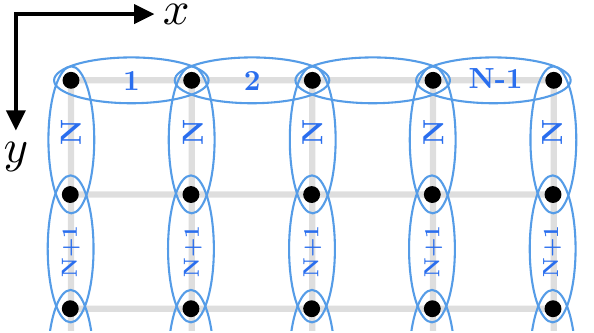}
    \caption{Sequential preparation of the GHZ state in 2d.}   \label{2d_GHZ_sequential}
\end{figure}

The 2d GHZ state can be prepared with a sequential circuit \cite{chen2024sequential} illustrated in Fig.\ref{2d_GHZ_sequential}: in the first part, one prepares the 1d GHZ state in the first row (at \( y = 1 \)) with the sequential application of the $u$ gate defined in Eq.\ref{eq:1dGHZ_before_swap}. In the second part, the $u$ gate is applied sequentially along the $y$ direction to order the qubits in every row sequentially, thereby generating the GHZ state in 2d.

To obtain a shallow circuit after spacetime rotation, we first need to rearrange the sequential circuit so that a fixed subsystem sequentially interacts with the other qubits. In the first part, the rearranged circuit has been discussed in Sec.\ref{sec:GHZ} (Eq.\ref{eq:1d_GHZ} specifically), which under an exchange between $x$ axis and time, one obtained a dual-$Q$ circuits that output a 1d cluster state, where every other qubit will be projected to $\ket{+}$, generating the 1d GHZ state in the remaining qubits.

In the second part, where one sequentially extends the GHZ-type order along the $y$ direction, we first treat each row of qubits as a \textit{unit cell} and rearrange the sequential circuit as:

\begin{align}\label{eq:2dghz}
   &\mathcal{U}=\prod_y U_{1,y}
\end{align}
Namely, the first unit cell sequentially interacts with the unit cells at various $y$ coordinates via the gate $U_{1,y}$: 
\begin{equation}\label{eq:2dse}
    U_{1,y}= \text{SWAP}_{1,y}\prod_x(e^{-i\frac{\pi}{4} X_{x,y}} e^{-i\frac{\pi}{4} Z_{x,1} Z_{x,y}}) 
\end{equation}
Here the operation $\text{SWAP}_{1,y}$ is the extensive applications of SWAP gates that exchanges the qubit at the first row and the $y$-th row.

    \begin{figure}[b]
\includegraphics[width=0.49\textwidth]{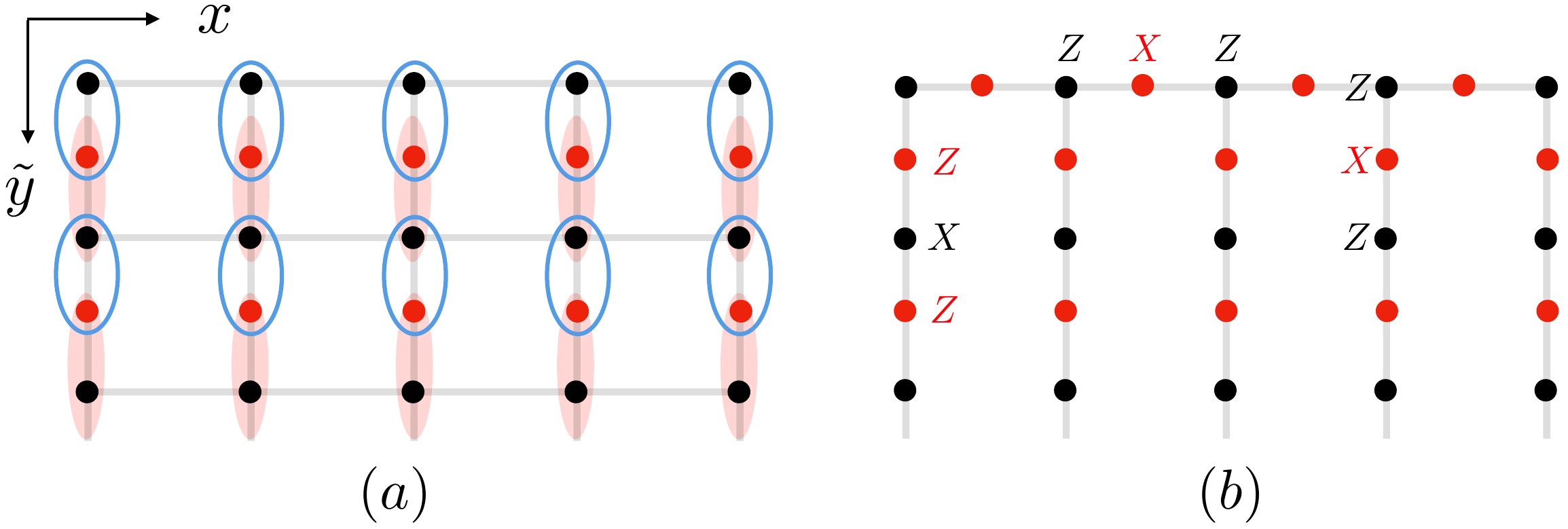}
    \caption{(a) The shallow circuit obtained by spacetime rotating the sequential circuit in Eq.\ref{eq:2dghz}: the bulk initially consists of Bell pairs (in pink ovals) extending along $\tilde{y}$ direction, and then one applies a depth-1 unitary circuit composed by the two-body gates in blue. Projecting the red qubit to $\ket{+}$ leads to a 2d GHZ state. (b) The short-range entangled state before projecting out the red qubits resembles a cluster state on a comb-like geometry. Projecting the red qubit to $X=1$ generates the GHZ order among the black qubits.}   \label{2dising}
\end{figure}

When exchanging the \( y \)-axis with the time axis, in addition to the original qubits on the vertices, one needs to introduce qubits on edges, which form Bell pairs (represented by filled pink ovals) with the neighboring vertices along $\tilde{y}$ direction (i.e. the original time direction in the sequential circuit); see Fig.\ref{2dising}(a). The unitary \(U_{1,i}\) in Eq.~\ref{eq:2dse} maps to the depth-1 unitary circuit (gate represented by blue ovals) that entangles the vertex and edge qubits. Finally, the red qubits emerging in this rotated circuit need to be projected to $\ket{+}$, which is fixed by the input state of the original sequential circuit. The detailed construction of this shallow circuit is presented in Appendix.\ref{app:2dghz}.

The process in the rotated circuit can again be thought of as $\mathbb{Z}_2$ symmetry gauging, where the dual-$Q$ circuit first prepares a 2d cluster state on a lattice with a comb-like geometry (Fig.\ref{2dising}(b)). The black qubits (on vertices) and red qubits (on edges) may be regarded as the matter fields and gauge fields, and three-body $ZXZ$ operators are the stabilizers of the cluster state (see Appendix.\ref{app:2dghz} for details). Projecting the gauge fields to $\ket{+}$ gives rise to the GHZ state with $\mathbb{Z}_2$ symmetry breaking. In particular, the gauge symmetry of the cluster state again implies that the projection can be deterministically implemented with a measurement and feedback based on the measurement outcomes, akin to the discussion in Sec.\ref{sec:gauge_symmetry}.

\subsection{2d toric-code topological order}\label{sec:2dtoric_main}
Here,  we first discuss a sequential unitary (SU) circuit for preparing the $\mathbb{Z}_2$ toric-code topological order. We then show that applying a spacetime rotation maps this SU circuit to a measurement-feedback (MF) circuit. This MF circuit mirrors the protocol discussed in Ref.\cite{Raussendorf_05, tantivasadakarn2021long,lu2022measurement} that transforms a $\mathbb{Z}_2$ 0-form $\times$ $\mathbb{Z}_2$ 1-form SPT to a $\mathbb{Z}_2$ toric code in constant depth. We also note that starting from a different MF circuit—motivated by the tensor-network representation of the toric code—a spacetime rotation yields an SU circuit distinct from the one presented here (see Appendix \ref{appendix:reverse_toric}).

We now begin by defining the toric code model. We consider a 2d lattice with each edge accommodating a qubit. The model Hamiltonian reads $H= - \sum_v A_v - \sum_p B_p$, where $A_v$ and $B_p$ are defined as
\begin{equation}
\includegraphics[width=5.5cm]{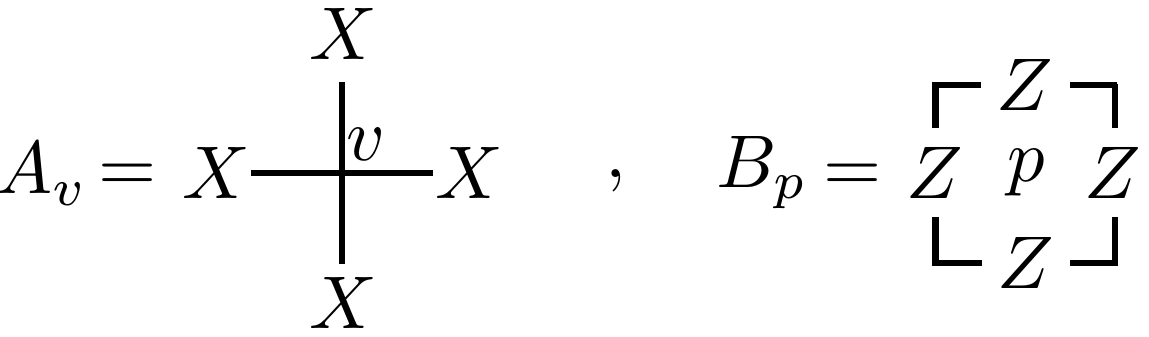}
\end{equation}
Namely, $A_v \equiv XXXX$ is the product of four Pauli-Xs around a vertex $v$ and  $B_p \equiv ZZZZ$ is the product of four Pauli-Zs around a plaquette $p$. Since every local term in the Hamiltonian commute, a ground state $\ket{\psi_{\text{TC}}}$ of this model satisfies $A_v \ket{\psi_{\text{TC}}}  =  B_p \ket{\psi_{\text{TC}}}  =  \ket{\psi_{\text{TC}}}$. Here we consider 

\begin{equation}\label{eq:toric_state}
 \ket{\psi_{\text{TC}}}   \propto  \prod_p (1+B_p) \ket{+++...},  
\end{equation}
which manifestly satisfies $A_v=B_p=1$. 

As discussed in the Refs.~\cite{chen2024sequential,chen2024quantum,hu2025preparing}, starting from $\ket{+++...}$, which automatically satisfies $A_v=1$ constraints, one can sequentially apply local controlled gates to enforce the $B_p=1$ constraint, thereby generating the toric-code state $\ket{\psi_{\text{TC}}}$. To illustrate the idea, let’s first focus on the four qubits in a single plaquette on the 2d lattice, and consider the four-body gate:  

\begin{equation}\label{eq:toric_U_p}
\includegraphics[width=8.2cm]{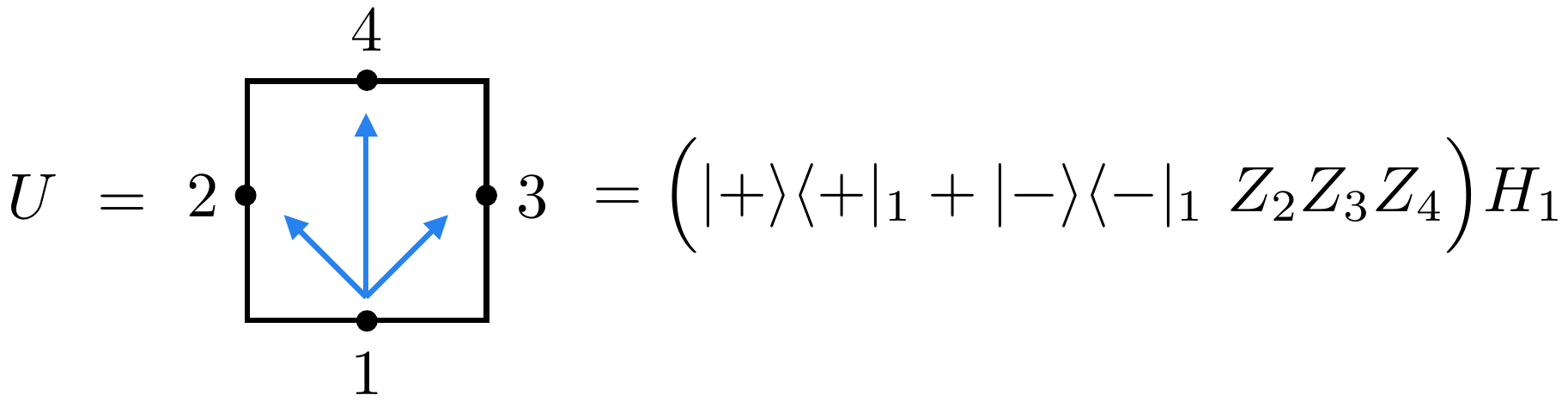}
\end{equation}
Namely, it is a Hadamard gate on qubit 1, followed by a controlled gate, where the application of $Z_2Z_3 Z_4$ depends on whether qubit 1 is in $\ket{+}$ or $\ket{-}$. As one can check explicitly, this unitary gate enforces the plaquette constraint $Z_1 Z_2Z_3Z_4 =1$ as long as the qubit $1$ is initialized at the Pauli-X eigenstate $\ket{+}$  
\begin{equation}
U \ket{+}_1  \ket{\phi}_{\overline{1}}\propto  \frac{1+Z_1 Z_2Z_3Z_4 }{2} \ket{+} _1 \ket{\phi}_{\overline{1}}
\end{equation} 
where $\ket{\phi}_{\overline{1}}$ is any state defined on the complement of the qubit $1$.

Therefore, by initializing all the qubits to $\ket{+}$, the toric-code state (Eq.\ref{eq:toric_state}) can be generated by sequentially applying the above $U$ gate to enforce the plaquette constraints on every plaquette: 
\begin{equation} 
\includegraphics[width=5cm]{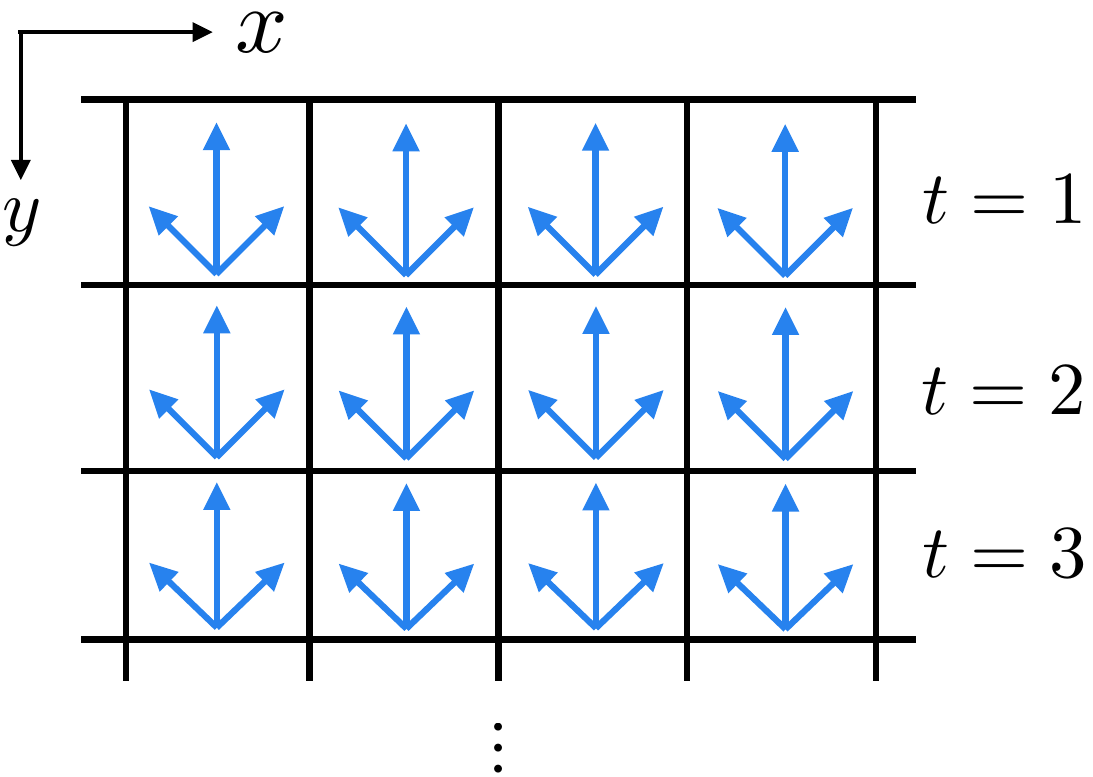}
\end{equation}
Note that since $U$ gates along the horizontal ($x$) direction commute, they can be implemented on a row of plaquettes simultaneously. As such, the circuit is only sequential in the $y$ direction, and the depth of the circuit scales linearly with the size of the lattice in the $y$ direction.

To discuss the corresponding spacetime-rotated circuit, we first rearrange the sequential circuit, so that the first unit cell sequentially interacts with the remaining unit cell. Specifically, we divide the 2d lattice into unit cells labeled by $i$ as follows. 
\begin{equation}
\includegraphics[width=5.1cm]{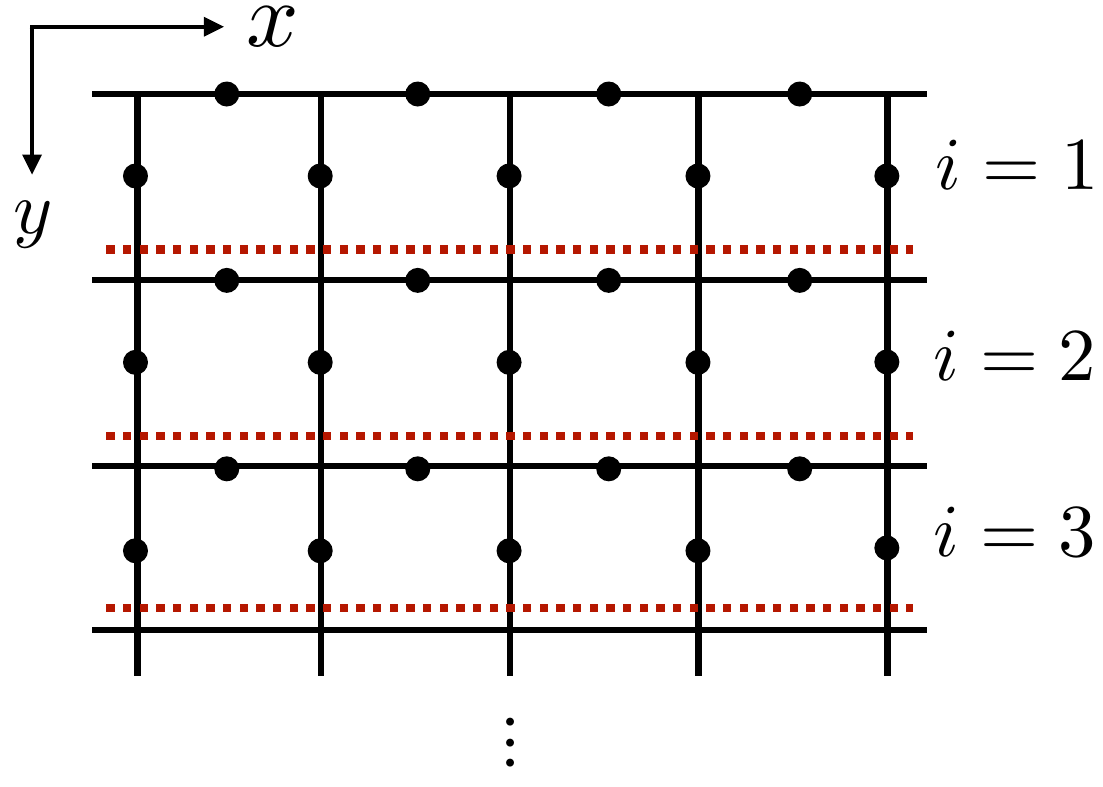}
\end{equation} 
where a unit cell consists of a row of $x$-oriented bonds and a row of $y$-oriented bonds. The sequential circuit is
\begin{equation}\label{eq:toric_sequential_circuit_main}
\mathcal{U} = \prod_{i=2}^{L_y} U_{1,i}
\end{equation} 

with 
\begin{align*}
\includegraphics[width=7.1cm]{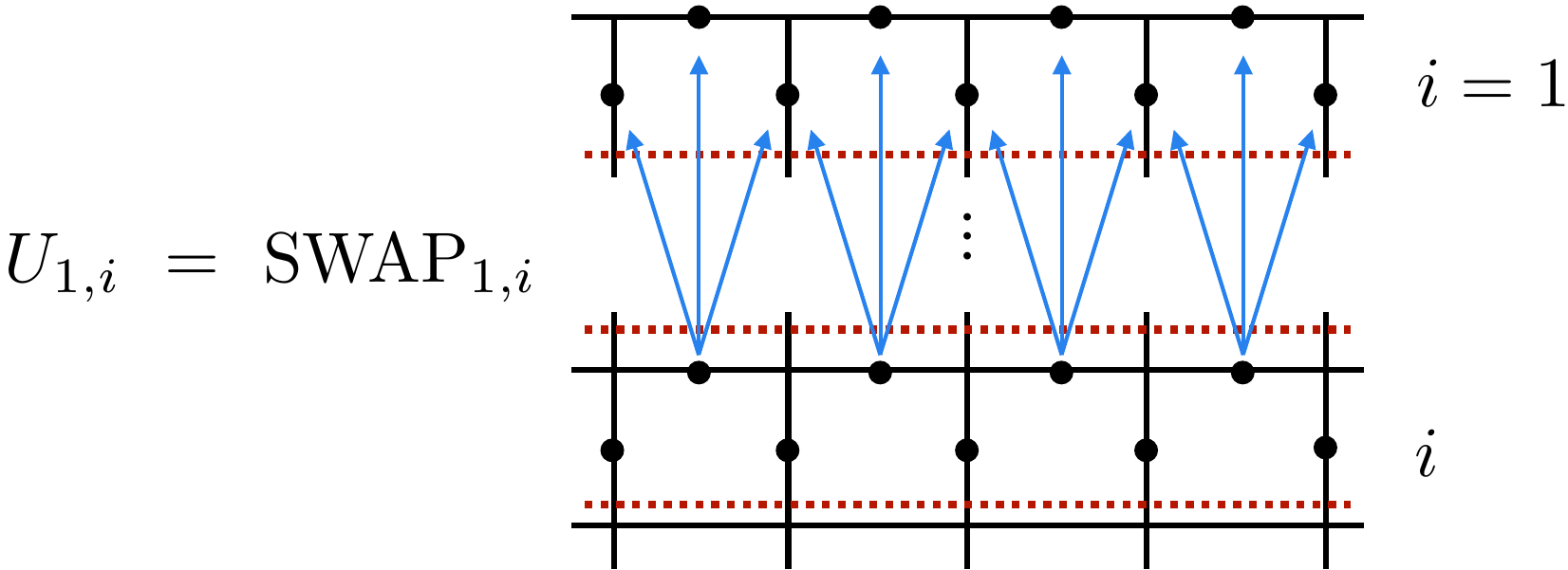}
\end{align*}

Namely, $U_{1,i}$ entangles the first unit cell and $i$-th unit cell with the gate defined by Eq.\ref{eq:toric_U_p}, followed by $\text{SWAP}_{1,i}$ that exchanges the two unit cells.

The sequential circuit Eq.\ref{eq:toric_sequential_circuit_main} performs the \textit{``generalized'' Kramers-Wannier duality} that can map a \(\mathbb{Z}_2\) paramagnetic state (e.g. $\ket{+++...}$) onto the deconfined phase of the \(\mathbb{Z}_2\) gauge theory (e.g. the toric-code state).  In particular, this duality implements the following mapping of bulk operators 
\begin{equation} \label{eq:toric_mapping}
\includegraphics[width=4.7cm]{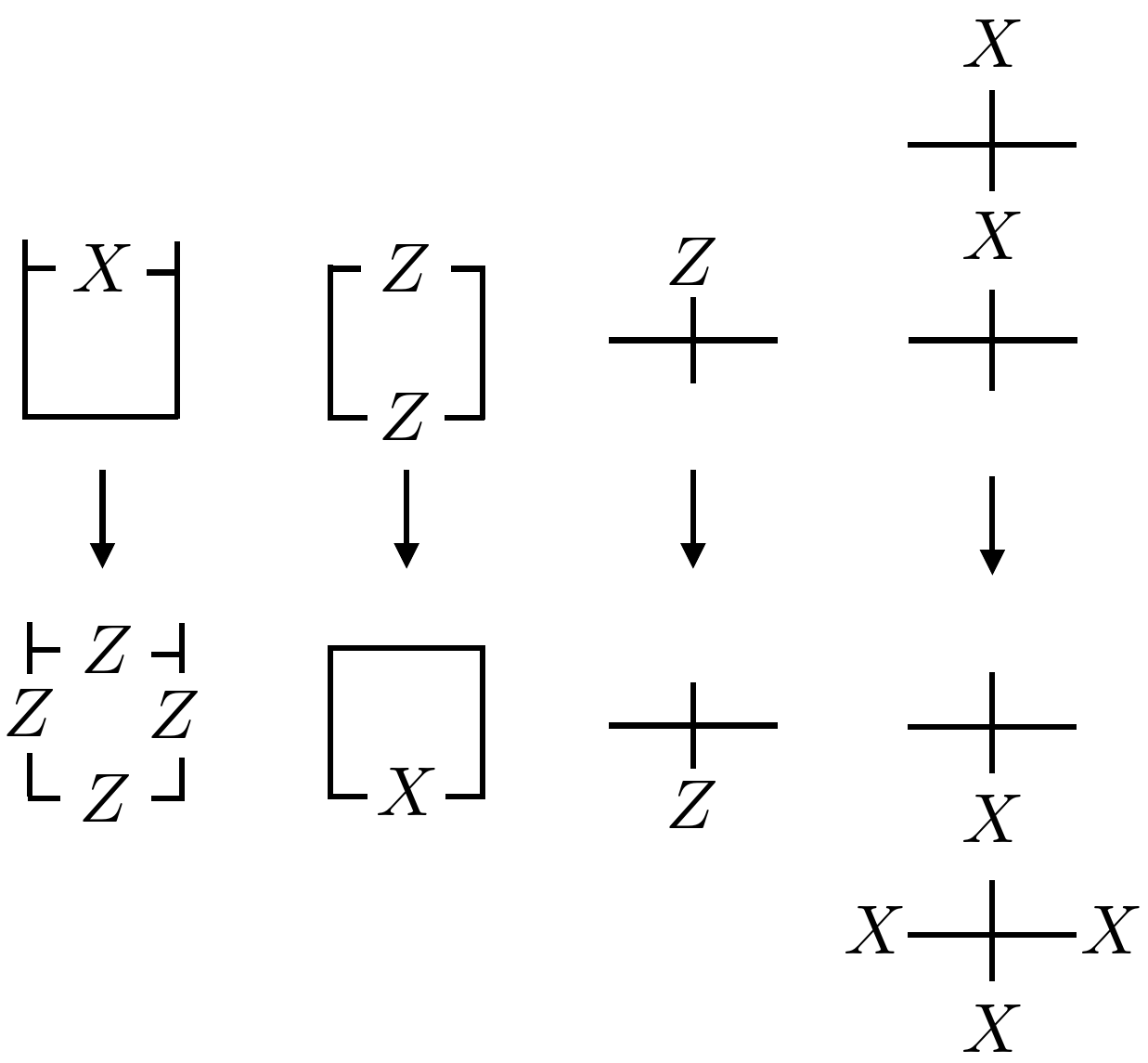}
\end{equation} 
These operator mapping rules also provides a simple way to understand the emergence of the toric-code state by evolving the trivial state $\ket{+++...}$ with the sequential circuit: the $X=1$ condition of the horizontal bond maps to the plaquette constraint $ZZZZ=1$ in the output, and the $XX=1$ condition of two vertical bonds on a vertex maps to the vertex constraint $XXXX=1$ in the output. These plaquette and vertex constraints then indicate the emergence of the toric code in the output.

\begin{figure*}[t]
		\centering
		\begin{subfigure}{\textwidth}
		\includegraphics[width=0.88\textwidth]{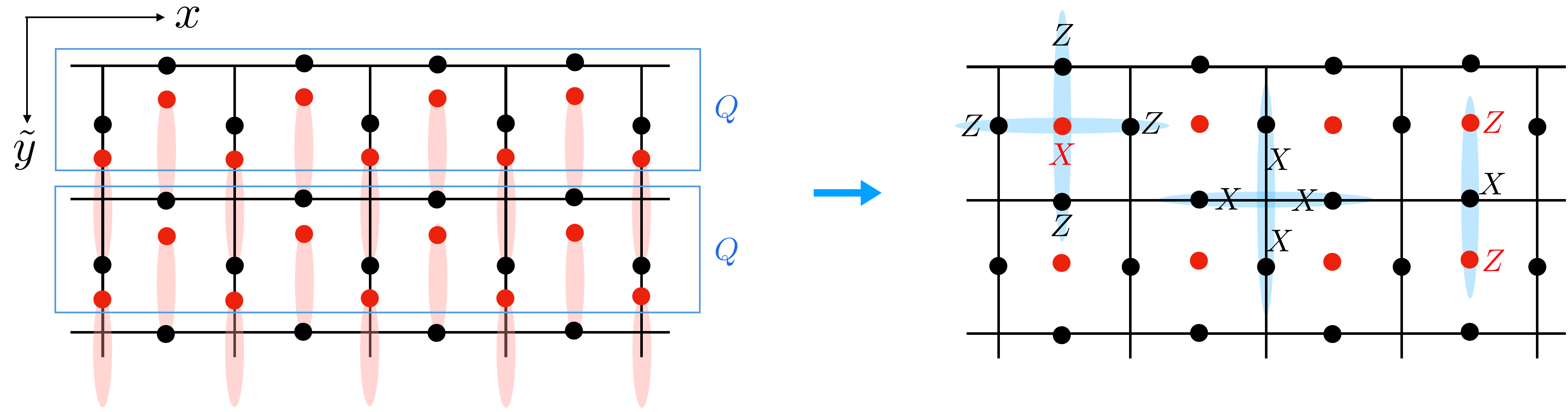}
	\end{subfigure}\caption{Left panel: by exchanging the $y$ direction and time direction of the sequential circuit (Eq.\ref{eq:toric_sequential_circuit_main}), one obtains a shallow circuit that prepares the toric code. The number of qubits is doubled, and the initial states are formed by decoupled Bell pairs in the bulk (marked by pink ovals). Under a shallow circuit of $Q$ gates (see Appendix.\ref{app:2dtoric}), followed by projecting the red qubit to $\ket{+}$, one obtains the toric-code state on the black qubits. Right panel: the overall process in the rotated, shallow circuit amounts to preparing a short-range entangled state $\ket{\psi_{\text{Lieb}}}$ on the Lieb lattice, followed by projecting the red qubit at the plaquette center to $\ket{+}$. Notably, $\ket{\psi_{\text{Lieb}}}$ resembles a cluster state with $\mathbb{Z}_2$ 0-form $\times \mathbb{Z}_2$ 1-form SPT order, which is known to generate the toric-code state upon measurement \cite{Raussendorf_05, tantivasadakarn2021long,lu2022measurement}.}
    \label{fig:toric_main}
	\end{figure*}

With the rearranged sequential circuit (Eq.\ref{eq:toric_sequential_circuit_main}), we now discuss the essential features of the spacetime-rotated (dual-$Q$) circuit obtained by exchanging the $y$ direction and time direction, and the detailed construction of the circuit can be found in Appendix.\ref{app:2dtoric}. In the spacetime-rotated description, the initial state of the dual-$Q$ circuit consists of Bell pairs in the bulk (see Fig.\ref{fig:toric_main}), which are spacetime-dual to the worldline of the first unit cell in the sequential circuit. Note that in the dual-$Q$ circuit, the number of qubits is doubled, and half of them will be projected to $\ket{+}$ since they are fixed by the initial state of the sequential circuit. After the application of the shallow dual-$Q$ circuit with $Q$ gate (i.e. the spacetime rotation of $U_{1,i}$ in Eq.\ref{eq:toric_sequential_circuit_main}), one obtains a short-range entangled (stabilizer) state, and notably, the stabilizers are fixed by the operator mapping rule of the generalized Kramers-Wannier duality (Eq.\ref{eq:toric_mapping}): 

\begin{equation}\label{eq:toric_dual_stabilizer}
\includegraphics[width=8cm]{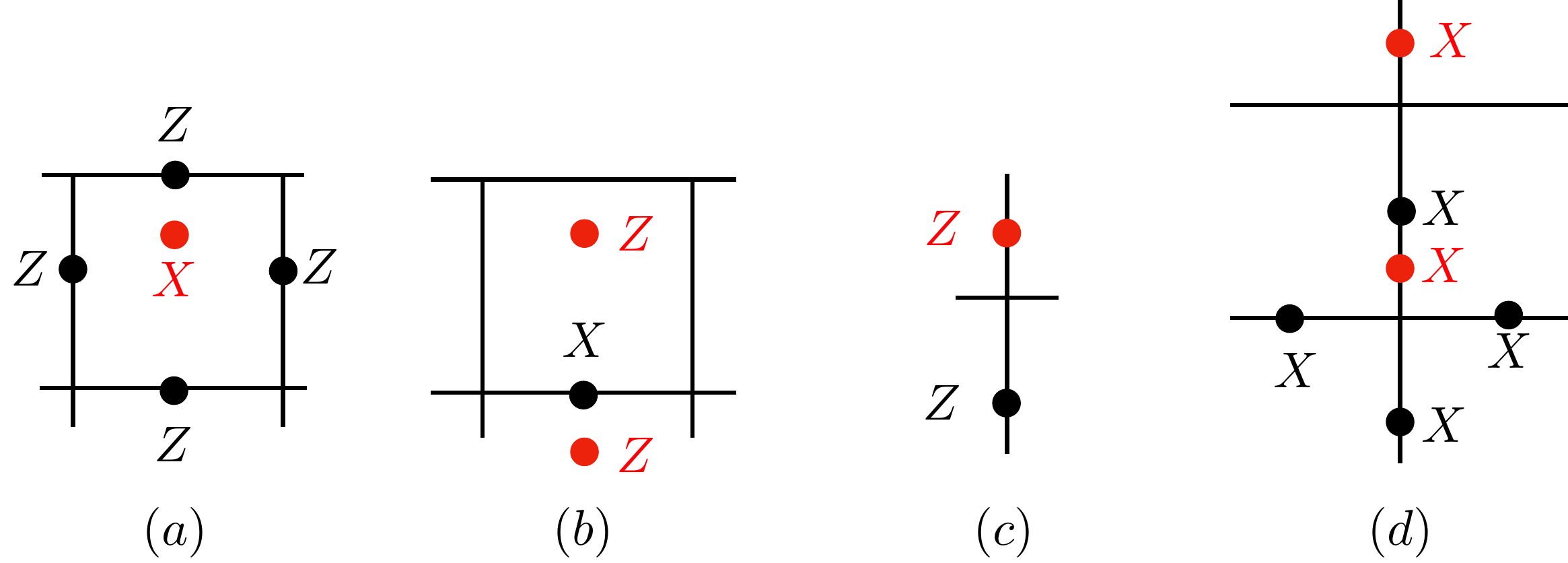}
\end{equation}
Alternatively, these stabilizers can be derived using the explicit dual-$Q$ circuit in Appendix.\ref{app:2dtoric}.

From conditions (a) and (d), we see that by projecting the red qubit to $\ket{+}$, the resulting state of black qubits will have the constraints that the product of four Zs around a plaquette equals one and the product of four Xs around a vertex equals one, indicating the toric-code order. These projections can be implemented by the single-site Pauli-X measurement and a subsequent feedback to effectively correct unwanted measurement outcomes. As such, the sequential circuit that prepares the toric-code state is related to a constant-depth measurement-feedback (MF) circuit via a spacetime duality. 

Notably, such MF circuit is akin to the well-known MF circuit, which first prepares a $\mathbb{Z}_2$ 0-form $\times \mathbb{Z}_2$ 1-form SPT $\ket{\psi_{\text{SPT}}}$ and applies single-site measurement and feedback to prepare the toric code in constant depth \cite{Raussendorf_05, tantivasadakarn2021long,lu2022measurement}. To illustrate the resemblance, we recall the definition of $\ket{\psi_{\text{SPT}}}$: consider a 2d square lattice with one qubit on every edge and one qubit on every vertex with periodic boundary conditions along both horizontal and vertical directions, $\ket{\psi_{\text{SPT}}}$ is the unique ground state of the stabilizer Hamiltonian $-\sum_v X_v \prod_{ e, ~  v  \in \partial e      }  Z_e -  \sum_{e}X_e \prod_{v,~ v   \in \partial e}Z_v$. The first term takes the form $X_v ZZZZ$, i.e. the product of a Pauli-X at the vertex $v$ and its four neighboring Pauli-Zs on edges, and the second term takes the form $X_e ZZ$, i.e. the product of a Pauli-X at the edge $e$ and its two neighboring Pauli-Zs on vertices. These two types of terms stabilizes $\ket{\psi_{\text{SPT}}}$, and we note that taking the product of $X_e \prod_{v,~ v   \in \partial e}Z_v  =1 $ around a single plaquette $p$ gives the constraint that $\prod_{e \in \partial p }X_e= 1$.

These stabilizers of $\ket{\psi_{\text{SPT}}} $ resemble the stabilizers emerging from our shallow dual-$Q$ circuit (Eq.\ref{eq:toric_dual_stabilizer}). To see this, we note that Eq.\ref{eq:toric_dual_stabilizer}(c) implies the red and black qubit on two adjacent vertical edges are locked in the Pauli-Z eigenbasis, so they can be regarded as a single effective qubit with the effective Pauli-Z and Pauli-X defined as $\tilde{Z}= Z_{\text{red}} = Z_{\text{black}}$ and $\tilde{X}= X_{\text{red}}X_{\text{black}}$. Consequently, every vertical edge effectively only accommodates a qubit, and the six-body stabilizer in Eq.\ref{eq:toric_dual_stabilizer}(d) can be reduced to four-body stabilizers $XX\tilde{X} \tilde{X}$. As such, the output of the shallow dual-$Q$ circuit enjoys the stabilizers as depicted in Fig.\ref{fig:toric_main} right panel, which resembles $\ket{\psi_{\text{SPT}}}$ defined on the dual lattice, and the subsequent $X=1$ projection of the red qubit leads to the toric-code state.

\section{Applications of spacetime duality}\label{applications}

As discussed in previous sections, spacetime duality reveals deep connections between sequential and measurement-feedback circuits. Here, we leverage these insights to demonstrate practical advantages for measuring unconventional physical properties of quantum states, such as disorder parameters, strange correlators, and measurement-induced long-range order. Specifically, spacetime duality facilitates two complementary measurement approaches: one minimizing spatial overhead and the other minimizing temporal overhead. This duality thus provides significant flexibility, enabling measurement protocols to be tailored to the strengths and constraints of diverse quantum hardware platforms.

\subsection{Probing spontaneous symmetry breaking}

Here, we explore how spacetime duality offers a new perspective on spontaneous symmetry breaking (SSB) in many-body states and motivates novel measurement protocols for probing SSB.

For concreteness, let's consider the 1d Ising model $-\sum_{i=1}^{N} Z_iZ_{i+1}$, which exhibits spontaneous symmetry breaking (SSB) of the global $\mathbb{Z}_2$ symmetry $\overline{X}=\prod_{i=1}^N X_i$. The SSB allows for a long-range entangled ground state - GHZ state $\propto(\ket{0}^{\otimes N}+\ket{1}^{\otimes N})$, which exhibits a symmetry-breaking order that manifests by the nonzero order parameter: $\langle Z_i Z_j\rangle =1 $. The non-vanishing order parameter is a universal feature of the SSB phase, whereas in the disordered phase (e.g. $-\sum_{i=1}^{N} Z_iZ_{i+1} - g \sum_{i=1}^N X_i $ with $g>1$ ), one finds a vanishing order parameter at long distances.

\begin{figure}[t]
    \centering
\includegraphics[width=0.45\textwidth]{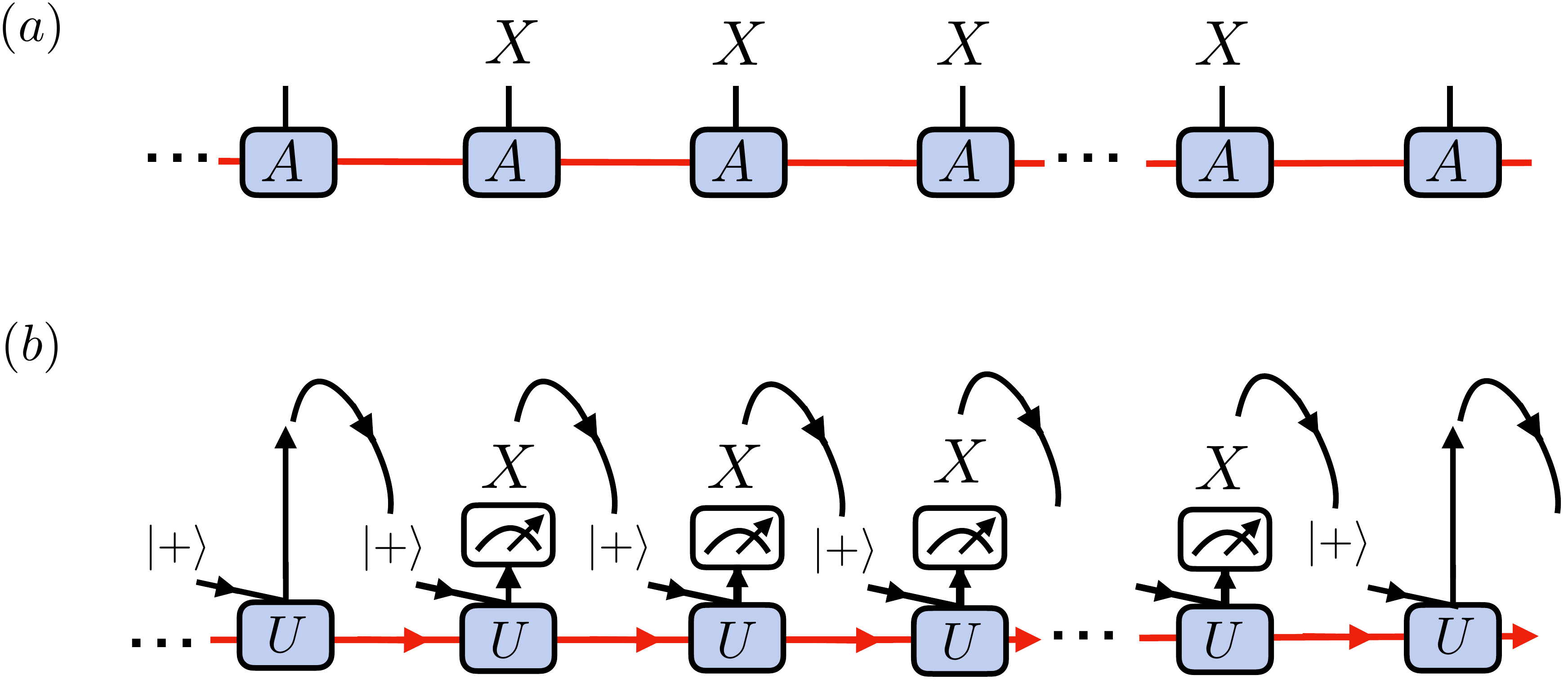}
    \caption{(a) A disorder operator is a string of Pauli-Xs acting on the physical legs of an MPS. (b) The MPS can be mapped to a sequential circuit under a spacetime duality. Two qubits (worldline depicted in red and black) are sufficient to measure the disorder operator of a 1d many-body state because the qubit with black worldline can be reset to $\ket{+}$ and reused to represent all the physical legs of the MPS.}\label{disorder}
\end{figure}

A complementary probe for the SSB is via the disorder parameter \cite{Fradkin_disorder_2017}, defined as the expectation value of the truncated symmetry generator in a subregion  \(\langle \prod_{a=1}^l X_{i+a} \rangle\), see Fig.\ref{disorder}(a). Since this operator creates domain walls (the location where $Z_jZ_{j+1} = -1$) on the boundary of the subregion, a non-zero expectation value implies the proliferation of domain walls, which leads to a disordered phase. In contrast, in the SSB phase, where the domain walls are suppressed, the disorder parameter decays exponentially with the string length $l$.

The spacetime duality provides a protocol for measuring the disorder parameter of a 1d state, using a sequential protocol that only requires $O(1)$ number of qubits. The central idea is performing a spacetime rotation of the MPS for the 1d state, so that it can be regarded as a sequential protocol, in which the first qubit sequentially interacts with the other qubits. Importantly, this only requires two qubits at a time, so that the second qubit, after interacting with the first qubit, can be reused as the third qubit to entangle with the first qubit as shown in Fig.\ref{disorder}(b). This provides a way to measure any operator of this MPS with a constant number of qubits, albeit increasing the temporal overhead. We note that this idea is essentially the same as the holographic simulation discussed in Ref.\cite{PhysRevA.103.042613,PhysRevResearch.3.033002,PRXQuantum.4.030334}. Below we will provide a distinct perspective for understanding the disorder parameter via the MF symmetry introduced in Eq.\ref{eq:MF_symmetry} and discuss the corresponding measurement protocol.

\begin{figure}[t!]
    \centering
\includegraphics[width=0.6\textwidth]{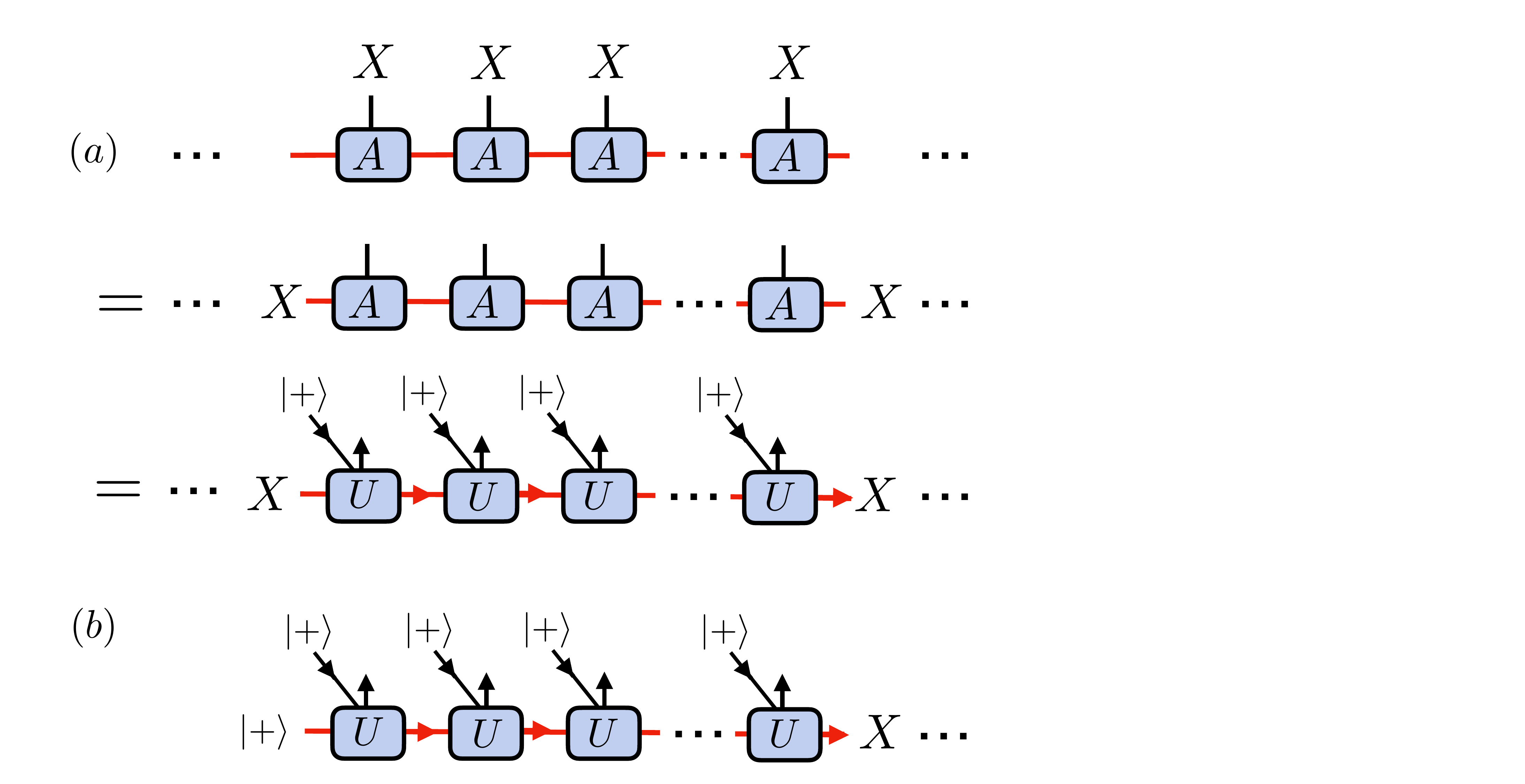}
    \caption{(a) A disorder operator acting on the physical legs of an MPS, with the MF symmetry, can be mapped to the two Pauli-Xs on the two virtual legs, which can be regarded as the autocorrelation function $\langle X(t_1)X(t_2) \rangle$ of the red qubit in the sequential circuit under a spacetime rotation. (b) This autocorrelation can be probed by initializing the red qubit to $\ket{+}$ and measuring its Pauli-X expectation value at another time slice.}\label{disorder_defect}
\end{figure}

If the local tensor has the MF symmetry where a single Pauli-X on the physical leg can be pushed to the two Pauli-Xs on the two virtual legs (Eq.\ref{eq:MF_symmetry}):

\begin{align*}
\includegraphics[width=3.3cm]{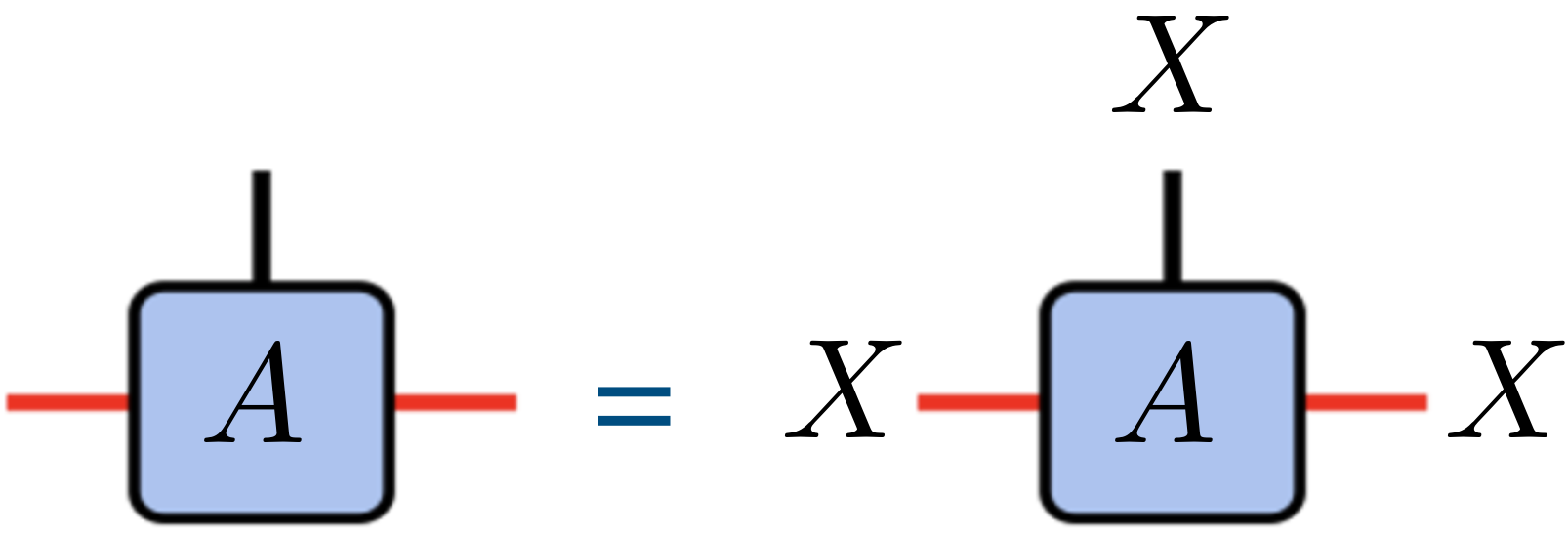}
\end{align*} 
then the disorder operator of a string of Xs is equivalent to inserting \(X\) operators at the virtual legs of the left and right endpoints of the string (Fig.\ref{disorder_defect}(a)), which can be regarded as inserting two defects to the original MPS. As such, the expectation value of the disorder operator \(\langle \prod_{a=1}^l X_{i+a} \rangle\) essentially measures the fidelity of the wavefunction under the insertion of such twisted defects. In the disordered phase, charges are localized (i.e. no long-distance correlations), meaning the insertion of defects has little effect on the wavefunction, thereby implying a high fidelity, and hence a non-vanishing value of the disorder operator. In contrast, in the SSB phase, charges exhibit strong spatial correlation, causing the insertion of defects to qualitatively alter the wavefunction, so the fidelity (and hence disorder parameter) decays exponentially with the separation between the two defects.

The spacetime duality provides a dual perspective for the disorder parameter. Specifically, the defect along the spatial direction of the MPS maps to the defect on the first (bond-space) qubit along the temporal direction of the sequential unitary circuit, as shown in Fig.\ref{disorder_defect}(a).  As such, the disorder parameter can be understood as the autocorrelation of the first qubit between two distinct time slices $\langle X_1(t)X_1(0) \rangle$. Physically, this autocorrelator measures the temporal evolution of the charge operator of the first qubit. When the sequential circuit generates a disordered wavefunction, charges remain localized, and the charge at the first site does not significantly spread with time. In contrast, for an SSB state, the sequential circuit necessarily promotes the spreading and fluctuations of charges across different sites. This spreading results in a distinct temporal behavior of the charge, causing the auto-correlator to decay exponentially over time. Thus, the temporal decay of the autocorrelator provides a clear indicator of the symmetry-breaking nature of the state. To probe the temporal correlation, one can initiate the first qubit at $\ket{+}$, and evolve under the sequential circuit. The autocorrelator $\langle X_1(t)\,X_1(0)\rangle$, i.e. the correlation of Pauli-Xs at two different time slices, is then reduced to $\langle X_1(t)\rangle$, which can be measured efficiently (Fig.\ref{disorder_defect}(b)). Alternatively, one can initialize an ancilla qubit together with the first qubit in the symmetric EPR state $\frac{|++\rangle + |--\rangle}{\sqrt{2}}$ so both qubits are aligned along the $S_x$ direction at $t=0$. During the sequential circuit the ancilla is left untouched, while unitary gates act only on the system. Under this setup the autocorrelator of the first spin, $\langle X_1(t)\,X_1(0)\rangle$, is mapped to the equal-time two-point correlator $\langle X_1(t)\,X_{\text{anc}}(t)\rangle$ that measures the correlation between the first qubit and its ancilla at time $t$.
The MF symmetry, along with a spacetime duality, therefore provides an efficient protocol for measuring the disorder operator of a 1d MPS with two qubits and a single Pauli-X measurement.

\subsection{Probing SPT via measurement-induced long-range order and strange correlators}
Here we leverage spacetime duality to design protocols for probing certain properties of symmetry-protected topological (SPT) orders \cite{chen11a,chen11b}, using the 1d cluster state as an example.

\begin{figure}[t!]
    \centering
\includegraphics[width=0.47\textwidth]{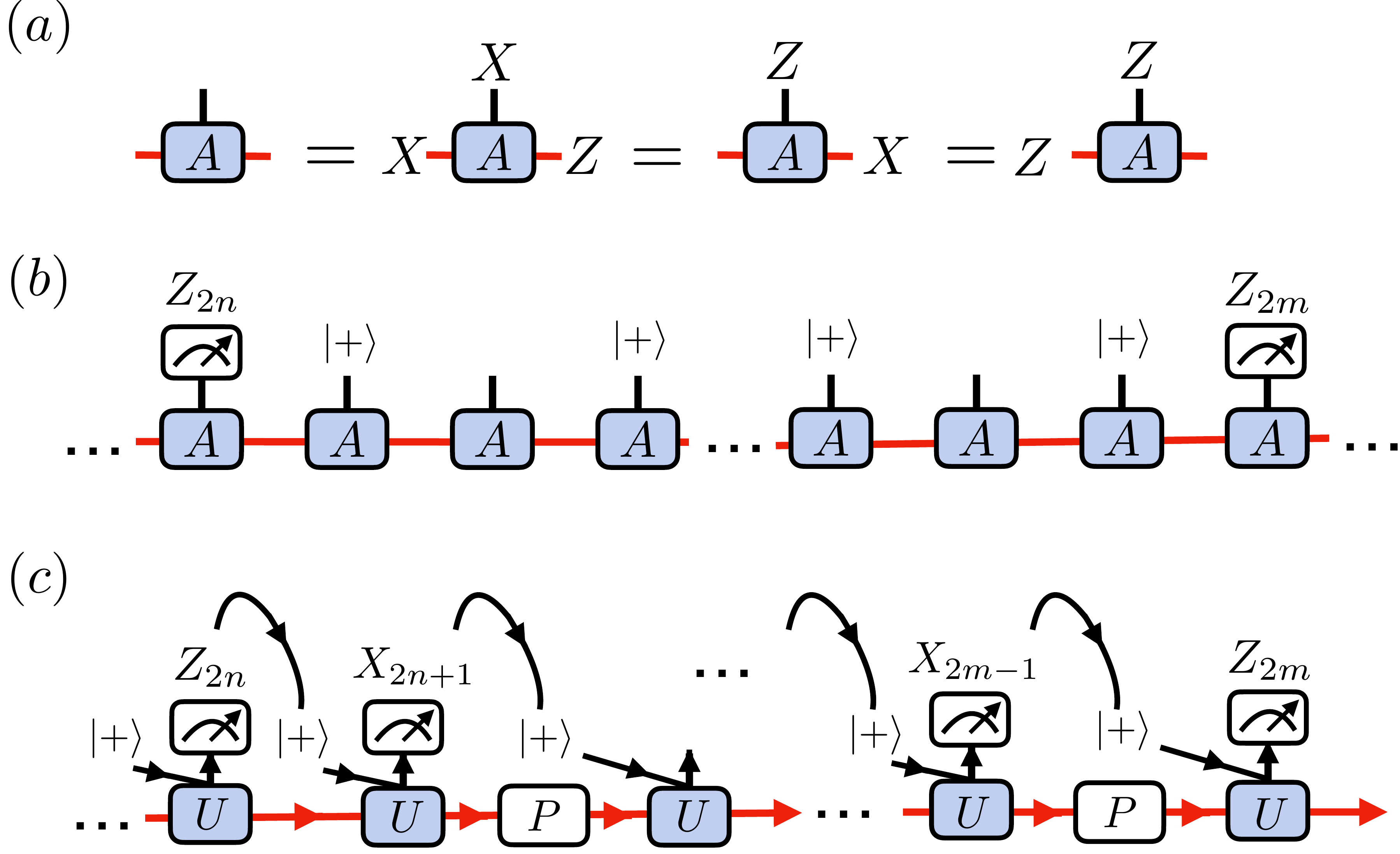}
\caption{(a) The MPS tensor of the cluster state possesses three independent stabilizer symmetries. (b) projecting all the odd qubits to $\ket{+}$ in between two even sites induces a long-range correlation $\langle Z_{2n}Z_{2m} \rangle=1$ w.r.t. the projected wave function. (c) A sequential protocol that measures the measurement-induced long-range order using two physical qubits, whose worldlines are marked by red and black. The qubit with black worldline is always reset to $\ket+$ before participating in the two-qubit unitary gate specified by the tensor $A$. The projection on odd sites can be implemented by Pauli-X measurements. The unwanted measurement outcome, can be compensated by a Pauli-$X$ feedback on the red qubit, due to the symmetry condition in (a).}
\label{fig:cluster}
\end{figure}

Consider a 1d lattice with periodic boundary condition, 1d cluster state is the ground state of the Hamiltonian $-\sum_{i=1}^N  Z_{i-1}X_i Z_{i+1}$, which can be expressed as $\ket{\text{cluster}} = \prod_{i=1}^{N} \text{CZ}_{i,i+1} \ket{+++...}$\footnote{$\text{CZ}_{i,i+1}$ is a two-qubit controlled-Z gate with the following action in the computational basis: $\text{CZ}_{i,i+1} \ket{\sigma_i,\sigma_{i+1}} = (-1)^{\sigma_i \sigma_{i+1}}\ket{\sigma_i,\sigma_{i+1}} $ with $\sigma_i, \sigma_{i+1}  \in \{ 0,1  \} $.}. It is an SPT protected by $\mathbb{Z}_2 \times \mathbb{Z}_2$ symmetries, generated by $\prod_{i \in \text{even}} X_i   $ and $\prod_{i \in \text{odd}} X_i$ defined on even sites and odd sites, respectively. It is an MPS with bond dimension 2, and the local tensor $A$ is uniquely determined by three symmetries depicted in Fig.\ref{fig:cluster}(a).  Below we will discuss the protocols for probing the two salient features of the state: the measurement-induced long-range order and the strange correlators. In particular, we will show that the former, in fact, provides a lower bound for the latter, which to our knowledge, was not discussed in the literature.

\textit{The measurement-induced long-range order}: 
the two qubits on the same sublattice (e.g. even sites) exhibit a long-range correlation when projecting the qubits at odd sites in between to the +1 Pauli-X eigenstate \cite{tantivasadakarn2021long}; see Fig.\ref{fig:cluster}(b). This follows from the string order parameter of the cluster state, i.e.  $\langle Z_{2n} X_{2n+1}   X_{2n+3} ...X_{2m-1} Z_{2m}   \rangle= 1$, and projection to $X=1$ in the bulk of the string leads to the long-range correlation $\langle Z_{2n} Z_{2m} \rangle= 1$ in the projected wave function.  To probe such long-range order experimentally, a natural approach is to prepare the $N$-qubit cluster state, perform single-qubit Pauli-X measurements on the intermediate sites, and apply appropriate feedback to correct for undesired outcomes. Finally, one measures the ZZ correlation between the two qubits.

The spacetime duality provides a protocol that only require two qubits for probing such a long-range order: in the spacetime-dual picture, the MPS of the cluster state can be interpreted as a sequential circuit, where a single (bond-space) qubit sequentially entangles with all the other qubits at different sites on the physical legs of the MPS; see Fig.\ref{fig:cluster}(c) \footnote{Technically, this sequential protocol is probing the cluster state in open boundary condition, which nevertheless will exhibit the same measurement-induced long-range order as in the case of periodic boundary condition.}. As in the previous protocols for measuring the disorder parameter, a single qubit (depicted with a black worldline) can be reset and reused to serve as the physical legs at different sites to entangle with the bond-space qubit (i.e. the qubit with a red worldline). When it serves as the physical leg at odd sites, one can perform Pauli-X measurement, aiming to project the qubit to $\ket{+}$. Crucially, the unwanted measurement outcome (i.e. $\ket{-}$), which corresponds to inserting a Pauli-Z defect on the physical leg, can immediately be compensated by a Pauli-X feedback acting on the bond-space qubit, using the symmetry condition in Fig.\ref{fig:cluster}(a). This therefore provides a qubit-efficient measurement protocol, in which one can repeat the protocol to collect the measurement outcome of $Z_{2n}, Z_{2m}$ and estimate the expectation value.

\textit{Strange correlators}: strange correlators offer an alternative diagnostic for detecting SPT order, in which standard correlation functions between local operators fail to distinguish the topological phase from a trivial phase \cite{you2014wave}. It reveals the hidden topological structure via the cross-correlation between the SPT wavefunction $\ket{\text{SPT}}$ and a trivial state $\ket{\text{Trivial}}$:
\begin{align}
&C(r)= \frac{ \bra{\text{Trivial}} O^{\dagger}_{r'} O_{r+r'}\ket{\text{SPT} }}{ \braket{\text{Trivial}|\text{SPT} }}. 
\end{align}
As shown in Ref.\cite{you2014wave}, the strange correlator exhibits long-range or quasi-long-range order in the SPT phase. For example, the cluster state, an SPT protected by a $\mathbb{Z}_2\times \mathbb{Z}_2$ symmetry, exhibits a long-range strange correlator 
\begin{align}
C =  \frac{ \bra{ +++...  } Z_{2n} Z_{2m}\ket{\text{cluster}} }{ \braket{+++...|\text{cluster} }} =1.
\end{align}
This can be derived using the $Z_{i-1}X_i Z_{i+1}$ stabilizer of the cluster state:

\begin{equation}
\begin{split}
    &~~~~~\bra{ +++...  } Z_{2n} Z_{2m}\ket{\text{cluster}}\\
    &=  \bra{ +++...    } Z_{2n}X_{2n+1} ... X_{2m-1} Z_{2m}\ket{\text{cluster}}  \\
&= \braket{ +++...|\text{cluster}}.
\end{split}
\end{equation}

Now we discuss a qubit-efficient protocol for measuring the strange correlators, inspired by spacetime duality. The protocol is a straightforward generalization of the protocol in Fig.\ref{disorder}(c) for measuring the disorder operator. First, we notice that the denominator of the strange correlator can be written as $ \braket{+++... |\text{cluster} } =  \bra{+++...} \prod_i \text{CZ}_{i,i+1} \ket{+++...}$ so it can be estimated by measuring the SPT entangler $\prod_i\text{CZ}_{i,i+1}$ w.r.t. $\ket{+++...}$. Instead of preparing the state $\ket{+++...}$ of $N$ qubits, we only require two qubits denoted by $A$ and $B$: \\

\noindent (i) Initialize $A, B$ in $\ket{+}$ and measure $\text{CZ}_{AB}$, obtaining a measurement outcome $\alpha_{12}$. \\

\noindent (ii) Reset  qubit $A$ to $\ket{+}$, and measure $\text{CZ}_{AB}$, obtaining a measurement outcome $\alpha_{23}$. \\

\noindent (iii) Reset qubit $B$ to $\ket{+}$, and measure $\text{CZ}_{AB}$, obtaining a measurement outcome $\alpha_{34}$. \\

\noindent  (iv) Repeat the steps (ii) and (iii) to obtain the collection of measurement outcomes $\{\alpha_{12}, \alpha_{23}, ...  \alpha_{N-1,N}\}$. The product $\prod_{i=1}^{N-1} \alpha_{i,i+1}$ then constitutes a single measurement data of $\prod_{i=1}^{N-1} \text{CZ}_{i,i+1}$ w.r.t. $\ket{+++...}$.  \\

Similarly, to measure the numerator of the strange correlator, one can write it as $\bra{ +++...  } Z_{2n} Z_{2m}\ket{\text{cluster}}  =\bra{ +++...  } Z_{2n} Z_{2m} \prod_i \text{CZ}_{i,i+1}\ket{+++...}$. One can then adopt the sequential protocol above to measure the expectation value, with the only difference being the two additional single $Z$ measurements at two different time slices.

By measuring the numerator and denominator separately, one can take the ratio to estimate the strange correlators. We note that since both the numerator and the denominator decay exponentially with system size $N$, obtaining an accurate estimate of the ratio may incur exponentially large sampling overhead. To address this issue, we now introduce an \textit{`averaged' strange correlator}. Interestingly, this quantity exactly corresponds to the measurement-induced long-range order discussed earlier. Moreover, it provides a lower bound on the strange correlator.

To begin, we first define the strange correlator  $C_{\alpha\beta}$ with the reference trivial state $\ket{\text{trivial}_{\alpha,\beta}}\equiv \ket{\alpha}_{2n}\ket{\beta}_{2m} \otimes_{i \neq 2m\neq 2n }\ket{+}$; $\alpha, \beta \in \{ +, - \}$ label the $\pm $ Pauli-X eigenstate: 
\begin{equation}
C_{\alpha\beta}=\frac{ \bra{\text{trivial}_{\alpha,\beta}}Z_{2n} Z_{2m}\ |\text{SPT}   \rangle }{\langle  \text{trivial}_{\alpha,\beta}| \text{SPT}\rangle },
\end{equation} 
Then we define $P_{\alpha,\beta}$: 

\begin{equation}
    P_{\alpha \beta} = \frac{  |   \langle  \text{trivial}_{\alpha,\beta}     |\text{SPT}\rangle|^2  }{ |  (\langle +   |^{\otimes N-2}) |\text{SPT}\rangle|^2   }, 
\end{equation}
which is the conditional probability of observing $\ket{\alpha}, \ket{\beta}$ at site $2n, 2m$, given the rest $N-2$ qubits are in the state $\ket{+}$. These ingredients allow us to define the `averaged' strange correlator

\begin{equation}
\bar{C}=\sum_{\alpha,\beta} P_{\alpha\beta} C_{\alpha\beta}
\end{equation} 
In fact, it is straightforward to check that this is exactly the expectation value of $Z_{2n } Z_{2m}$ w.r.t. the projected state $(\otimes_{i \neq 2n \neq 2m} \ket{+}\bra{+} ) \ket{ \text{SPT}}$ discussed earlier. Therefore, it takes a non-zero constant value even when the two sites are far separated, and it can be measured efficiently using the aforementioned protocol. We also note that it provides a lower bound for a strange correlator referenced to a particular trivial product state since 

\begin{equation}
    |\bar{C}|=|\sum_{\alpha,\beta} P_{\alpha\beta} C_{\alpha\beta} | \leq \sum_{\alpha,\beta} P_{\alpha\beta} |C_{\alpha\beta} | \leq |C_{\text{max}}|. 
\end{equation}
Here $|C_{\text{max}}|$ is defined as the maximal strange correlator over the four possible reference states $\ket{\text{trivial}_{\alpha,\beta}}$, with $\alpha, \beta \in \{ +, - \}$. The only caveat is that this maximal strange correlator does not need to be given by the reference state $\ket{\text{trivial}_{+,+}} = \ket{+}^{\otimes N}$ as in the conventional choice of strange correlators.

\section{Summary and outlook}\label{discuss}

In this work, we introduced a spacetime duality between sequential unitary (SU) circuits and measurement-feedback (MF) circuits. Within this framework, we demonstrated the equivalence between the Kramers-Wannier non-invertible transformation in SU circuits and the symmetry-gauging operation in MF circuits. Leveraging this duality, we developed efficient protocols for measuring nonlocal observables using sequentially generated quantum states. The discovery of such dualities naturally raises several intriguing questions for future investigation.

Firstly, our analysis primarily considered circuits that generate long-range order associated with spontaneous symmetry breaking (SSB) of Abelian higher-form symmetries, such as the 1d GHZ state and the 2d toric code. Extending our formalism to more general classes of quantum many-body states presents an important next step. Specifically, we anticipate that fixed-point topological orders described by non-Abelian quantum doubles of solvable groups will exhibit exact spacetime dualities between SU and MF circuits. This expectation arises because such states can be deterministically prepared using constant-depth MF circuits, reflecting the MF symmetry discussed in Sec. \ref{sec:reverse} \cite{tantivasadakarn2021long,bravyi2022adaptive,lu2022measurement,verresen2021efficiently,tantivasadakarn2023hierarchy,nat_shortest_nonabelian_2023}.

Secondly, given our focus on circuits generating fixed-point states, it is essential to investigate whether the observed long-range order and entanglement persist under circuit perturbations. For MF circuits, previous studies have partially explored stability aspects \cite{lee2022decoding,chen2023realizing,zhu2023nishimori,lu2023mixed,sahay2024finite,sahay2025classifying}; however, a comprehensive understanding remains lacking. Similarly, the stability of SU circuits warrants further exploration. We anticipate that the established spacetime duality could serve as a unifying framework, systematically addressing stability considerations for both SU and MF circuits.

A further promising direction opened by our work involves leveraging the equivalence between sequential circuits—a subclass of projected entangled pair states (PEPS)—and infinite-depth quantum channels~\cite{PhysRevLett.128.010607,gopalakrishnan2023push}, to establish connections between quantum channel classification and spacetime PEPS. Under these equivalences, the PEPS transfer matrix can be interpreted as a repeatedly applied quantum channel, with the boundary reduced density matrix corresponding to the steady state of this channel. The structure of this steady state is fully determined by the dominant eigenvector of the transfer matrix. By viewing dissipation and decoherence as intrinsic properties of the transfer matrix, this approach opens novel avenues for studying open quantum systems, such as exploring Lieb-Schultz-Mattis (LSM)-type constraints \cite{kawabata2024lieb} and spontaneous strong-to-weak symmetry breaking (SWSSB) in dissipative settings \cite{sala2024spontaneous,fidpaper}.

Finally, spacetime duality may provide an approach to overcoming the postselection bottleneck in measurement-based circuits lacking feedback mechanisms. Within this duality framework, postselection in measurement-based circuits translates into selecting specific initial states in sequential unitary circuits. Thus, the inherent scalability issues due to expensive postselection operations in measurement-based protocols could be mitigated by employing dual unitary protocols that significantly reduce this overhead. This strategy potentially enables the engineering of critical states exhibiting divergent correlation lengths in sequential unitary circuits by systematically adjusting the input qubit states \cite{lee2022decoding,zhu2023nishimori,haller2023quantum,PhysRevResearch.6.043256}.

\acknowledgments 
We thank Tarun Grover, Rahul Sahay, Ruben Verresen, and Nat Tantivasadakarn for their helpful discussions and comments. 
This work was performed in part at the Kavli Institute for Theoretical Physics (YY, SG), which is supported by NSF PHY-2309135. YY acknowledges support from NSF under award number DMR-2439118. TCL acknowledges the support of the RQS postdoctoral fellowship through the National Science Foundation (QLCI grant OMA-2120757).

\appendix

\section{Non-invertible KW duality vs. SU circuits}\label{appendix:KW}
Following in Ref.\cite{KW_Sahand_2024}, we here provide a self-contained discussion/clarification on (i) why Kramers-Wannier (KW) duality is a non-invertible transformation and (ii) why KW duality can be implemented by a sequential unitary (SU) circuit, which is nevertheless invertible. 

Consider a 1d lattice of size $L$ with a periodic boundary condition and one qubit per site. The KW duality implements the following operator mapping: 
\begin{equation}\label{eq:KW_appendix}
\begin{split}
X_i  &\rightsquigarrow  Z_iZ_{i+1}, \\
    Z_i Z_{i+1} &\rightsquigarrow   X_{i+1}
\end{split}
\end{equation}
The notation $\rightsquigarrow$ indicates the fact that this operator mapping can not be implemented by any invertible transformation. This can be proven by a contradiction: assuming that Eq.\ref{eq:KW_appendix} is implemented by an invertible operator $V$ (i.e. the inverse $V^{-1}$ exists), so that $VX_i V^{-1}   = Z_iZ_{i+1}$. Then by taking the product of this equation over all sites, one has $V (\prod_iX_i) V^{-1}   = \prod_{i=1}^L Z_iZ_{i+1}  =1$, meaning $V$ can map $\prod_i X_i$ to an identity operator, leading to a contradiction. As such, the KW duality cannot possibly be implemented by any invertible operator, and is hence non-invertible. This non-invertibility also manifests in the 1d transverse-field Ising model. Consider the fixed point of the spontaneous symmetry breaking (SSB) phase, i.e. $-\sum_i Z_i Z_{i+1}$, under the KW duality, it can be mapped to the fixed point of the symmetry unbroken phase, i.e. $-\sum_i X_i$. They have distinct ground-state degeneracies, so this mapping cannot be performed by any invertible transformation, which would have preserved the eigenspectrum. 

Given the above discussion, how should one reconcile the fact that there is a sequential unitary circuit that implements the KW duality? The resolution lies in the fact that while KW is non-invertible in the entire Hilbert space, it is invertible, and even unitary in the $\mathbb{Z}_2$ symmetric subspace specified by $\prod_i  X_i =1$. As such, there is a unitary mapping between the unique ground state in the symmetry-unbroken phase (e.g. $\ket{+}^{\otimes L}$), which is $\mathbb{Z}_2$ symmetric, and the $\mathbb{Z}_2$ symmetric superposition state in the SSB phase (e.g. $\ket{\text{GHZ}} \propto \ket{0}^{\otimes L}   + \ket{1}^{\otimes L}$), and this unitary mapping is exactly realizable by a sequential unitary circuit. Finally, we remark that this sequential circuit implements the KW duality only in the $\mathbb{Z}_2$ symmetric subspace; the operators not respecting the $\mathbb{Z}_2$ symmetry will be mapped to non-local operators (see Ref.\cite{chen2024sequential} for the explicit mapping).

\section{Spacetime rotation of unitary gates}\label{app:dualcircuit}
Here we discuss the general structure of the $Q$ gate that is obtained by a spacetime rotation on the $U$ gate. In particular, we will show that while $Q$ gate is generically non-unitary, it can be implemented by a unitary acting on the extended Hilbert space with an extra Bell pair, followed by a Bell projection (i.e. Bell-basis measurement with post-selection to the Bell pair $\ket{00}+ \ket{11}$).  

First, as discussed in Sec.\ref{sec:1d_dual_Q}, each unitary gate $U$ in the sequential circuit can be decomposed to a unitary gate  $u$ followed by a SWAP gate:  $U = \text{SWAP} ~u$. This can be diagrammatically represented as 
\begin{equation}\label{eq:appendix:u}
\includegraphics[width=3cm]{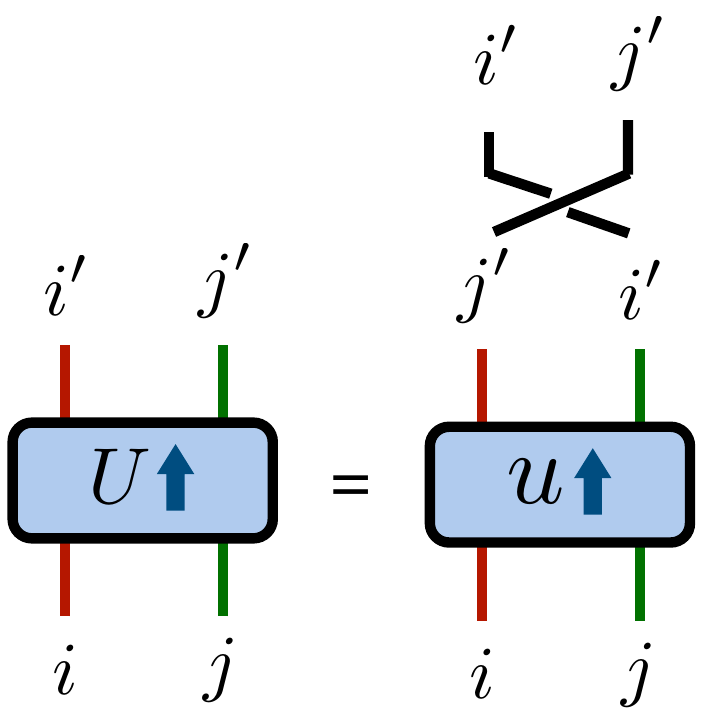}
\end{equation}
where the worldlines of the first and second qubits are colored in red and green, respectively.  In the GHZ preparation protocol with a sequential circuit, $u$ takes the form $e^{-i\frac{\pi}{4} X_{i}} e^{-i\frac{\pi}{4} Z_1 Z_{i}}$ (see Eq.\ref{unitary1}).

Under a spacetime rotation, the four-leg tensor $U_{i'j';ij}= \bra{ i',j'}U  \ket{i,j}$ can be used to define the $Q$ gate: $U_{i'j';ij}=  Q_{ j ,j' ; i ,i'} = \bra{j ,j'} Q  \ket{i,i'}$, which can be represented as 

\begin{equation}\label{eq:appendix:q}
\includegraphics[width=4.3cm]{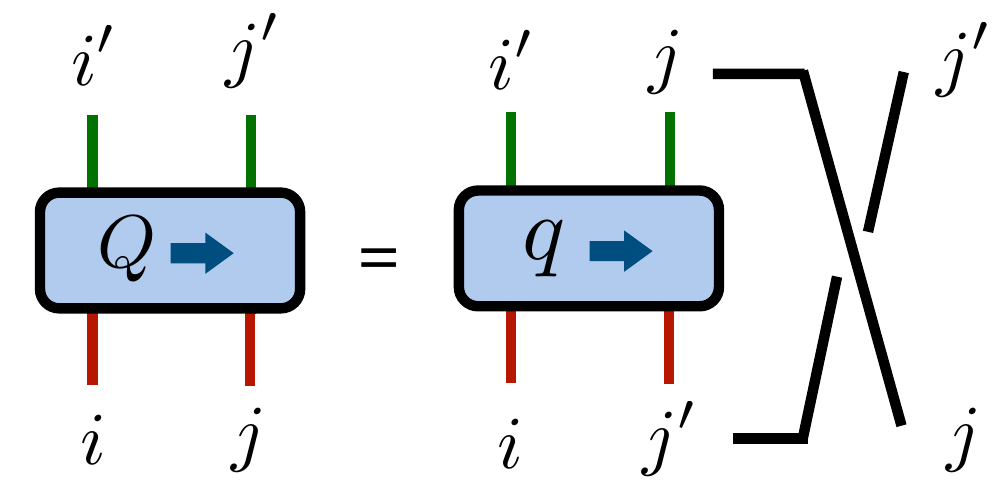}
\end{equation}
where the $Q$ gate is also decomposed to a $q$ gate followed by a SWAP gate. By comparing the $u$ gate and $q$ gate in Eq.\ref{eq:appendix:u} and Eq.\ref{eq:appendix:q}, one finds 

\begin{equation}
u_{j',i';i,j}    =  q_{ j',j  ;i,i'} 
\end{equation}
so that the $q$ gate is simply the partial transpose of the $u$ gate acting on its second qubit. Namely, 

\begin{equation}
     q = u^{T_2} 
\end{equation}

As a sanity check, one can consider the GHZ-state preparation protocol in Sec.\ref{sec:GHZ}, in which case, taking a partial transpose on the $u$ gate (Eq.\ref{unitary1}) exchanges the order of $ZZ$ rotation and $X$ rotation in the $q$ gate (Eq.\ref{eq:1d_q}).

Given this insight, now we show how to implement the $q$ gate, or equivalently, the partial-transpose gate $u^{T_2}$.  To begin, we note that $u^{T_2}$ can be represented as

\begin{equation}\label{}
\includegraphics[width=4.3cm]{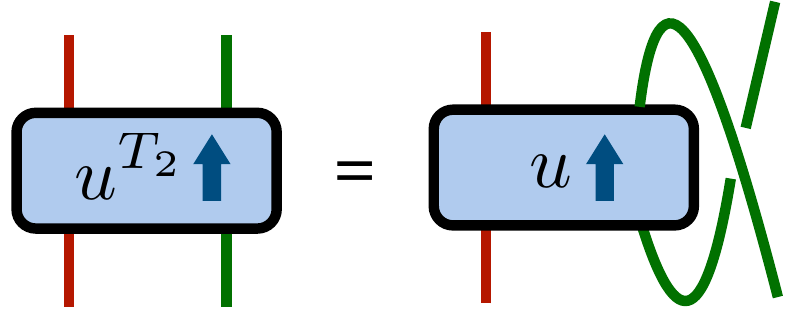}
\end{equation}
where the input and output of the second qubit are exchanged due to the partial transpose. This diagram immediately implies the following protocol for implementing  $u^{T_2}$:

\begin{equation}\label{}
\includegraphics[width=2cm]{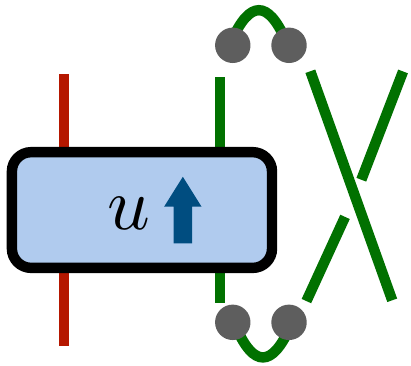}
\end{equation}
In this protocol, in addition to the two input physical qubits, we introduce two ancilla qubits in a Bell pair $\ket{00}+ \ket{11}$ (denoted by the two gray dots connected by a short green curve); one of the ancilla qubits participates in the $u$ gate and the other participates in the SWAP gate. Finally, the two ancilla qubits are projected again to the Bell pair $\ket{00}+ \ket{11}$. Note that the projection can be achieved by Bell-basis measurement, which has four possible outcomes: $Z^{a}X^{b}(\ket{00}+ \ket{11}  ) $ with $a,b \in \{ 0,1\}$, and we need to post-select the measurement outcome $a=b=0$. Completing the above steps implements the $q$ gate, and with an additional SWAP gate, the $Q$ gate (defined in Eq.\ref {eq:appendix:q}) can be implemented as well.

To summarize, given any sequential unitary circuit, a spacetime rotation leads to a shallow unitary-projective circuit that can be decomposed into local unitary gates, Bell-pair projection, and the single-site projection on a subset of qubits corresponding to the initial state of the original sequential circuit.

\section{From MF circuits back to SU circuits: 2d topological order as an example}\label{appendix:reverse_toric}
Here we extend the framework in Sec.\ref{sec:reverse} (the reversed side of the duality) to the preparation of the 2d toric code. To this end, we first review an MF circuit introduced in Ref.\cite{lu2022measurement}, which prepares the toric code based on its tensor-network representation.  

To begin, one can prepare copies of four-qubit clusters $A$ on every vertex of a 45-degree rotated square lattice:

\begin{equation}
\includegraphics[width=5cm]{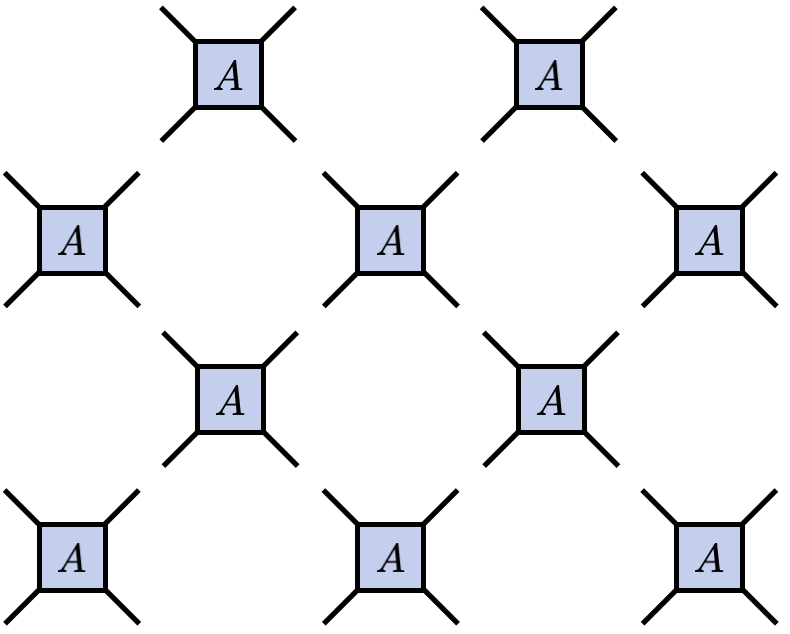}
\end{equation}
Each cluster is described by the state $\sum_{ijkl}  A_{ijkl}\ket{i,j,k,l}$, where $i,j,k,l \in \{ 0,1  \}$ denote the computational (Pauli-Z) basis, and the local tensor $A_{ijkl}= 1$ for $i+j+k+l= 0  \text{ mod  }2$ and $A_{ijkl}=0$ otherwise. Namely, the cluster is a uniform superposition of the configuration with even number of $1$, which is also a stabilizer state with the ZZZZ and XX stabilizers

\begin{equation}\label{eq:2d_toric_symmetry}
\includegraphics[width=5cm]{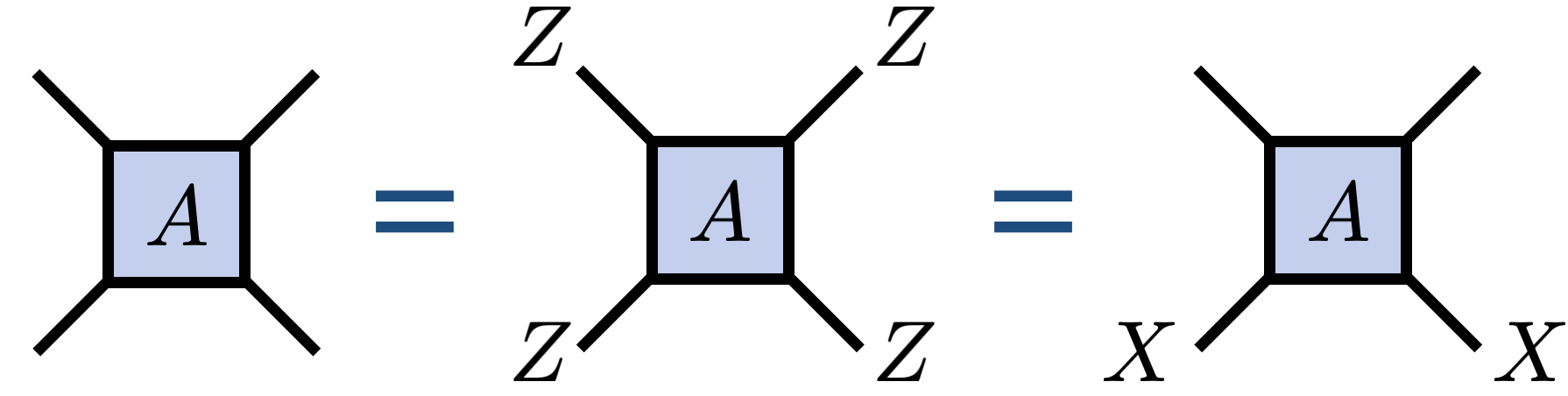}
\end{equation}
Note that the product of two Pauli-Xs on any two legs is a valid stabilizer, and we only plot one such XX stabilizer for 
brevity.

Given this initial state, one can extensively measure the two-body $ZZ$ operators (represented by the red ovals)  on edges connecting two neighboring clusters:

\begin{equation}
\includegraphics[width=5cm]{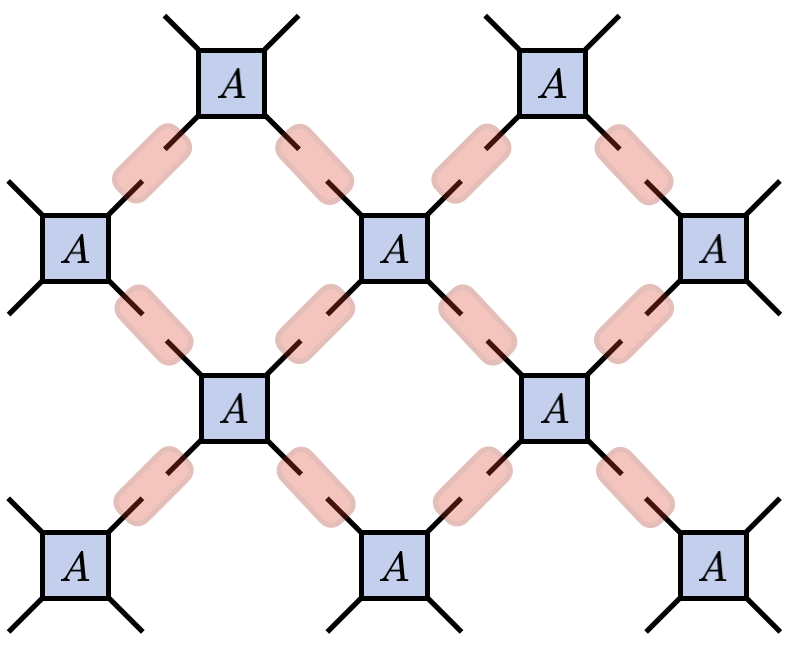}
\end{equation}

When a $ZZ$ measurement gives the outcome one, the two measured qubits take the same value in the Z basis so that they are effectively fused to a single qubit. Therefore, if the outcome is one for all measurements, the post-measurement pure state is exactly a toric-code ground state, a uniform superposition of computational basis state satisfying $ZZZZ=1$ constraint on all vertices; namely, it is a superposition of closed strings, where the absence or presence of strings corresponds to 0 or 1 configuration of a qubit. When an unwanted measurement outcome (i.e. $ZZ=-1$) appears, that amounts to inserting a defect of Pauli-X, which can annihilated using the XX symmetry in Eq.\ref{eq:2d_toric_symmetry} \footnote{If imposing periodic boundary condition on the 2d lattice to have the topology of a torus, then the outcomes $ZZ=-1$ appear in pairs, which can be paired up using the XX symmetry. If considering the open boundary condition, then the XX symmetry allows us to push the X-defects (associated with $ZZ= -1$ outcomes) to the boundary of the 2d lattice, which can then be annihilated as well.}.

We note that in this protocol, all the measured qubits are part of the long-range entangled state. This is in contrast to the MF circuit in Sec.\ref{sec:reverse}, where all the measured qubits decouple from the target long-range entangled state due to the use of Bell-pair measurements. To put the current protocol into the same framework, each local tensor can be added with two extra legs (qubits) by copying the bottom-right qubit and the bottom-left qubit. The previous extensive ZZ measurements can then be replaced with Bell-basis measurements:

\begin{equation}
\includegraphics[width=5cm]{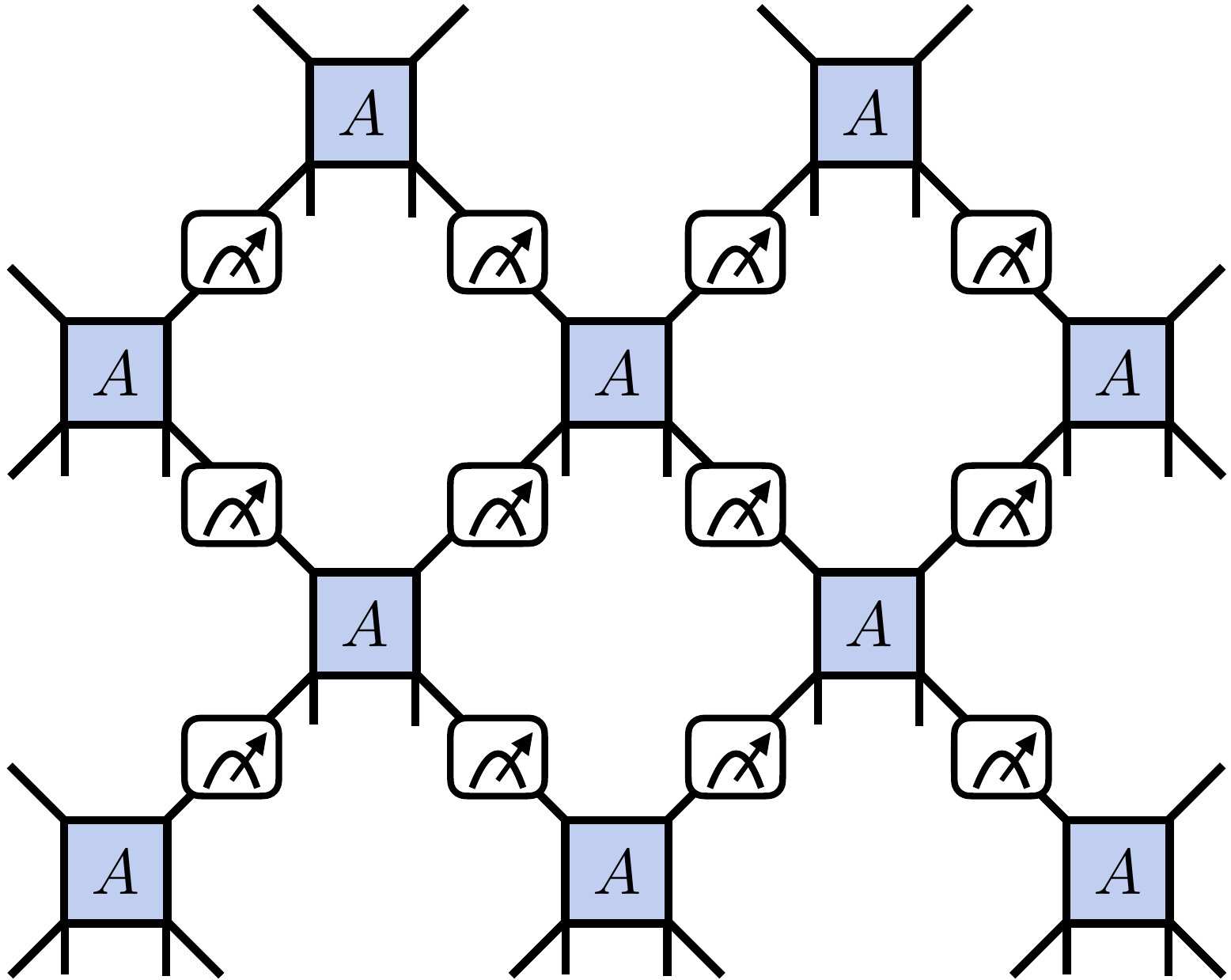}
\end{equation}

In other words, the newly added legs will be the physical qubits and the original four legs will be the virtual qubits, which are projected out after the measurements. When all measurements project to the Bell pair ($\ket{00} + \ket{11}$), all tensors can be contracted correctly, leading to the toric-code state on the unmeasured (physical) qubits. Importantly, the three other unwanted measurement outcomes, corresponding to $Z,X, ZX$ defects on edges, of a single Bell-basis measurement can be corrected using the symmetry of the local tensor: 

\begin{equation}\label{eq:2d_toric_added_symmetry}
\includegraphics[width=6cm]{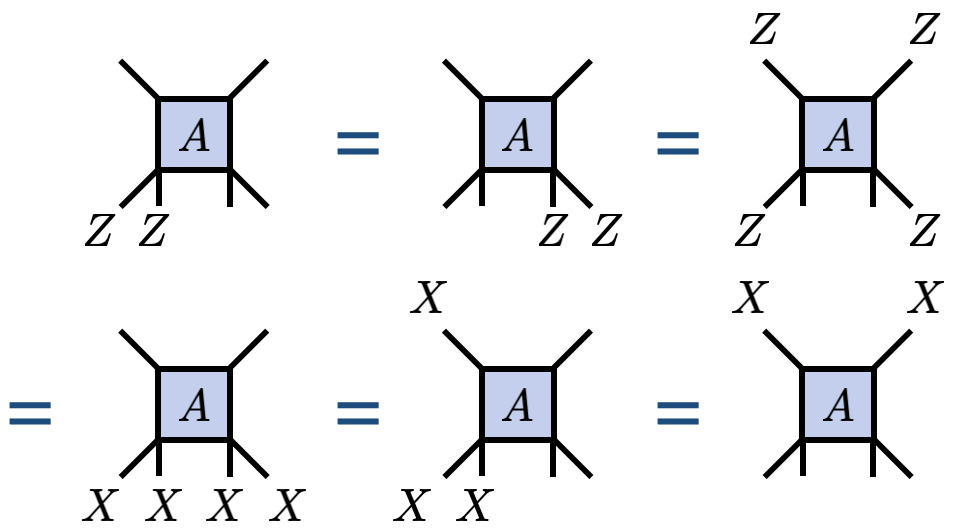}
\end{equation}

Notably, these symmetries, which allow for the feedback correction in the MF protocol, also imply that the 6-leg tensor can be regarded as a four-qubit unitary gate $U$, with two of the input legs fixed at $\ket{0}$:

\begin{equation}
\includegraphics[width=3.5cm]{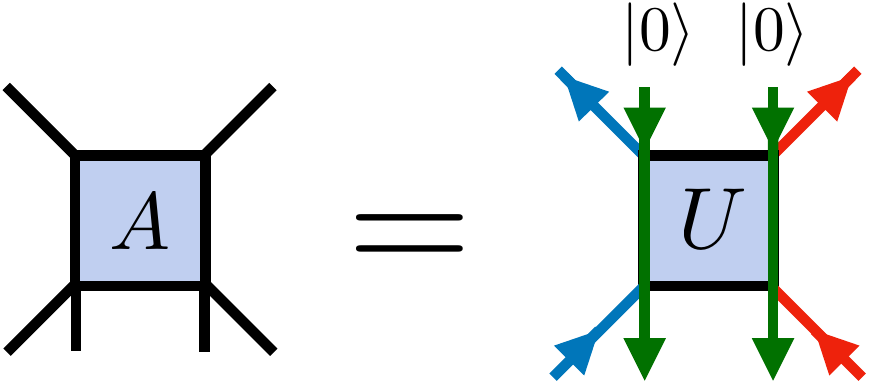}
\end{equation}
where the ingoing and outgoing arrows denote the inputs and outputs, respectively\footnote{This follows from the following observation: using the MF symmetry in Eq.\ref{eq:2d_toric_added_symmetry}, one can check that any single Pauli operator on the red or blue input can be pushed to Pauli operators on the red/blue/green output legs. This implies the tensor $A$ is an isometry from $ (\mathbb{C}^2)^{\otimes 2} \to  (\mathbb{C}^2)^{\otimes 4} $, which can be further extended as a unitary by introducing two extra input legs with a fixed initial state.}.

As such, under a spacetime rotation, the MF protocol can be viewed as a sequential unitary circuit for generating the toric-code state, depicted as follows:

\begin{equation}
\includegraphics[width=6.6cm]{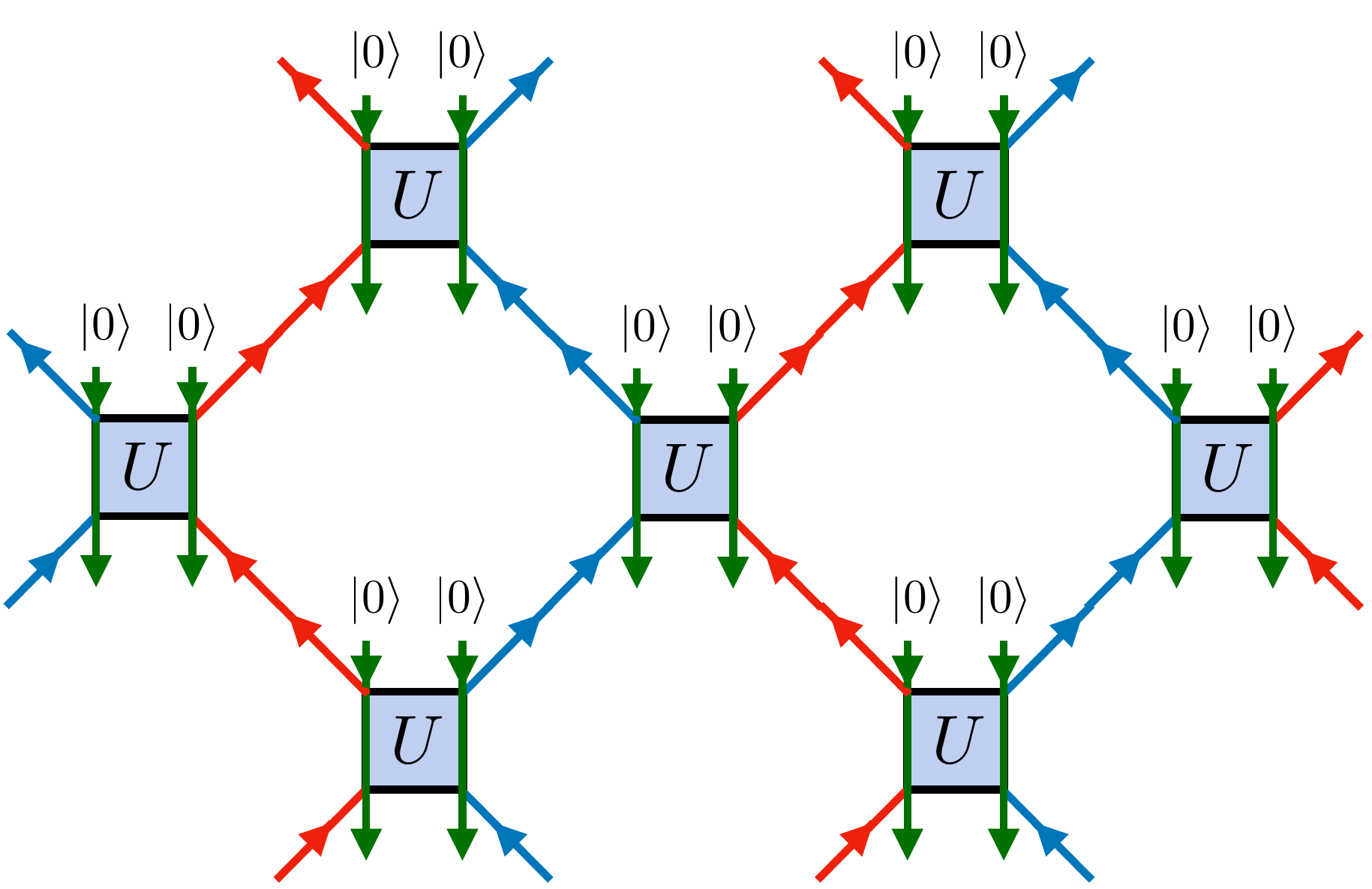}
\end{equation}

Here, the blue/red trajectories represent the world lines of the qubits that sequentially entangle the two qubits (in the state $\ket{0}$) on edges via the four-qubit gate $U$. In particular, the symmetry of a local tensor (Eq.\ref{eq:2d_toric_added_symmetry}) implies the output of the sequential circuit is a toric-code ground state satisfying $A_v = ZZZZ =1$ on every vertex $v$ and $B_p=XXXX=1 $ on every plaquette $p$:

\begin{equation*}
\includegraphics[width=8.5cm]{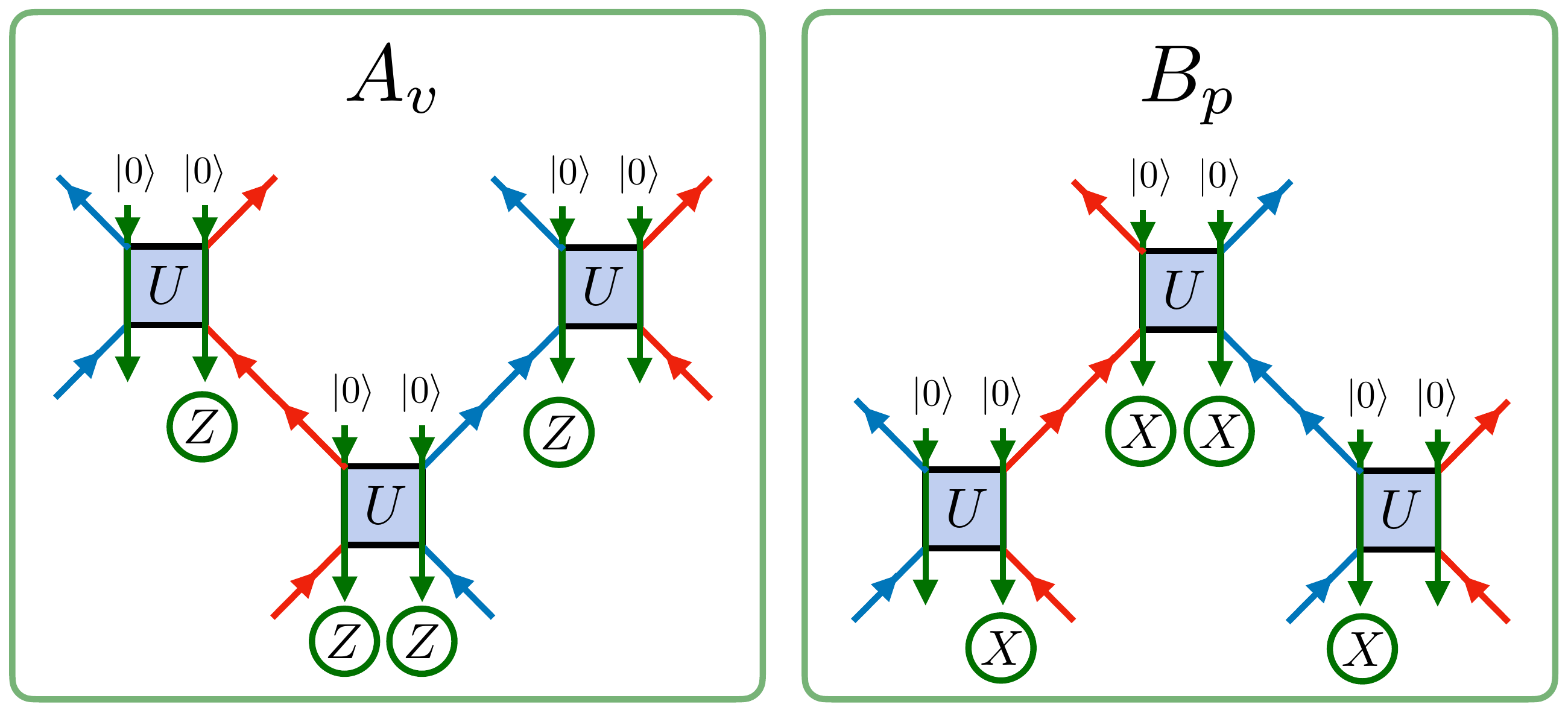}
\end{equation*}

To summarize, the MF symmetry in the MF circuits implies that a local tensor can be viewed as an isometry, which can be further extended to a unitary gate by inserting extra input product states. As such, various MF circuits (described by tensor-network formalism) in Ref.\cite{lu2022measurement} can be understood as the sequential unitary circuits for preparing long-range entangled states. For instance, following a similar circuit structure for the 2d toric code, one can construct the sequential circuit for preparing 2d abelian quantum double models \cite{Kitaev:1997wr} by spacetime-rotating the corresponding MF circuits.

\section{The dual Q-circuit for 2d GHZ state}\label{app:2dghz}

Here we provide the details of the dual-Q circuit by spacetime-rotating the sequential circuit that prepares the 2d GHZ state. As discussed in the main text, starting from the 1d GHZ state along the $x$ direction, the 2d GHZ state can be prepared by a sequential unitary circuit that extends the ferromagnetic order along the $y$ direction. Specifically, by treating each row of qubits (along the $x$ direction) as a \textit{unit cell}, the sequential circuit can be rearranged as 

\begin{align}\label{}
   &\mathcal{U}=\prod_y U_{1,y}
\end{align}
where the first unit cell sequentially interacts with unit cells with different $y$ via the gate $U_{1,y}$: 
\begin{equation}
    U_{1,y}= \text{SWAP}_{1,y}\prod_x(e^{-i\frac{\pi}{4} X_{x,y}} e^{-i\frac{\pi}{4} Z_{x,1} Z_{x,y}}) 
\end{equation}
with $\text{SWAP}_{1,y}$ being the extensive applications of SWAP gates that exchange the qubit at the first row and $ y$-th row. This circuit can be schematically depicted as follows:

\begin{equation}\label{}
\includegraphics[width=6cm]{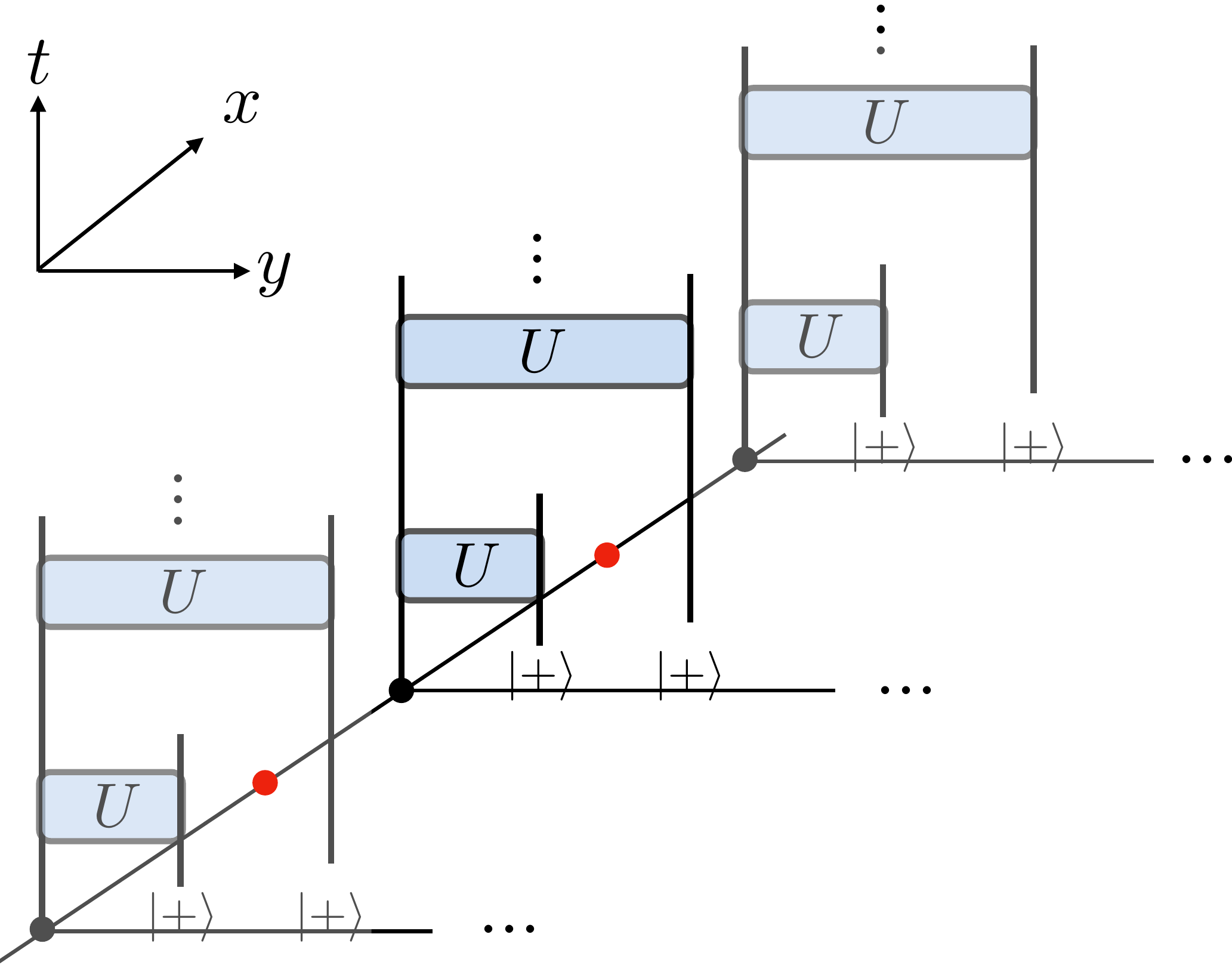}
\end{equation}  

Note that the 1d GHZ state at the first unit cell can be thought of as the 1d cluster state involving the black and red qubits along the 1d line, followed by projecting the red qubits to $\ket{+}$. 

By exchanging the \( y \)-axis with the time axis of the figure above, one obtains a shallow dual-$Q$ circuit acting on an initial state involving Bell pairs extending along the $\tilde{y}$ direction (i.e. the original time direction):

\begin{equation}\label{}
\includegraphics[width=5cm]{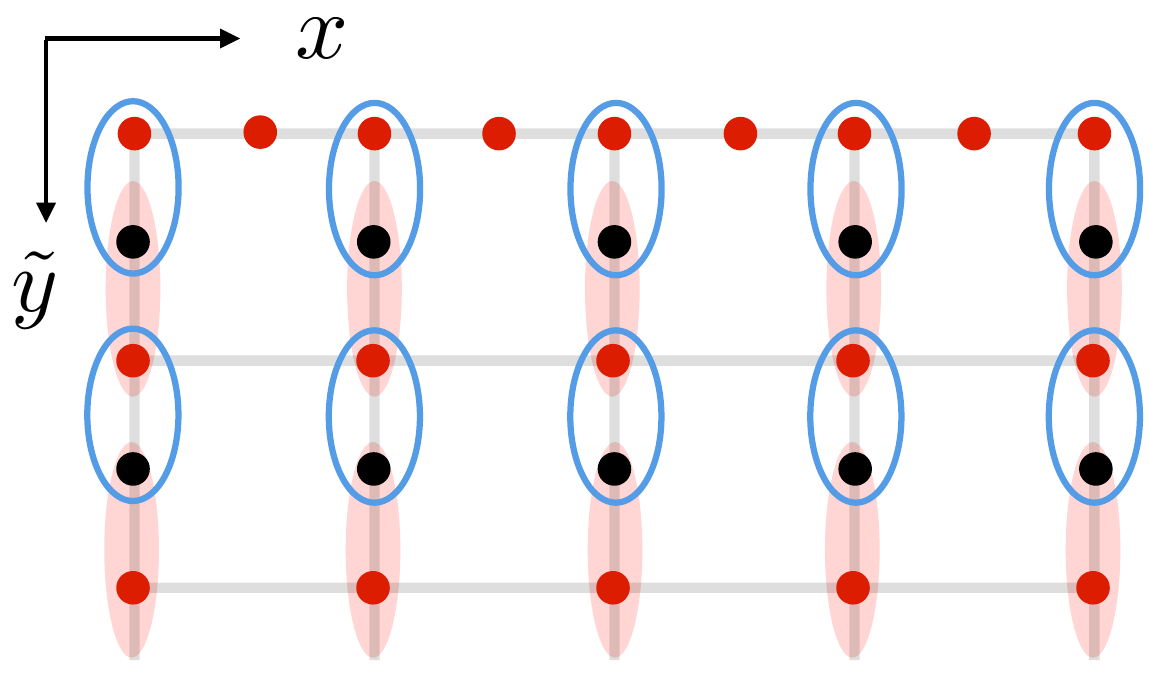}
\end{equation}  
The dual-$Q$ circuit is a depth-1 unitary circuit composed by the $Q$ gate defined as

\begin{equation}\label{}
\includegraphics[width=5.5cm]{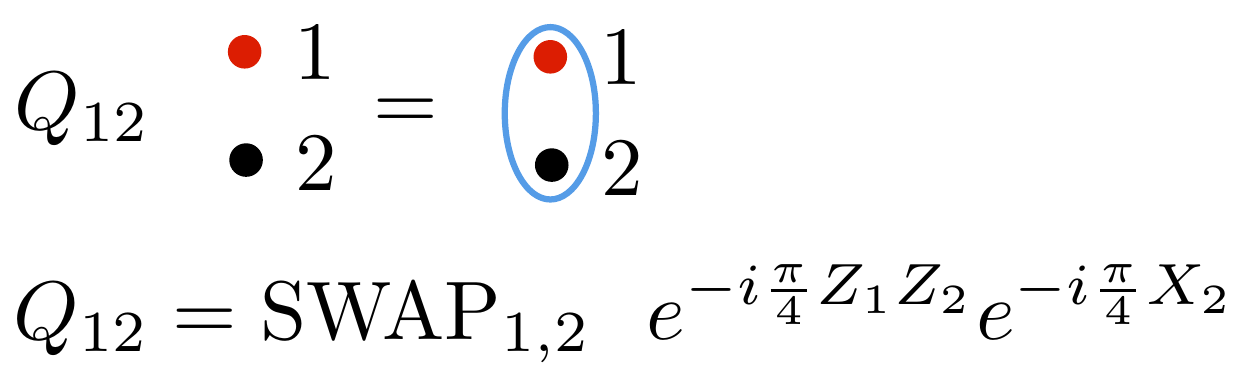}
\end{equation}  
After applying this circuit, all the red qubits need to be projected to $\ket{+}$ (fixed by the initial state in the sequential unitary circuit), producing the 2d GHZ state. Alternatively, one can use the SWAP gates in the $Q$ gate to swap the locations of the black and red qubits so that after the depth-1 circuit composed by the $q$ gate (green ovals) depicted below, projecting the red qubit (now on living on edges) to $\ket{+}$ leads to the GHZ state on vertices:

\begin{equation}\label{}
\includegraphics[width=5cm]{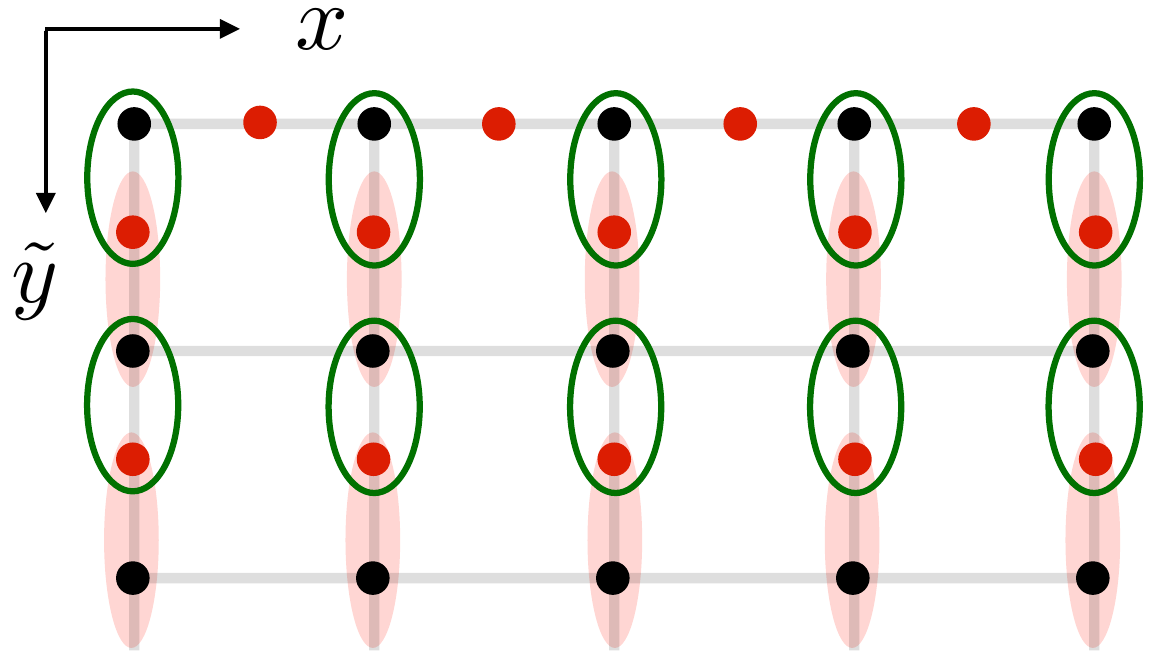}
\end{equation}  
where the $q$ gate is defined as 
\begin{equation}\label{}
\includegraphics[width=4cm]{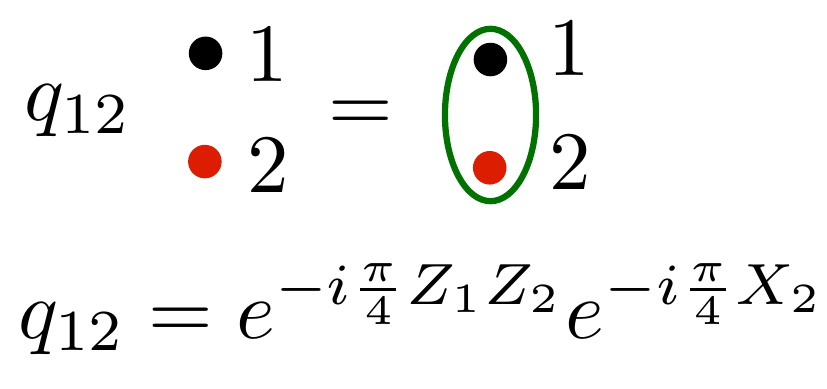}
\end{equation}  
Using the explicit form of this circuit, one finds that the output state is specified by the following stabilizers

\begin{equation}\label{2d_ghz_cluster}
\includegraphics[width=4.5cm]{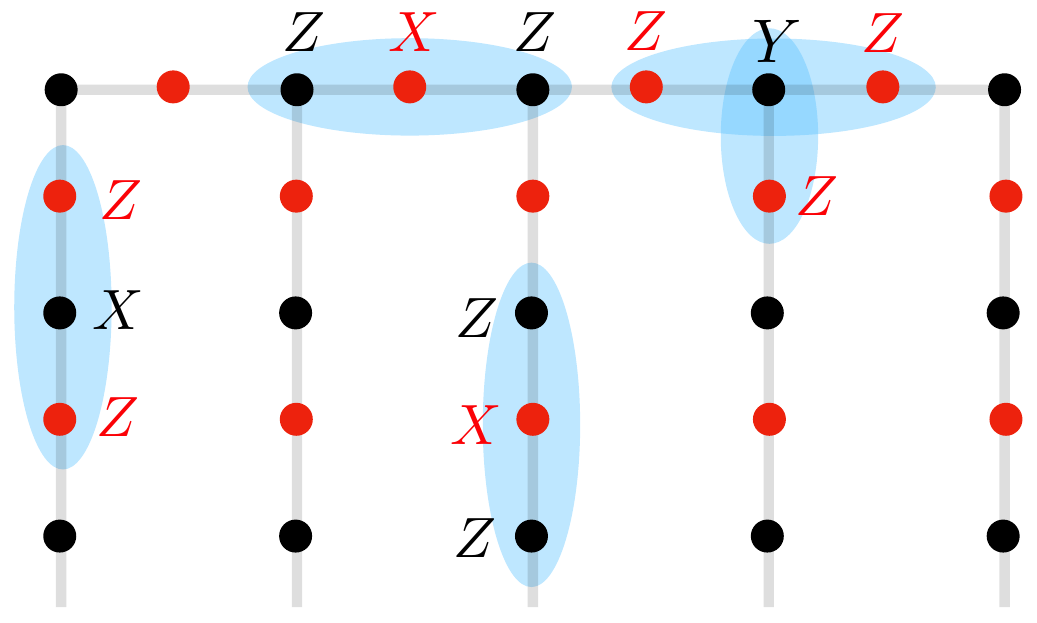}
\end{equation}

Note that in addition to the $ZXZ$ type stabilizer, there is a four-body $YZZZ$ stabilizer at junctions. A subsequent projection of the red qubits to $\ket{+}$ then gives rise to the 2d GHZ state. 

The above process resembles the well-known protocol for generating the 2d GHZ state by measuring a 2d cluster state \cite{Raussendorf_05,tantivasadakarn2021long,lu2022measurement}. Specifically, consider a 2d lattice with periodic boundary conditions and every vertex and every edge hosting a qubit, one defines a cluster state uniquely specified by the vertex stabilizers $X_v \prod_{ e, ~  v  \in \partial e      } Z_e$ and the edge stabilizers $X_e \prod_{v,~ v   \in \partial e}Z_v$. The first term takes the form $X_v ZZZZ$, i.e. the product of a Pauli-X at the vertex $v$ and its four neighboring Pauli-Zs on edges, and the second term takes the form $X_e ZZ$, i.e. the product of a Pauli-X at the edge $e$ and its two neighboring Pauli-Zs on vertices. Projecting all the edge qubits to $\ket{+}$ exactly leads to the 2d GHZ state on the vertices. In fact, one can first disentangle the qubits on the horizontal-oriented edges by projecting them to $\ket{0}$ (except for the edges along the green line): 
\begin{equation}\label{}
\includegraphics[width=6cm]{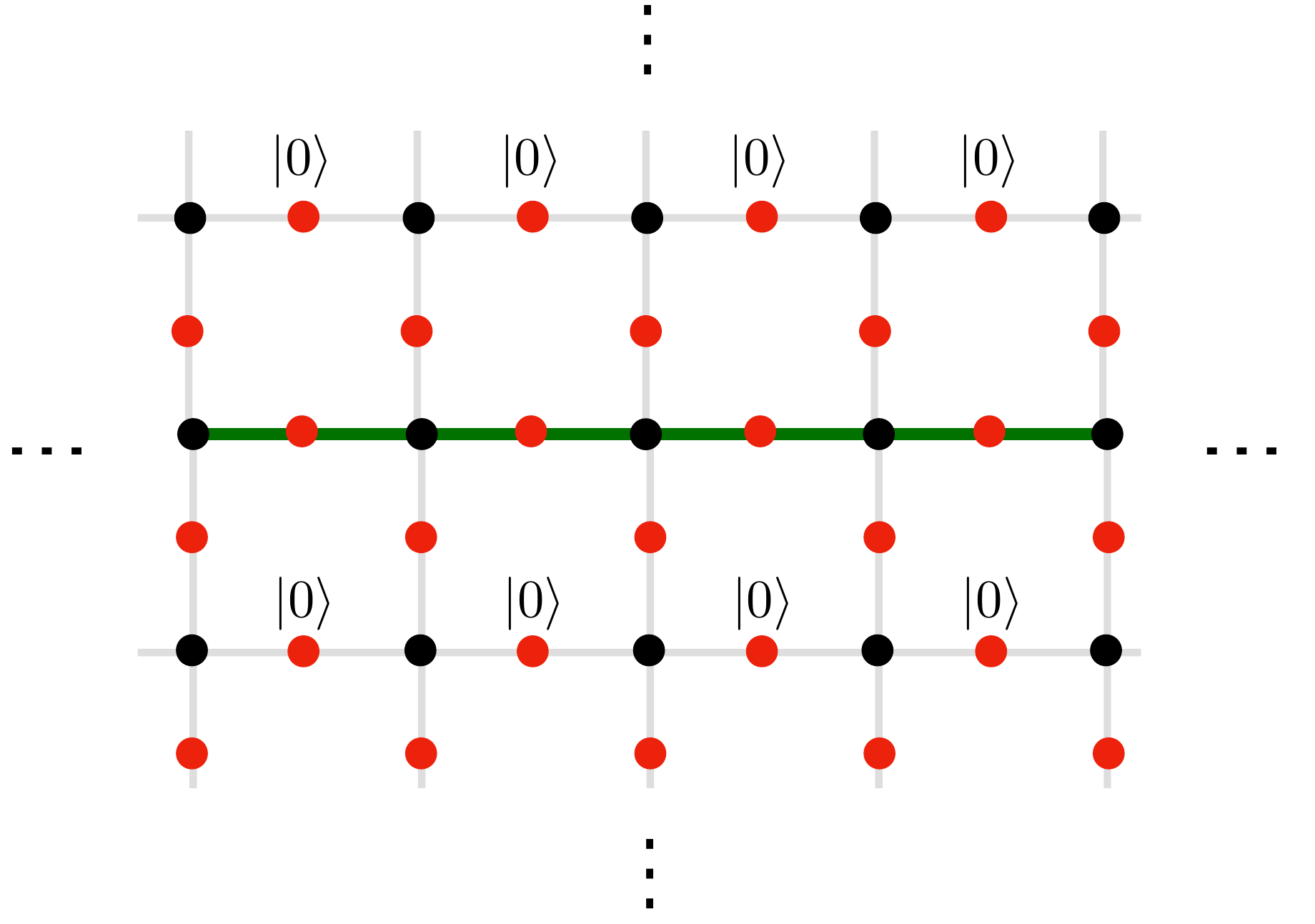}
\end{equation} 
The remaining qubits then form a cluster state with a comb-like geometry: 
\begin{equation}\label{}
\includegraphics[width=6cm]{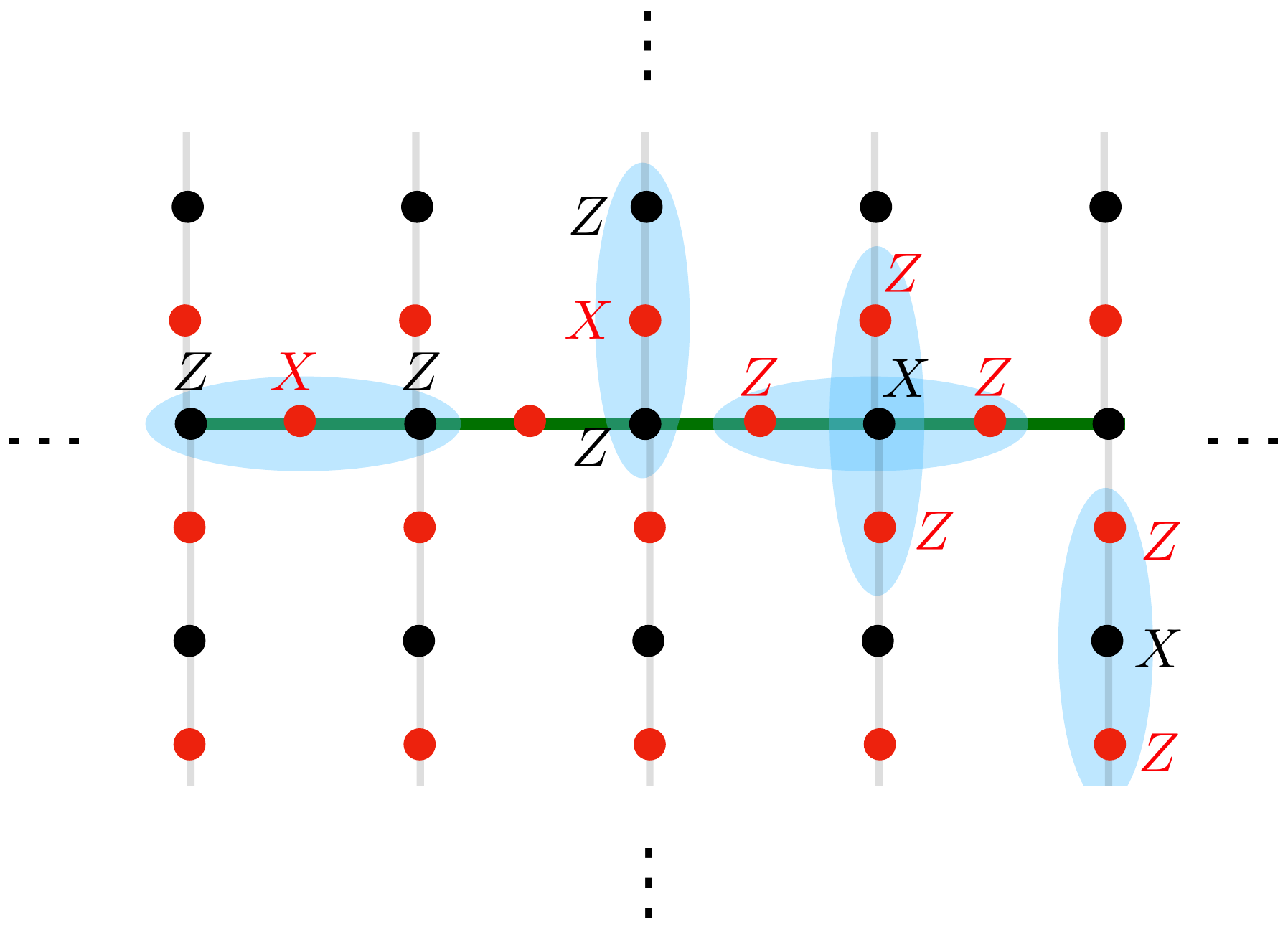}
\end{equation} 
This comb-like cluster state resembles the output state of our shallow circuit (Eq.\ref{2d_ghz_cluster}), and it is sufficient to generate the 2d GHZ state upon projecting edge (red) qubits to $\ket{+}$.

\section{The dual Q-circuit for 2d toric-code state}\label{app:2dtoric}
In the main text, we have shown that the toric code state can be prepared with the following circuit, where the first unit cell sequentially interacts with the $i$-th unit cell: 

\begin{equation}\label{}
\mathcal{U} = \prod_{i=2}^{L_y} U_{1,i}
\end{equation} 

with 
\begin{equation}\label{append_toric_gate}
\includegraphics[width=7.5cm]{toric_unit_gate.pdf}
\end{equation} 
Now we show how to obtain the corresponding dual-$Q$ circuit by exchanging the $y$ direction and the time direction. For this purpose, it may be useful to recall the mapping in the 1+1D spacetime, depicted in the following figure:
\begin{equation*}
\includegraphics[width=8cm]{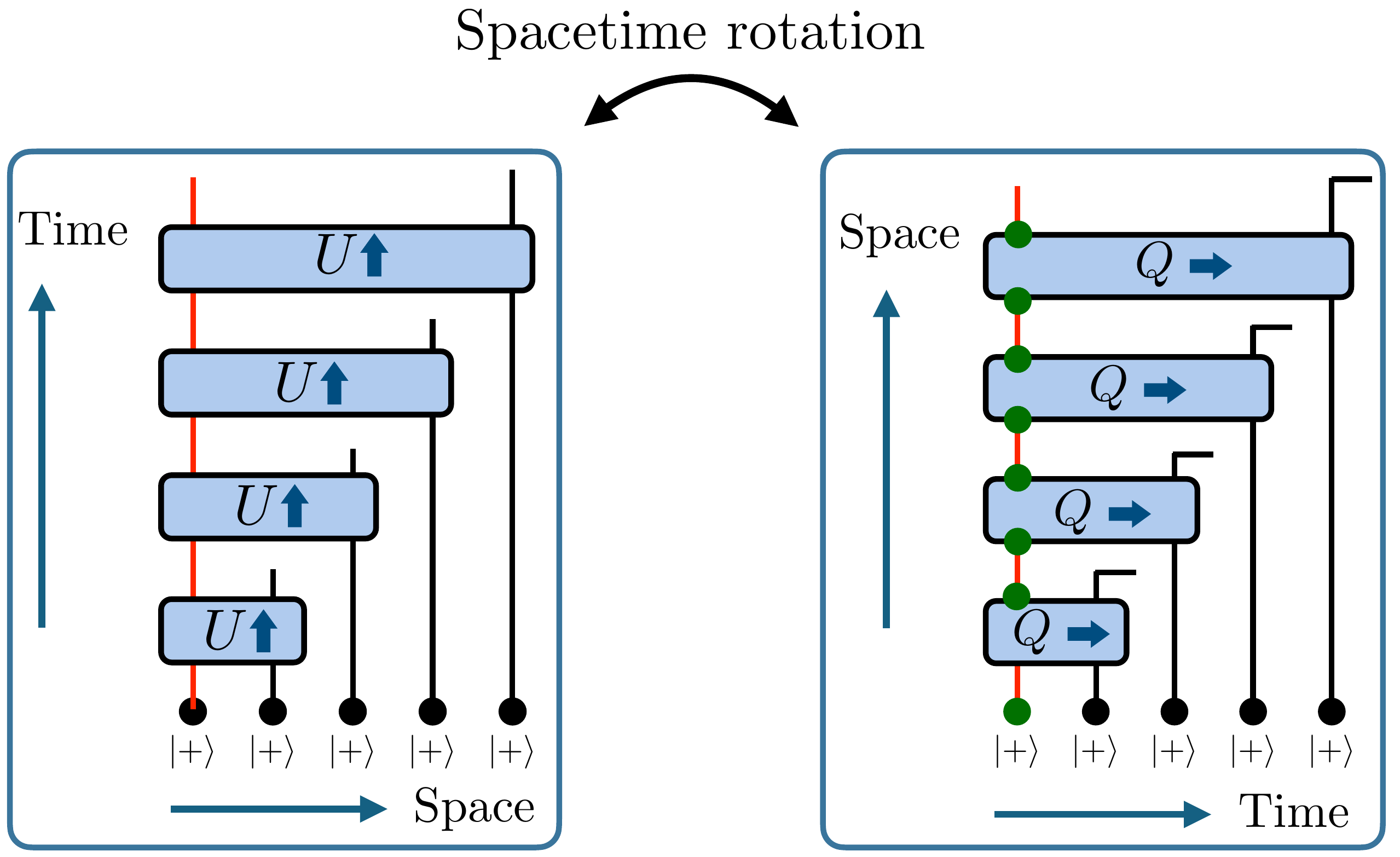}
\end{equation*} 
where the worldline of the first qubit in the sequential circuit maps to decoupled Bell-pairs in the bulk and dangling qubits on the boundary in the dual-$Q$ circuit. By treating each leg as a unit cell consisting of qubits along the $x$ direction, one can obtain the following shallow dual-$Q$ circuit in (2+1)-dimensional spacetime for the toric code:

\begin{align*}
\includegraphics[width=6cm]{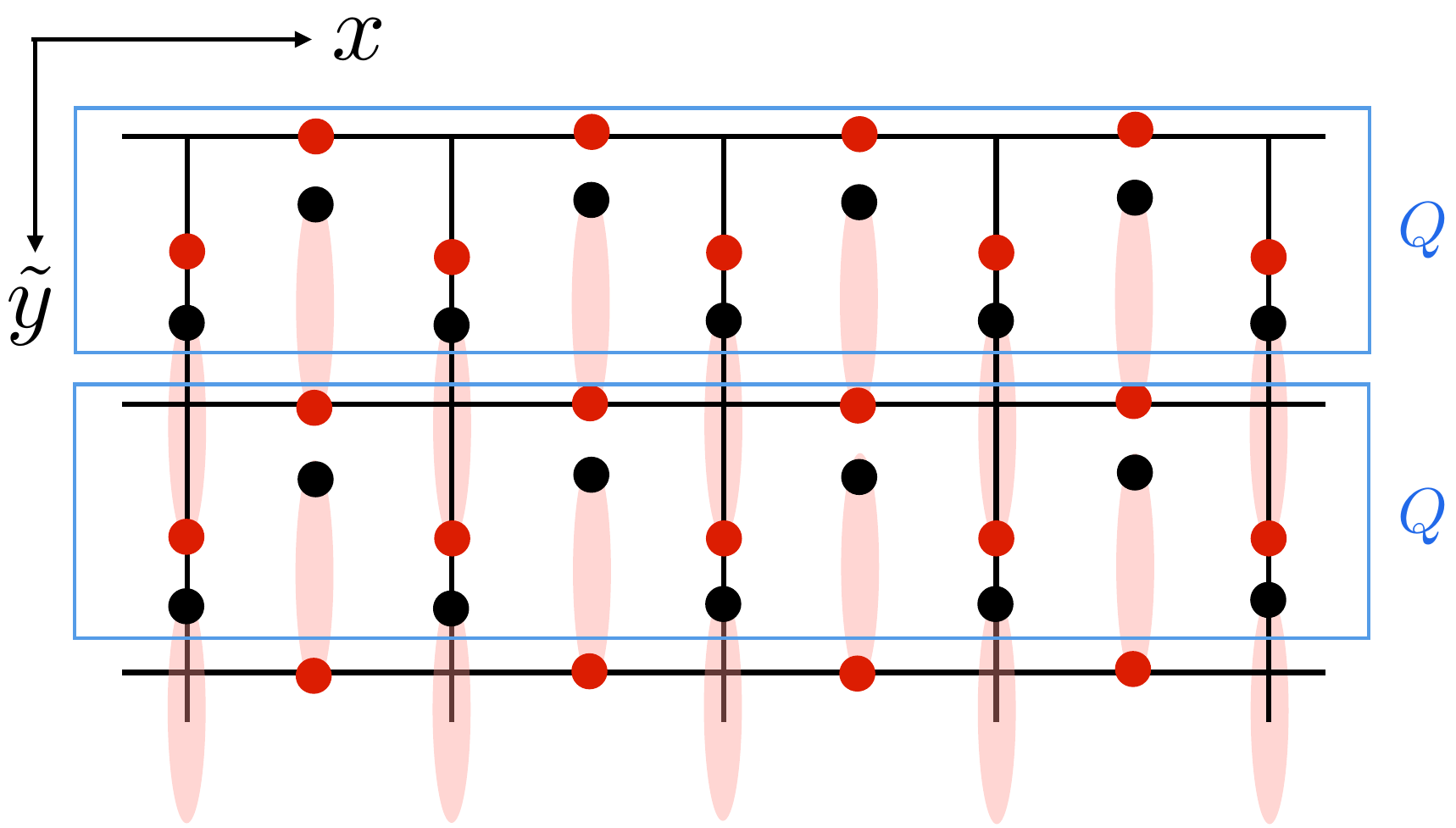}
\end{align*} 
Specifically, the top boundary consists of qubits in $\ket{+}$ and the bulk is given by decoupled Bell pairs (marked by the pink ovals) extending along $\tilde{y}$ direction (i.e., the original time direction). The qubits in red color eventually will be projected to $\ket{+}$ to match the initial state $\ket{+}$ in the sequential circuit. As discussed in Appendix.\ref{app:dualcircuit}, the $Q$ gate within the unit cell can be written as 

\begin{equation}
Q= (\text{SWAP}_{\text{red,black}}) ~~ q, 
\end{equation}
where $q$ is given by the partial transpose of the unitary gate $\text{SWAP}_{1,i} U_{1,i}$ defined in Eq.\ref{append_toric_gate}. Specifically, the $q$ gate is expressed as follows:

\begin{equation}
    \includegraphics[width=6.8cm]{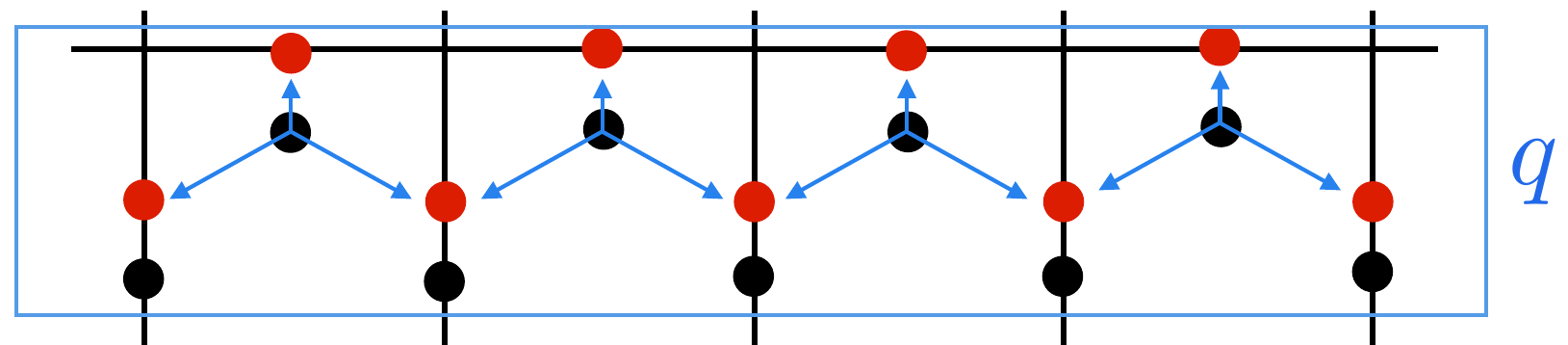}
\end{equation} 
which is the product of the four-body gate along the $x$ direction: 
\begin{equation}
    \includegraphics[width=7.8cm]{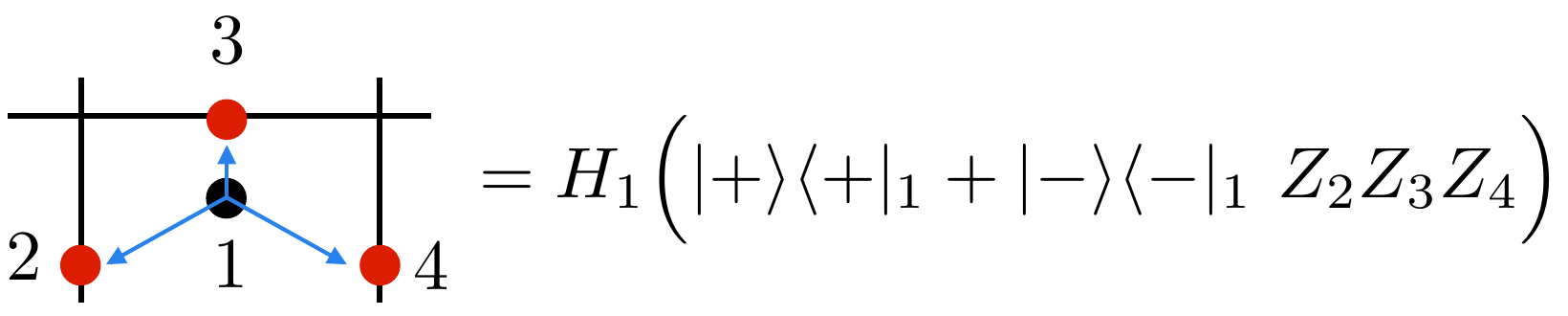}
\end{equation} 
Note that in contrast to the gate in the sequential circuit, where the Hadamard gate is applied before the controlled operation (e.g. see Eq.\ref{eq:toric_U_p}), here the Hadamard gate in the dual-$Q$ circuit is applied after the controlled operation, which is the consequence of taking a partial transpose. With the application of the $q$ gate, followed by swapping the neighboring red and black qubits, subsequently projecting the red qubit to $\ket{+}$ leads to the toric code for the black qubits, which now live on the edges.

\section{Spacetime duality for fractal symmetry-breaking states}\label{sec:fractal}
\begin{figure}
\includegraphics[width=0.45\textwidth]{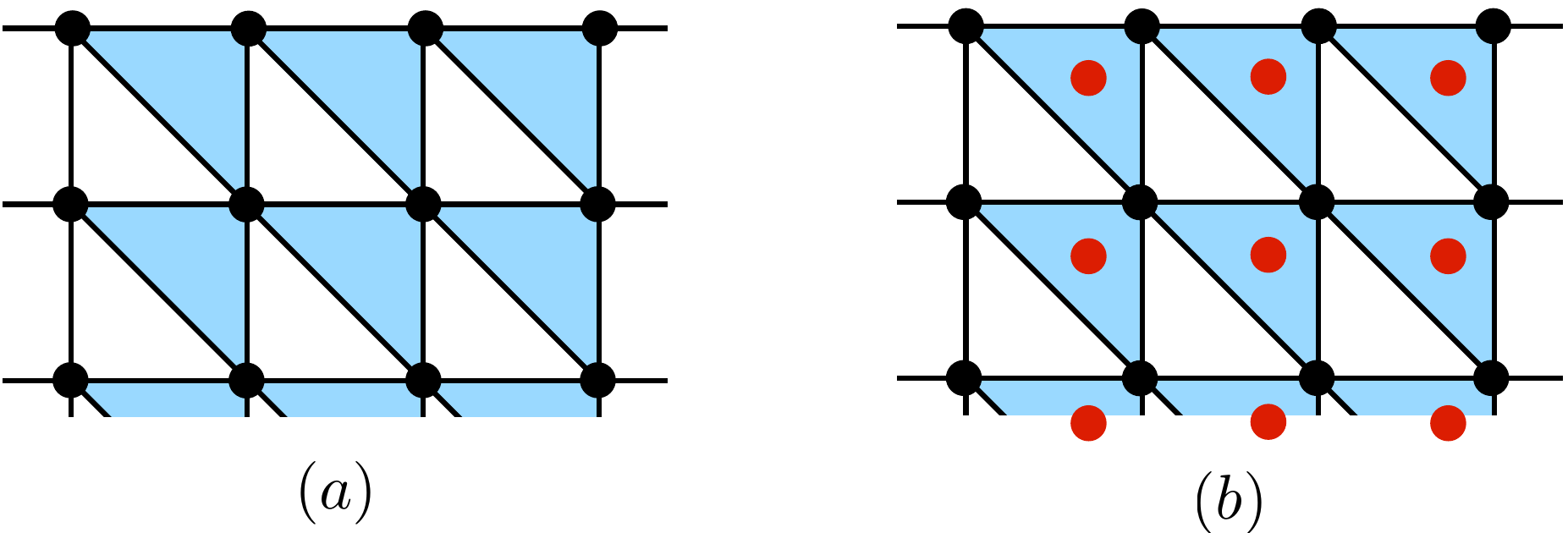}
    \caption{(a) The Newman–Moore model defined on a triangular lattice with one qubit per site, and qubits interact through the three-body Pauli-Z terms on every downward (blue) triangle (see Eq.\ref{eq:newman_moore}). (b) Under a spacetime duality, a sequential circuit is mapped to a shallow dual-$Q$ circuit that outputs a fractal-symmetric SPT state with the stabilizers shown in Eq.\ref{eq:fractal_stabilizer}. It can be interpreted as the gauged version of the Newman–Moore model, where the red qubits at the centers of downward triangles represent the gauge fields, and the black qubits at the vertices represent the matter fields. Projecting the gauge fields to $\ket{+}$ then gives rise to the fractal symmetry-breaking order on the matter fields.}
    \label{fractal}
\end{figure}

In this section, we discuss the spacetime duality between the sequential circuit and the corresponding measurement-feedback circuit that generates a fractal symmetry-breaking state. Specifically, we focus on the Newman–Moore Hamiltonian on a triangular lattice \cite{newman1999glassy} - a canonical example that exhibits the fractal symmetry breaking in the ground states:

\begin{equation}\label{eq:newman_moore}
    H=-\sum_{(ijk)_\triangledown}Z_i Z_j Z_k,
    \end{equation}
The stabilizer is a three-body Ising interaction on each downward triangle indicated by the blue regions in Fig.\ref{fractal}(a). The ground states exhibit a pattern in which every downward triangle has an even number of down-spin in the computational basis \cite{newman1999glassy}. While the three-body breaks the global $\mathbb Z_2$ symmetry, the model displays an exotic fractal symmetry generated by a cellular automaton. In particular, a symmetry operator on a Sierpinski gasket $\prod_{i\in\mathrm{ST}} X_i$ commutes with the three-body interaction in Eq.\ref{eq:newman_moore} (up to corner terms). Thus, the Hamiltonian displays a fractal $\mathbb{Z}_2$ symmetry with the Ising charge conserved on the subextensive manifold on Sierpinski triangles.

Starting from the trivial product state  $\ket{+++...}$, we can obtain a symmetry-breaking ground state using local unitary gates to enforce the constraint $Z_i Z_j Z_k=1 $ on every downward triangle. Drawing inspiration from the sequential circuit for the toric code (see Eq.\ref{eq:toric_U_p}), we here define the three-body gate on every downward triangle
\begin{equation}
\includegraphics[width=7.5cm]{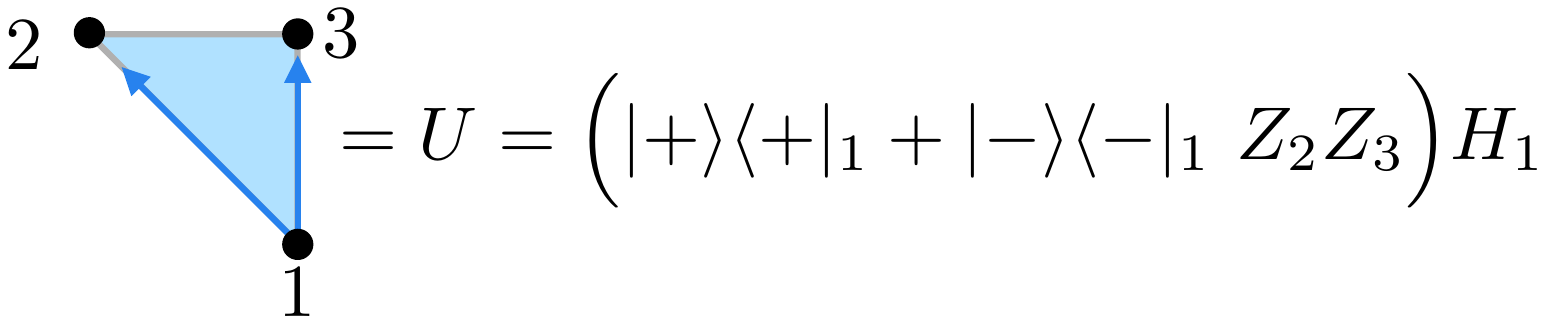}
\end{equation} 
Namely, the qubit $1$ is the controlled qubit, and the qubits 2 and 3 are the target. Under the application of this gate, one finds

\begin{equation}
\ket{+}_1   \ket{\phi}_{\overline{1}} ~~\to~ ~\frac{1+Z_1 Z_2Z_3}{2}\ket{+}_1   \ket{\phi}_{\overline{1}}
\end{equation}
so that the triangle constraint can be satisfied. Importantly, these gates on the same row commute, so they can be applied simultaneously along a horizontal strip. Iterating this procedure along the vertical direction then gives a sequential circuit that generates the fractal symmetry pattern in 2d. As in the previous example, we can rearrange the sequential circuit to $\mathcal{U} = \prod_{i=2}^{L_y}  U_{1,i}$ so that the first unit cell sequentially interacts with the remaining unit cell, with each unit cell defined as a 1d line of qubits along horizontal direction. In particular, the sequential circuit $\mathcal{U}$ implements the following operator mapping:

\begin{equation}\label{eq:fractal_mapping}
\includegraphics[width=7.8cm]{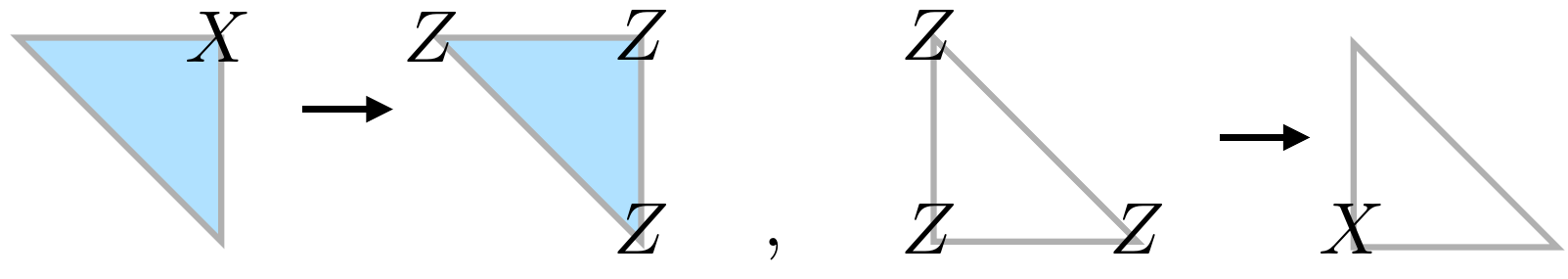}
\end{equation} 
which realizes the Kramers-Wannier duality associated with the fractal symmetry \cite{devakul2019fractal,zhou2021fractal}.

Now we discuss the corresponding measurement-feedback circuit by exchanging the vertical spatial direction and the time direction. The input and output of a qubit in the sequential circuit are mapped to two qubits under a spacetime rotation (see Fig.\ref{fractal}(b)), and the sequential unitary gate maps to a dual-$Q$ circuit that outputs a short-range entangled stabilizer state $\ket{\psi_{\text{out}}}$ with following stabilizers:
\begin{equation}\label{eq:fractal_stabilizer}
\includegraphics[width=4.5cm]{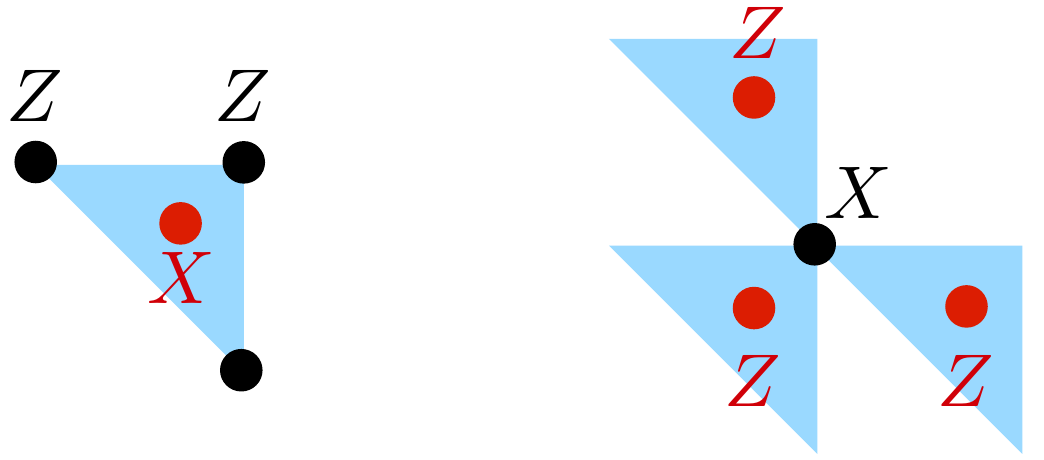}
\end{equation} 
These stabilizers can be found using the explicit form of the dual-$Q$ circuit. Alternatively, they are immediately given by the Kramers-Wannier duality associated with the fractal symmetry (Eq.\ref{eq:fractal_mapping}). Notably, these stabilizers indicate that the $\ket{\psi_{\text{out}}}$ exhibits a fractal symmetry-protected topological (SPT) order as discussed in Ref.\cite{devakul2019fractal}. It can also be viewed as a gauged version of the Newman–Moore model, where gauge fields (red qubits) reside at triangle centers and couple minimally to the original $ZZZ$ interactions of the Newman–Moore model.   Projecting the gauge qubits onto the $X=1$ state effectively Higgses the gauge field, driving the system into a fractal symmetry-breaking phase.

 \end{document}